\documentclass[aip,amsmath,amssymb,reprint]{revtex4-1}

\usepackage{graphicx}
\usepackage{dcolumn}
\usepackage{bm}

\usepackage[utf8]{inputenc}
\usepackage[T1]{fontenc}
\usepackage{mathptmx}

\usepackage{hyperref} 

\newcommand\mnras{{MNRAS}}     
\newcommand\aap{{A\&A}}        
\newcommand\pasj{{PASJ}}       
\newcommand\apjl{{ApJ}}        
\newcommand\apjs{{ApJS}}       

\newcommand\solphys{{Sol Phys}}

\newcommand*\jgr{J. Geophys. Res}  

\usepackage[dvipsnames]{xcolor}

\graphicspath{{./}{figures/}}

\begin{document}

\preprint{AIP/123-QED}

\title[Coalescence Instability in Chromospheric Partially Ionised Plasmas]{Coalescence Instability in Chromospheric Partially Ionised Plasmas}

\author{Giulia Murtas}
 \email{gm442@exeter.ac.uk}
 \affiliation{College of Engineering, Mathematics and Physical Sciences, Harrison Building, Streatham Campus, University of Exeter, North Park Road, Exeter, UK, EX4 4QF}
 
\author{Andrew Hillier}
\affiliation{College of Engineering, Mathematics and Physical Sciences, Harrison Building, Streatham Campus, University of Exeter, North Park Road, Exeter, UK, EX4 4QF}

\author{Ben Snow}
\affiliation{College of Engineering, Mathematics and Physical Sciences, Harrison Building, Streatham Campus, University of Exeter, North Park Road, Exeter, UK, EX4 4QF}

\date{\today}

\begin{abstract}
Fast magnetic reconnection plays a fundamental role in driving explosive dynamics and heating in the solar chromosphere. The reconnection time scale of traditional models is shortened at the onset of the coalescence instability, which forms a turbulent reconnecting current sheet through plasmoid interaction. In this work we aim to investigate the role of partial ionisation on the development of fast reconnection through the study of the coalescence instability of plasmoids. Unlike the processes occurring in fully ionised coronal plasmas, relatively little is known about how fast reconnection develops in partially ionised plasmas of the chromosphere. We present 2.5D numerical simulations of coalescing plasmoids in a single fluid magnetohydrodynamic (MHD) model, and a two-fluid model of a partially ionised plasma (PIP). We find that in the PIP model, which has the same total density as the MHD model but an initial plasma density two orders of magnitude smaller, plasmoid coalescence is faster than the MHD case, following the faster thinning of the current sheet and secondary plasmoid dynamics. Secondary plasmoids form in the PIP model where the effective Lundquist number $S = 7.8 \cdot 10^3$, but are absent from the MHD case where $S = 9.7 \cdot 10^3$: these are responsible for a more violent reconnection. Secondary plasmoids also form in linearly stable conditions as a consequence of the non-linear dynamics of the neutrals in the inflow. In the light of these results we can affirm that two-fluid effects play a major role on the processes occurring in the solar chromosphere.

\end{abstract}

\maketitle

\section{Introduction}
\label{sec:intro}

Magnetic reconnection is a process responsible for explosive events in astrophysical, space and laboratory plasmas allowing plasmas to break free from the frozen-in constraint imposed by ideal magnetohydrodynamics (MHD) \citep{2000mrmt.conf.....P}. The magnetic field lines reconnect through a narrow diffusion region, enabling the  conversion of stored magnetic energy into kinetic and thermal energy, and particle acceleration \citep{2000mrp..book.....B, 2000mrmt.conf.....P}.

A classical description for reconnection is provided by the Sweet-Parker model \citep{1957JGR....62..509P,1958IAUS....6..123S}, which describes steady-state reconnection in a long diffusion region of length $L$ and width $\delta$. The reconnection time scale goes as $\tau_{ _{SP}} \sim L / \delta \sim S^{1/2}$, where $S = L v_{A} / \eta$ is the Lundquist number, $\eta$ is the diffusivity and $v_A$ is the Alfv\'en speed.

The presence of localised, transient outflows in the solar chromosphere (such as the chromospheric jets \citep{2007Sci...318.1591S,2011ApJ...731...43N}) is believed to be the result of the onset of magnetic reconnection. Further evidence for magnetic reconnection is provided by the identification of bubbles of plasma in the outflow of chromospheric jets and UV bursts \citep{2011PhPl...18k1210S,2012ApJ...759...33S,2020arXiv200911475G}, which are generally interpreted to be plasmoids.

Plasmoids are commonly present in reconnecting systems \citep{1963PhFl....6..459F, 2001ApJ...551..312T, 2001EP&S...53..473S, 2009PhRvL.103j5004S, 2007PhPl...14j0703L, 2012PhPl...19d2303L, 2016PPCF...58a4021L} and they are believed to play a major role in speeding up reconnection by having an influence on the variation of the current sheet size \citep{1989ApJ...340..550Z}. Under the formation of plasmoids, the current sheet breaks into fragments and the resulting high current densities in each of these sections facilitate a high reconnection rate \citep{2015ApJ...799...79N}. Plasmoid formation due to the instability of Sweet-Parker current sheets has been extensively examined through numerical studies \citep{1984PhFl...27..137P, 1984PhFl...27.1207S, 1986PhFl...29.1520B, 1986JGR....91.6807L, 1991PhFlB...3.1927J, 1995PhPl....2..388U, 2005PhRvL..95w5003L, 2016PPCF...58a4021L}, and found in direct imaging observations of solar flares \cite{2012ApJ...745L...6T}. Several works proved that in fully ionised plasmas it is possible to trigger plasmoid formation in thin current sheets ($\delta/L \ll 1$) for a critical Lundquist number $\sim (10^3 - 10^4)$ \citep{1986PhFl...29.1520B, 2005PhRvL..95w5003L, 2009PhRvL.103j5004S, 2009PhPl...16k2102B, 2009PhPl...16l0702C, 2010PhPl...17f2104H, 2012PhPl...19d2303L, 2010PhPl...17e2109N, 2012PhPl...19g2902N, 2013PhPl...20f1206N, 2015ApJ...799...79N, 2016PPCF...58a4021L}. Numerical studies \cite{2009PhPl...16k2102B} found an upper limit for the critical Lundquist number $\sim 3 \cdot 10^4$. Above this critical value of $S$ the current sheet becomes rapidly unstable to the resistive tearing instability, forming a chain of plasmoids along the current sheet \cite{2007PhPl...14j0703L}.

In astrophysical plasmas with very large $S$ the onset of the tearing mode takes place in current sheets that collapse to a thickness of the order of $S^{-1/3}$, larger than the Sweet-Parker thickness of $S^{-1/2}$. The trigger of fast reconnection before a Sweet-Parker-type configuration can form was examined in detail by several studies \cite{2014ApJ...780L..19P, 2015ApJ...801..145T, 2015ApJ...813L..32T,2018PhPl...25c2113P}.

Many works proved that the critical Lundquist number and aspect ratio for the onset of the tearing instability are affected by changes in the initial setup, which might result in a discrepancy of orders of magnitude in the critical Lundquist number \cite{2013PhPl...20f1206N, 2017ApJ...849...75H, 2018ApJ...868..144N}. A major role is played by the initial current sheet configuration \cite{2012PhPl...19g2902N,2013PhPl...20f1206N,2017ApJ...849...75H} as well as by the amplitudes of viscosity and perturbation noises \cite{2016PhPl...23c2111C,2016PhPl...23j0702C,2017ApJ...850..142C,2017ApJ...849...75H}, and the plasma $\beta$\cite{2012PhPl...19g2902N,2013PhPl...20f1206N,2017ApJ...849...75H,2018ApJ...868..144N}. The variation of the critical $S$ as a function of the initial noise, investigated in some works\cite{2017ApJ...849...75H}, covers several orders of magnitude (from $S \sim 10^3$ to $S \sim 10^6$). The role of the configuration of the simulation domain in affecting the threshold of the critical Lundquist number has been pointed out in studies of magnetic reconnection in fully ionised plasmas \cite{2012PhPl...19g2902N,2013PhPl...20f1206N}. In these works it was showed that a 2.5D simulation of magnetic reconnection with a force-free current sheet and uniform plasma pressure as initial conditions lead to a much lower critical $S$ than those obtained in 2D cases with an initial Harris current sheet and nonuniform plasma pressure. Their results are consistent with the findings of a recent study \cite{2018ApJ...868..144N}, where magnetic reconnection is examined on different spatial scales in weakly ionized plasmas by using a reactive 2.5D multi-fluid plasma-neutral model\cite{2012ApJ...760..109L,2013PhPl...20f1202L}.

The nonlinear tearing mode shows the development of an important secondary instability called the coalescence instability \cite{2000mrp..book.....B}. This instability is driven by the coalescence of neighboring plasmoids sharing an X-point and results from the attractive forces between parallel currents. The coalescence instability is characterised by two different phases:  ideal, and resistive. The ideal phase has a growth rate that is almost independent of $\eta$ \citep{1986ITPS...14..929T}. The resistive phase is driven by the current sheet reconnection \citep{2000mrp..book.....B}. In a single-fluid MHD approach, the coalescence instability produces a reconnecting current sheet by driving plasmoid interaction, and is a key process that might explain fast reconnection without the need of anomalous resistivity terms to be added into the system \citep{2015ApJ...799...79N}.

In the solar chromosphere plasmas, as well as many other plasmas found in the universe, are partially ionised, and their ionisation degree falls in the range $10^{-4}-10^{-1}$ (\citep{1981ApJS...45..635V,1986A&A...154..231P,Khomenko2008, 2015ApJ...799...79N}). Multi-fluid effects linked to the different behaviour of the particle species must be taken into account for a correct physical description of this atmospheric layer. The low chromospheric densities do not allow a complete collisional coupling between ions and neutral species: the low ion fraction allows the fewer charged particles to be coupled to the neutrals, but the neutrals may not be completely coupled to the ions. A partial coupling between the two species results in the presence of relative motions. While ideal MHD equations are applicable to fully ionised plasmas, two-fluid effects should be described by taking into account the relative motions between plasma and neutral components in the solar chromosphere. The role of partial ionisation on the onset of magnetic reconnection and development of the resistive tearing instability was investigated in many studies \cite{1989ApJ...340..550Z,2011PhPl...18k1210S,2015PASJ...67...96S,2011PhPl...18k1211Z,2012ApJ...760..109L,2013PhPl...20f1202L, 2015ApJ...799...79N}. In a system where the two fluids are coupled through elastic collisions and subject to ionisation and recombination, the reconnection rate in the coalescence process depends on the ion fraction \cite{2008A&A...486..569S, 2009ApJ...691L..45S}.

The role of fast magnetic reconnection in triggering events in the solar chromosphere is undoubtedly fundamental. An additional complexity comes from the partially ionised nature of the solar chromosphere. In this paper we discuss the role of partial ionisation on magnetic reconnection through the study of plasmoid coalescence, with the aim of understanding to what extent the two fluid effects influence such process. In order to do this, we first compare two reference MHD and PIP simulations (Section \ref{sec:simulations}), and then investigate in more detail how two-fluid properties affect the coalescence instability through a parameter survey (Section \ref{sec:parameters}). In Section \ref{sec:discussion} our results are connected back to the physical scales of reconnection in the solar chromosphere.

\section{Methods}
\label{sec:methods}

We perform simulations of the coalescence instability in fully (MHD) and partially ionised plasmas (PIP), using the (P\underline{I}P) code \citep{2016A&A...591A.112H}. The code makes use of a four-step Runge-Kutta and a fourth-order central difference scheme.
The fully ionised plasma consists of a single-fluid model of a hydrogen plasma. The partially ionised plasma environment is simulated through a two-fluid model consisting of two separate sets of equations describing a neutral fluid and a charge-neutral ion-electron plasma which are collisionally coupled. The equations are derived from those found in previous models \cite{1965RvPP....1..205B,2012ApJ...760..109L,2012PhPl...19g2508M}.

All sets of equations are non-dimensionalised. The choice of this particular normalisation is performed in order to have a dependency on characteristic length scales, which are comparable to the size of the plasmoids involved in the merging. This allows the model to be applied to plasmoids at different scales in the solar chromosphere, from a few meters to a few hundred kilometers, as all the quantities scale together with the plasmoid size. Such normalisation also affects the collisional coupling between the two fluids. As a characteristic dimensional time scale $\tau$ can be derived from the physical plasmoid size and the characteristic speed in the solar chromosphere (in our case this is the sound speed), the non-dimensional collisional frequency is easily re-scaled back to physical quantities by dividing it by $\tau$. Further details on the non-dimensionalisation are provided at the end of this Section and in Section \ref{subsec:initial_conditions}.

The neutral fluid is described by non-dimensional inviscid hydrodynamics equations:
\begin{equation}
    \frac{\partial \rho_n}{\partial t} + \nabla \cdot (\rho_n \textbf{v}_n) = 0,
\end{equation}
\begin{multline}
    \frac{\partial}{\partial t}(\rho_n \textbf{v}_n) + \nabla \cdot (\rho_n \textbf{v}_n \textbf{v}_n + p_n \textbf{I}) = \\
    - \alpha_c (T_n , T_p ,v_D ) \rho_n \rho_p (\textbf{v}_n - \textbf{v}_p ),
    \label{eq:force_neutral}
\end{multline}
\begin{multline}
    \frac{\partial e_n}{\partial t}  + \nabla \cdot [\textbf{v}_n (e_n + p_n)] = \\
    - \alpha_c (T_n , T_p ,v_D) \rho_n \rho_p \Bigg[ \frac{1}{2} (\textbf{v}_n ^2 -\textbf{v}_p ^2) + \frac{3}{2} \Bigg(\frac{p_n}{\rho_n} - \frac{p_p}{2 \rho_p} \Bigg)\Bigg],
    \label{eq:neutral_energy_2}
\end{multline}
\begin{equation}
    e_n = \frac{p_n}{\gamma -1} + \frac{1}{2} \rho_n v_{n}^{2},
    \label{neutral_energy}
\end{equation}
\begin{equation}
    T_n = \gamma \frac{p_n}{\rho_n},
    \label{eq:neutral_temperature}
\end{equation}
while inviscid resistive magnetohydrodynamics relations govern the plasma, here stated in non-dimensional form:
\begin{equation}
    \frac{\partial \rho_p}{\partial t} + \nabla \cdot (\rho_p \textbf{v}_p) = 0,
\end{equation}
\begin{multline}
    \frac{\partial}{\partial t}(\rho_p \textbf{v}_p) + \nabla \cdot \Bigg(\rho_p \textbf{v}_p \textbf{v}_p + p_p \textbf{I} - \textbf{B} \textbf{B} +  \frac{\textbf{B}^2}{2} \textbf{I} \Bigg) = \\
    \alpha_c (T_n , T_p ,v_D) \rho_n \rho_p (\textbf{v}_n - \textbf{v}_p ),
    \label{eq:force_plasma}
\end{multline}
\begin{multline}
    \frac{\partial}{\partial t} \Bigg( e_p + \frac{B^2}{2}\Bigg) + \nabla \cdot [ \textbf{v}_p (e_p + p_p) +\\
    -(\textbf{v}_p \times \textbf{B}) \times \textbf{B} + \eta (\nabla \times \textbf{B}) \times \textbf{B}] = \\
    \alpha_c (T_n , T_p,v_D ) \rho_n \rho_p \Bigg[\frac{(\textbf{v}_n^2 - \textbf{v}_p^2 )}{2} + \frac{3}{2} \Bigg(\frac{p_n}{\rho_n} - \frac{p_p}{2 \rho_p} \Bigg)\Bigg],
    \label{plasma_energy_equation}
\end{multline}
\begin{equation}
    \frac{\partial \textbf{B}}{\partial t} - \nabla \times (\textbf{v}_p \times \textbf{B} - \eta \nabla \times \textbf{B}) = 0,
\end{equation}
\begin{equation}
    e_p = \frac{p_p}{\gamma -1} + \frac{1}{2} \rho_p v_{p}^{2},
    \label{plasma_energy}
\end{equation}
\begin{equation}
    \nabla \cdot \textbf{B} = 0,
\end{equation}
\begin{equation}
    T_p = \gamma \frac{p_p}{2\rho_p}.
    \label{eq:plasma_temperature}
\end{equation}
In the equations above the subscripts $p$ and $n$ refer respectively to the ion-electron plasma and the neutral fluid, $\mathbf{v}$, $p$, $\rho$, $T$ and $e$ are the velocity, gas pressure, density, temperature and internal energy of each species, $\gamma= 5/3$ is the adiabatic index and \textbf{B} is the magnetic field. Both fluids follow the ideal gas law. The factor 2 in Equation (\ref{eq:plasma_temperature}) is to take into account the electron pressure. The parameter $\alpha_c$, given by Equation (\ref{alpha_c}), is associated to the two fluids collisional coupling. This particular formulation of $\alpha_c$ is new to models of magnetic reconnection in partially ionised plasmas, and it accounts for the increased amount of collisions at supersonic drift velocities. The non-dimensional expression for $\alpha_c$ \cite{1986MNRAS.220..133D} including charge exchange \cite{2018ApJ...869...23Z} is the following:

\begin{equation}
\alpha_c = \alpha_c (0) \sqrt{\frac{T_n +T_p}{2}} \sqrt{1 + \frac{9\pi}{64} \frac{\gamma}{2(T_n + T_p)} v_D^2},
\label{alpha_c}
\end{equation}
where $\alpha_c (0)$ is the initial coupling and $v_D$ = $\mid$ \textbf{v$_n$} - \textbf{v$_p$} $\mid$ is the magnitude of the drift velocity between the neutral components and the hydrogen plasma. When the drift velocity becomes bigger than the thermal velocity the particles are subject to a higher number of collisions as they are drifting past each other. The collisional coupling between ions and electrons is represented by setting a small finite diffusivity $\eta$ that is assumed to be spatially uniform and not varying with time. In this work we are not including the Hall effect.

The two systems of equations are non-dimensionalised \citep{2019PhPl...26h2902H} by a reference density $\rho_0$ and the total sound speed $c_s = \sqrt{\gamma (p_n + p_p )/(\rho_n + \rho_p)}$, initially set equal to 1. For the MHD simulation, where the plasma is fully ionised, the initial density and pressure are constant and equal to:
\begin{equation}
    \rho_p = \xi_p \rho_0 = 1,
\end{equation}
\begin{equation}
    p_p = p_0 = \gamma^{-1}.
\end{equation}
For the PIP simulations, the bulk density and pressure are equal to the MHD values:
\begin{equation}
    \rho_n + \rho_p = \xi_n \rho_0 + \xi_p \rho_0 = 1,
\end{equation}
\begin{equation}
    p_n + p_p =  \frac{\xi_n}{(\xi_n + 2 \xi_p)} p_0 + \frac{2 \xi_p}{(\xi_n + 2 \xi_p)} p_0 = \gamma^{-1},
\end{equation}
and they are uniform in all the domain. Initially the two fluids are in thermal equilibrium. 

\subsection{Initial conditions}
\label{subsec:initial_conditions}

\begin{figure}[htb]
    \centering
    \includegraphics[width=\columnwidth,clip=true,trim=1cm 2.7cm 1.5cm 3cm]{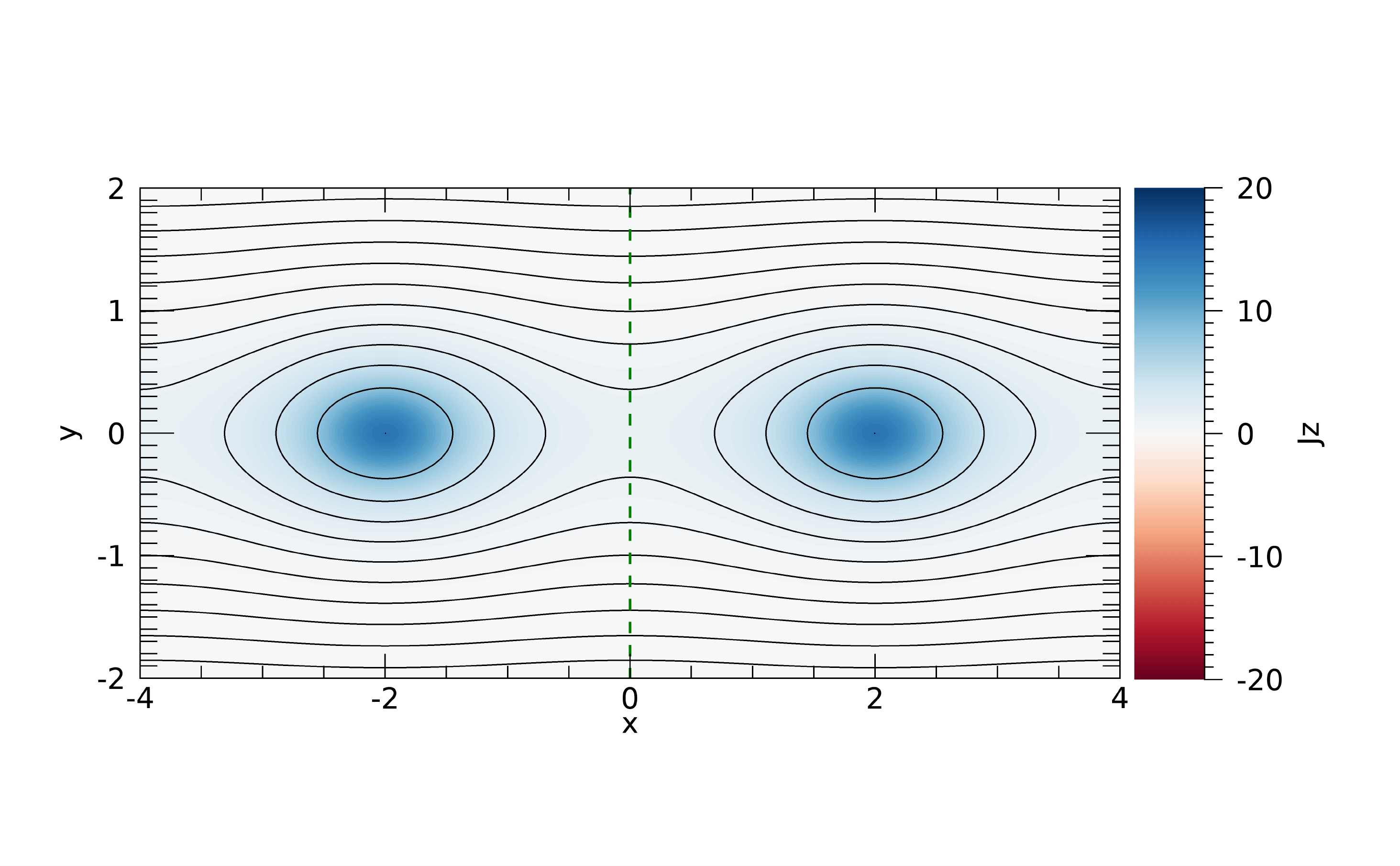}  
    \caption{Initial distribution for the current density $J_z$ ($t = 0$) for both the MHD and the PIP simulations. Two initial plasmoids (blue spots), which are concentrations of positive current, are positioned with their centre on the $x-$axis. The magnetic field lines are displayed in black. The left boundary of the domain for the reference MHD simulation and the cases discussed in the parameter survey is shown in green.}
    \label{fig:initial_conditions}
\end{figure}

The bulk initial conditions of the PIP case are equal to the initial conditions of the MHD case. The initial setup is provided by a force-free modified Fadeev equilibrium \citep{Fadeev_1965,2000mrp..book.....B}. The magnetic scalar potential of the classical 2D Fadeev equilibrium in the $xy-$plane is given by \citep{2000mrp..book.....B}:
\begin{equation}
    \psi(x,y)= \frac{B_{\infty}}{k} \ln[\cosh(ky) + \epsilon \cos(kx)],
    \label{fadeev_streamfunction}
\end{equation}
where $B_{\infty}$ is the field intensity for the limit $|y| \rightarrow \infty$. In our simulations $B_{\infty}$ is equal to $\sqrt{2 \gamma^{-1} \beta^{-1}}$, where plasma $\beta = 2 p/B^2$, $k = \frac{\pi}{2}$ and we set $\epsilon = 0.5$, which corresponds to a moderately peaked current localization at the plasmoid centre, shown in Figure \ref{fig:initial_conditions}. As $\epsilon \rightarrow 0$ there is a weaker localization and a weaker attraction between the plasmoids, while $\epsilon \rightarrow 1$ corresponds to a peaked localization and stronger attraction forces. At the upper limit ($\epsilon = 1$), the current distribution becomes the delta function.

In the mid to upper chromosphere the plasma $\beta$ may become very small. Although the photospheric magnetic field emerging from the convection zone is not force-free, its structure is rearranged by the time it reaches the corona as the non force-free components decay due to the action of chromospheric neutrals \citep{2009ApJ...705.1183A}. It is hence of interest to study the coalescence instability in a regime that is initially force-free. The magnetic field $B_x$ and $B_y$ components from the classic Fadeev equilibrium are not sufficient to satisfy the condition $\textbf{J} \times \textbf{B} = 0$ for a force-free field. Therefore we modify the traditional Fadeev equilibrium by including a component $B_z$, making it force-free. The magnetic field components are shown in Equations (\ref{fadeev_bx_1})-(\ref{fadeev_bz_1}).

\begin{equation}
    B_{x} = - \frac{\sqrt{2 \gamma^{-1} \beta^{-1}} \epsilon \sin(kx)}{\cosh(ky) + \epsilon \cos(kx)},
    \label{fadeev_bx_1}
\end{equation}

\begin{equation}
    B_{y} = - \frac{\sqrt{2 \gamma^{-1} \beta^{-1}} \sinh(ky)}{\cosh(ky) + \epsilon \cos(kx)},
    \label{fadeev_by_1}
\end{equation}

\begin{equation}
    B_{z} = \frac{\sqrt{2 \gamma^{-1} \beta^{-1}} \sqrt{1- \epsilon^2}}{\cosh(ky) + \epsilon \cos(kx)}.
    \label{fadeev_bz_1}
\end{equation}
Setting $\epsilon = 0$ in the equations above, the $B_y$ component leads to a current sheet with the characteristic tanh profile of the well known Harris sheet \citep{1962NCim...23..115H}. Our initial conditions for the current density are displayed in Figure \ref{fig:initial_conditions}, and are the same for both MHD and PIP cases.

The Fadeev equilibrium is unstable to the coalescence instability \citep{2000mrp..book.....B}. We hence choose a small perturbation in the velocity of both plasma and neutral components to break the initial equilibrium by pushing neighbouring plasmoids towards each other. The velocity perturbation is given by:
\begin{equation}
    v_{x,p} = v_{x,n} = -0.05 \sin \Bigg(\frac{kx}{2} \Bigg) e^{-y^2} + v_{\operatorname{noise}},
    \label{velocity_perturbation}
\end{equation}
where $v_{\operatorname{noise}}$ is a white noise component simulating small environmental perturbations. The sine term dependent on $x$ in the main perturbation results in a push on the pair of plasmoids so they move closer to each other. As the domain has periodic boundary conditions on the sides, there is an effective chain of plasmoids moving along the $x-$axis. The perturbation causes the coalescence to take place for each pair of plasmoids separately, while moving the other plasmoids away. The term dependent on $y$ localises the perturbation to a small region around the plasmoids centre. The white noise perturbation, which is set equal for plasma and neutrals, has a magnitude of 0.0005, two orders of magnitude smaller than the main perturbation in Equation (\ref{velocity_perturbation}). Choosing such value prevents the noise from dominating the motion of the two plasmoids during coalescence, but allows the development of dynamics at a smaller scale by breaking the symmetry of the system. The same random noise seed was used in all simulations.

The reference MHD simulation in Section \ref{sec:simulations} is resolved by $2062 \times 3086$ grid cells, corresponding to a cell size of $\Delta x = 1.95 \cdot 10^{-3}$ and $\Delta y = 2.6 \cdot 10^{-3}$. In the PIP case the partial ionisation effects lead to the thinning of the current sheet and the development of sharp small-scale magnetic structures, as a result of the neutrals decoupling from ions and leaving the current sheet \citep{1994ApJ...427L..91B, 1995ApJ...448..734B, 2009ApJ...705.1183A, 2015PASJ...67...96S}. Our simulations show the formation of a thinner current sheet in between the plasmoids merging due to this two-fluid effect. The reference PIP case in Section \ref{sec:simulations} was hence run at the higher resolution of $\Delta x = 1.2 \cdot 10^{-3}$ and $\Delta y = 1.6 \cdot 10^{-3}$ to ensure the current sheet is resolved by our grid. The grid in the PIP case is composed by 6478 points in the $x$ direction and 4862 points in the $y$ direction. In Section \ref{sec:parameters} we present a parameter study of the coalescence process. The resolution used for each simulation and the total number of grid points are detailed in that section.

The initial separation between the plasmoids (calculated from $O-$point to $O-$point, identified as blue spots in Figure \ref{fig:initial_conditions}) is equal to $4 L$, where $L$ is resolved by 515 grid points in the MHD case and by 809 grid points in the PIP case. The plasmoid width, calculated as the distance between top and bottom edges of the separatrix and which initial value is $1.66 L$ (resolved respectively by 638 grid points in the MHD case and by 1037 grid points in the PIP case), is determined by the conditions of the Fadeev equilibrium for the magnetic field.

The non-dimensional diffusion length scale is calculated as $L_{\operatorname{diff}} = \sqrt{4 \eta \tau}$ for both the reference cases and the simulations of the parameter survey. Taking $\tau = 1$, the diffusion length scale for the MHD and PIP reference cases is $4.5 \cdot 10^{-2}$ for a diffusivity $\eta = 5 \cdot 10^{-4}$, while for the cases in the parameter survey, who are characterised by $\eta = 1.5 \cdot 10^{-3}$, $L_{\operatorname{diff}} = 8 \cdot 10^{-2}$. The approximate number of grid cells per diffusion length scale at the lower resolution of the MHD case are $(23, 17)$ in $(x, y)$. These respectively increase to 41 grid cells along $x$ and 31 grid cells along $y$ for $L_{\operatorname{diff}}$ in the parameter survey. Therefore, the diffusion scale is resolved in all simulations. The collisional ion-neutral time scale $\tau_{\operatorname{col,pn}} = (\alpha_c \rho_n)^{-1}$ for the reference PIP case is $10^{-2}$, while the collisional neutral-ion time scale, defined as $\tau_{\operatorname{col,np}} = (\alpha_c \rho_p)^{-1}$, is 1. These values lead to the non-dimensional coupling length scales $L_{pn} = 10^{-2}$ and $L_{np} = 1$, which are both well resolved by our grid.

\subsection{Boundary conditions}
\label{sec:boundary_conditions}

While coalescing, the plasmoids move towards each other along the $x-$axis. Because of the symmetry of the problem, in the reference MHD simulation (Section \ref{sec:simulations}) and in the set of simulations performed for the parameter survey (Section \ref{sec:parameters}) we cut the computational domain at $x=0$ and use symmetric boundaries, where $v_x$ and $B_y$ change sign across each boundary and $v_y$ and $B_x$ remain the same. The left boundary is shown in Figure \ref{fig:initial_conditions} as a dashed green line. The computational domain size is chosen equal to $x = [0, 4]$ and $ y = [-4, 4]$.

The dynamics of the plasmoids merging in the reference PIP case (Section \ref{sec:simulations}) is evaluated in a full computational domain, with $x = [-4, 4]$ and $ y = [-4, 4]$. This arrangement was made to be able to better examine the dynamics in the region of the current sheet at higher resolution. In this case, the top and bottom boundaries are kept symmetric, while the side boundaries are chosen to be periodic.

\section{Results}
\label{sec:simulations}

First, we explore the coalescence instability in both a single-fluid fully ionised plasma and a two-fluid partially ionised plasma by comparing two simulations, an MHD case and a PIP case. The single-fluid case acts as a reference case for the more complex two-fluid simulation.

\begin{figure*}[htb]
    \centering
    \includegraphics[width=0.9\textwidth,clip=true,trim=0cm 1.3cm 0cm 0cm]{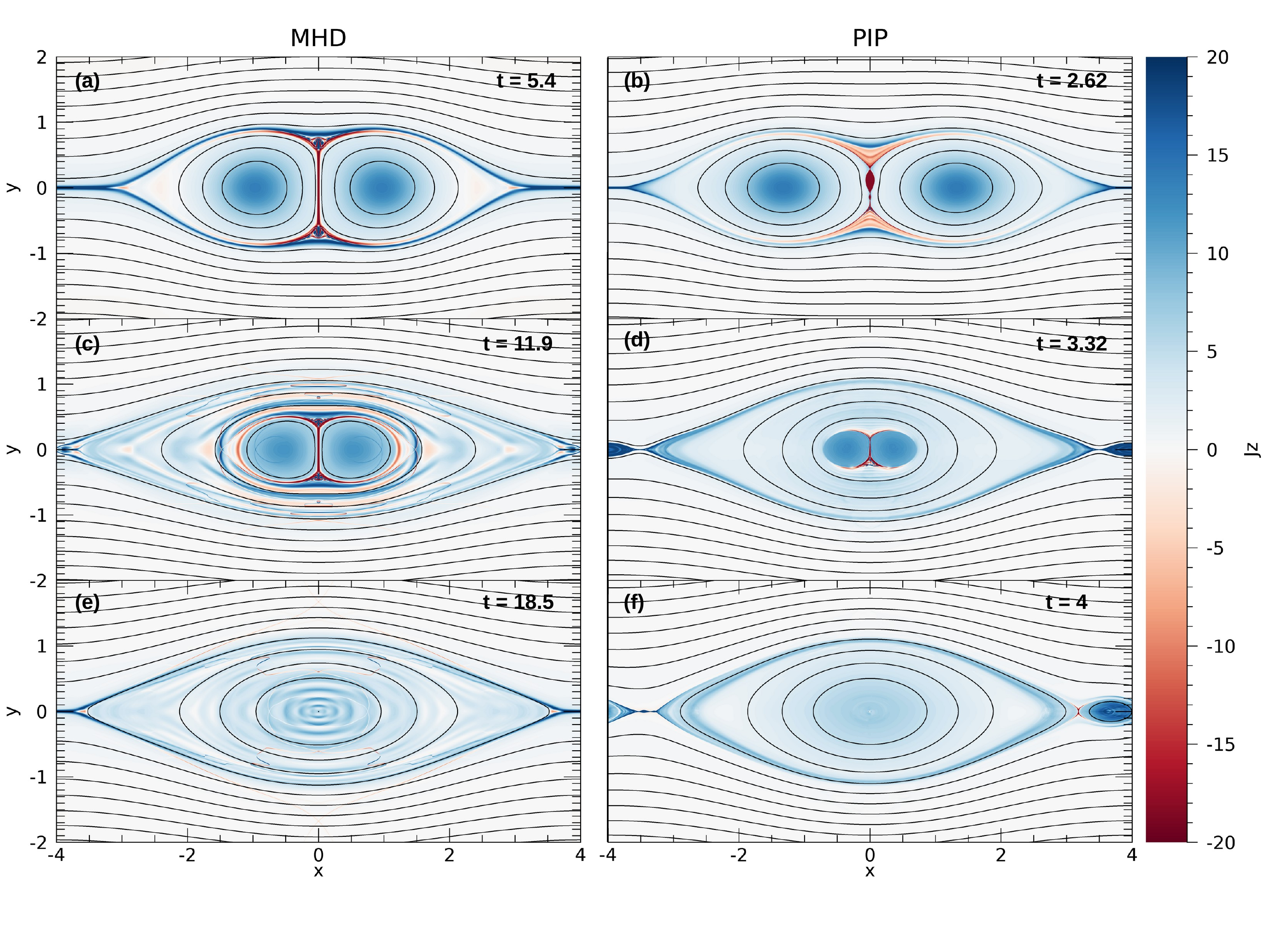}  
    \caption{Comparison of $J_z$ between the MHD case (left column) and the reference PIP case (right column). The frames identify different steps of the coalescence instability. Panels (a) and (b) show the initiation of the reconnection process. In panels (c) and (d) the evolution of coalescence is displayed at later stages. The final stage of coalescence is shown in panels (e) and (f), with the formation of the resulting plasmoid. The same magnetic field lines are displayed in black for all frames. Times are given in the same non-dimensional unit.}
    \label{fig:current_evolution_frames}
\end{figure*}

The initial parameters for both simulations are constant diffusivity $\eta = 0.0005$ and plasma $\beta = 0.1$. In the PIP simulation we set the collisional coefficient $\alpha_c = 100$ and the ion fraction $\xi_p = \rho_p / (\rho_p + \rho_n) = 0.01$, while the effective ion fraction in the MHD case is $\xi_p = 1$. The effects of the parameters variation on the coalescence in PIP simulations are investigated later on in Section \ref{sec:parameters}.

Figure \ref{fig:current_evolution_frames} displays a sequence of the evolution of the current density $J_z$, which is directed out of the plane. For a better comparison, the frames show times where similar physics takes place in both the MHD and PIP cases. As $\eta \neq 0$ the magnetic flux reconnects and leaves the current sheet that is formed in between the coalescing plasmoids: this is the region of strong negative current between the two plasmoids in panels (a), (b), (c) and (d) of Figure \ref{fig:current_evolution_frames}. In a first phase, the current sheet length rapidly increases when the plasmoids approach, then it progressively reduces with the size of the coalescing plasmoids (panels (c) and (d) of Figure \ref{fig:current_evolution_frames}). The reconnection results in the formation of a single large plasmoid, as shown in panels (e) and (f) of Figure \ref{fig:current_evolution_frames}.

The left-right symmetry in the PIP case is broken during the coalescence, as evident in particular from panels (d) and (f) of Figure \ref{fig:current_evolution_frames}. The asymmetry arises from the initial noise perturbation in equation (\ref{velocity_perturbation}) allowing the onset of small-scale dynamics in the central current sheet. The symmetry in the MHD case is also reinforced from the presence of a central boundary at $x = 0$, as introduced in Section \ref{sec:boundary_conditions}.

\begin{figure}[htb]
    \centering
    \includegraphics[width=\columnwidth,clip=true,trim=0cm 0cm 0cm 0cm]{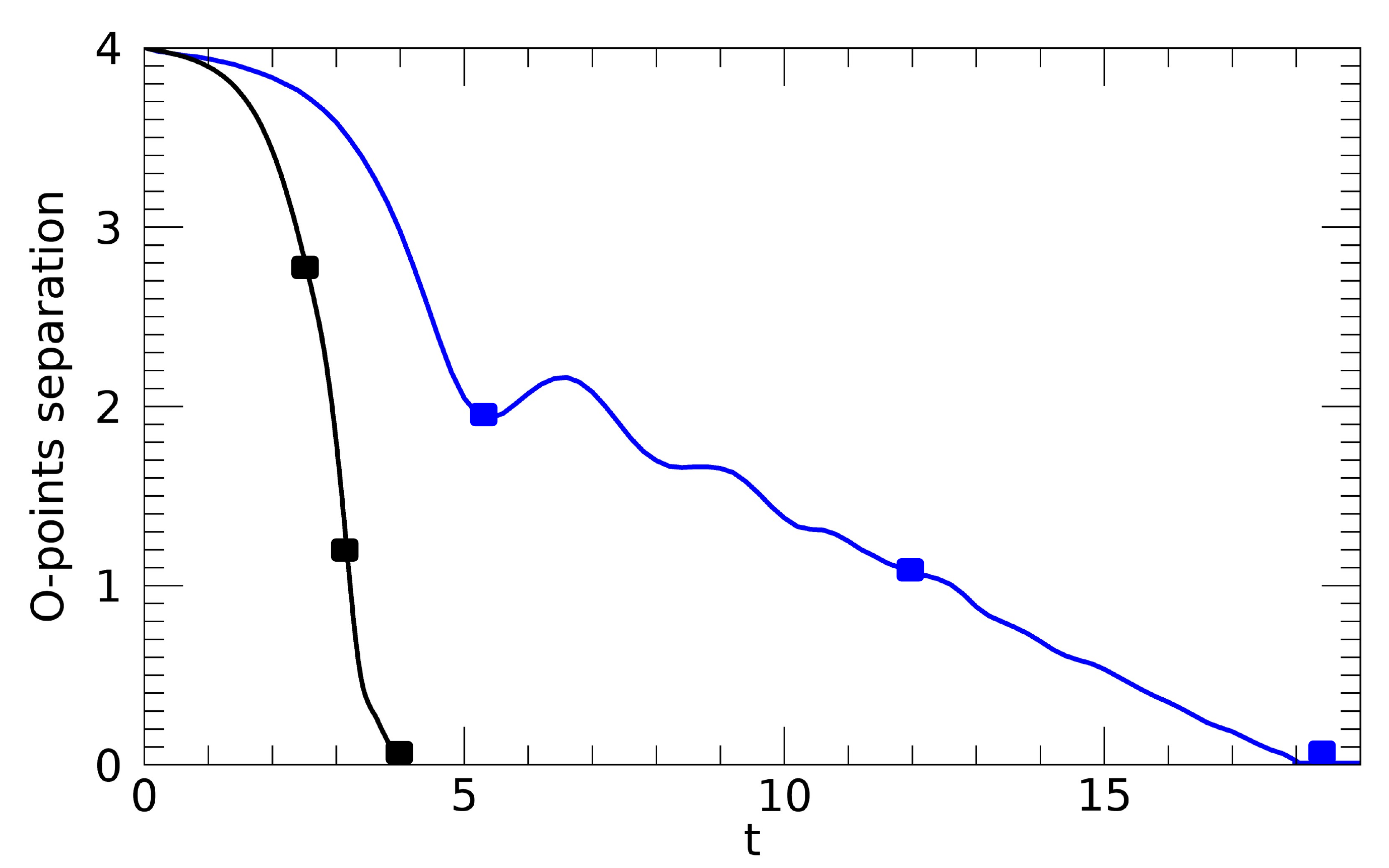}
    \caption{Time variation of the distance between the merging plasmoids, calculated as the distance between the $O-$points, for the MHD case (solid blue) and the PIP case (solid black). The blue squares refer to the times shown in panels (a), (c) and (e) of Figure \ref{fig:current_evolution_frames}. The black squares identify the times shown in panels (b), (d) and (f) of Figure \ref{fig:current_evolution_frames}.}
    \label{fig:L_O-points_vs_t}
\end{figure}

Figure \ref{fig:L_O-points_vs_t} displays in blue the separation of the two plasmoids in the MHD case, calculated as the distance between $O-$points. The squares along the curve identify the times of panels (a), (c) and (e) in Figure \ref{fig:current_evolution_frames}. This distance fluctuates in time, with peaks that appear regularly during the reconnection phase. The merging plasmoids accelerate towards each other, move slightly apart as they bounce on the current sheet and accelerate back again towards the centre. Such movement can be associated to the high gas pressure generated inside the current sheet.

As observed from the black curve in Figure \ref{fig:L_O-points_vs_t}, the distance between the two plasmoids reduces rapidly in the PIP case, following the faster reconnection process and without displaying the same oscillations that are remarkable in the MHD case. The reason could be that the reconnection is happening fast enough that the plasmoids do not need to rebound off each other. The squares here identify the times of panels (b), (d) and (f) in Figure \ref{fig:current_evolution_frames}.

The shortening of the coalescence timescale in the PIP case can be associated to the decoupling of ions and neutrals in the reconnection region. This results in a faster thinning of the current sheet as the lower ion density allows a stronger compression than in the MHD case. The decoupling is discussed in more detail in Section \ref{sec:inflow}. The sharpening of the magnetic field profile has already been observed in many studies \cite{1963ApJS....8..177P,1994ApJ...427L..91B,1995ApJ...448..734B,1999ApJ...511..193V,2003ApJ...583..229H,2017ApJ...842..117A} as a result of the ambipolar diffusion. The reduced single-fluid model for two-fluid effects provided by the ambipolar diffusion provides a good approximation for strongly coupled systems, and it is consistent with the effects that we record in our simulations. However, it might fail in describing the complexity of weakly and intermediate coupled systems, and could not be used to explain the different time scale of the first phase of coalescence, when the plasmoids move towards each other. The increased complexity of our system is therefore better investigated through a full two-fluid model such as the one used in this work.

At a first qualitative view, several differences are present between the fully ionised and the partially ionised cases. Firstly, the plasmoid merging occurs faster in the PIP simulation, ending at $t \sim 4$, while coalescence takes a longer time in the MHD case, ending at $t \sim 18.5$. The end is identified with the time when a single, large plasmoid is fully formed and the current density at its centre reaches a positive maximum, stabilizing to a constant value. The difference in the coalescence time scale is related to the reconnection rate, discussed in Section \ref{sec:reconnection_rate}. A second difference is that there is no clear sign of shocks in the PIP case, while the MHD case shows an abundance of fine structures in panels (c) and (e) of Figure \ref{fig:current_evolution_frames}. The identification and classification of these structures as shocks are investigated in detail in Section \ref{sec:shocks}.

Two unique features are present in the PIP simulation only. The first is the production of secondary plasmoids in the central current sheet, linked to the onset of additional instabilities that will be discussed in Section \ref{sec:secondary_plasmoids}. The second is the formation of an extended neutral jet in the reconnection region, whose properties are examined in Section \ref{sec:neutral_jet}.

\subsection{Ideal phase of coalescence}
\label{sec:inflow}

\begin{figure}[htb]
    \centering
    \includegraphics[width=\columnwidth,clip=true,trim=1.5cm 0cm 1.5cm 1cm]{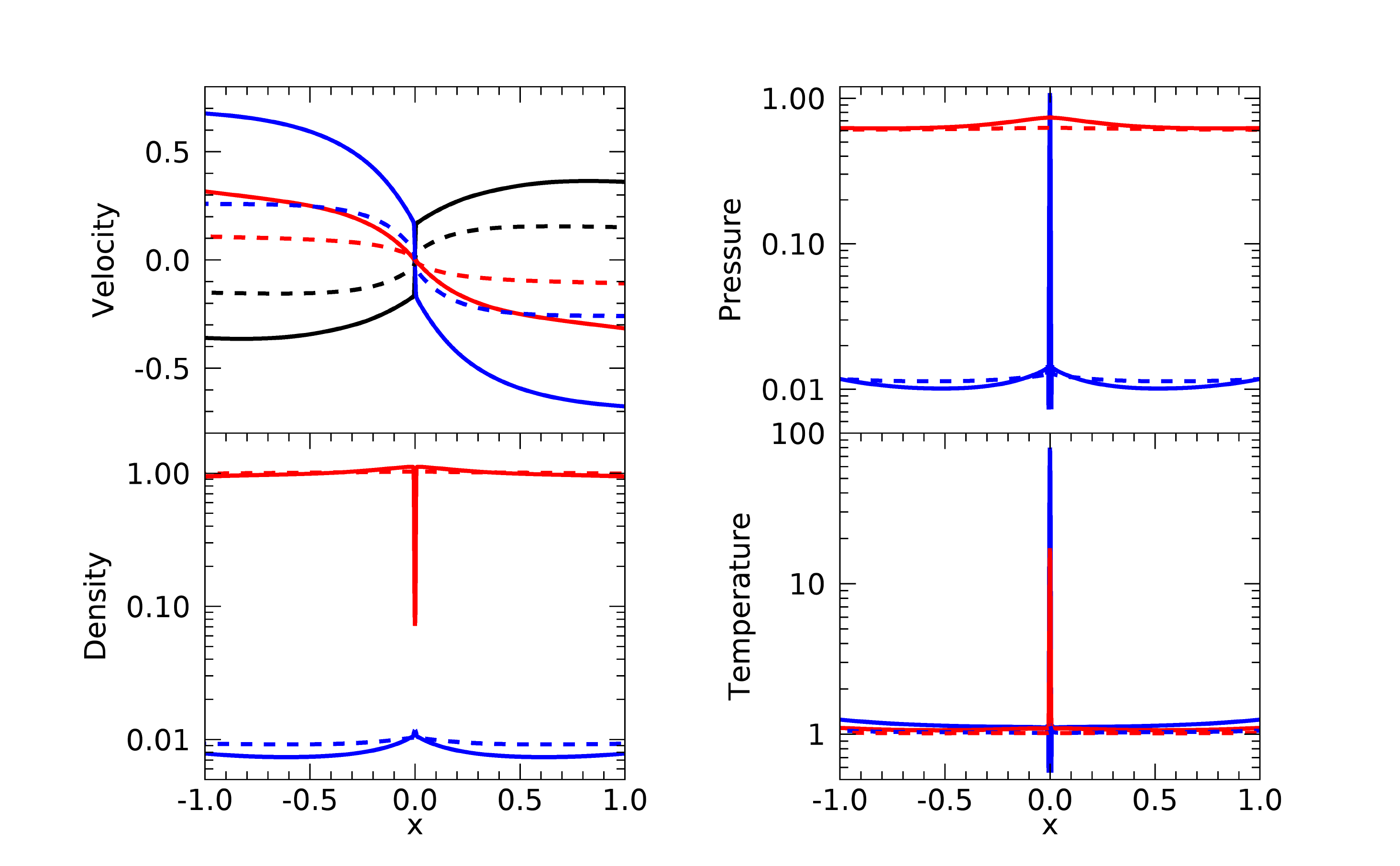}
    \caption{Plots of the velocity (top left), pressure (top right), density (bottom left) and temperature (bottom right) of ions (blue) and neutrals (red) at $t = 1.62$ (dashed lines) and $t = 2.42$ (solid lines) in the inflow of the PIP case. The drift velocity is displayed in black in the top left panel.}
    \label{fig:PIP_inflow_properties}
\end{figure}

The mutual attraction of parallel currents pulls the two initial plasmoids together, and a current sheet forms as a result of the anti-parallel magnetic field being pushed together. In the MHD simulation the ideal phase of coalescence is characterised by a plasma inflow forming along the $x-$axis that contributes to the formation of the central current sheet. The charged species in the PIP simulation are pulled to the centre of the domain by the Lorentz force and the neutrals are dragged by collisions resulting in a small drift velocity that can be seen in Figure \ref{fig:PIP_inflow_properties} for $t = 1.62$. The drift velocity ($\mathbf{v}_n - \mathbf{v}_p$) differs from zero in the inflow, which indicates that the two species are weakly coupled in this first phase of coalescence, and increase steadily in time with the acceleration of the plasmoids motion towards each other. After $t = 2.42$ the reconnection process change, as the tearing instability takes place with the formation of secondary plasmoids (panels (b) and (d) of Figure \ref{fig:current_evolution_frames}).

\begin{figure}[htb]
    \centering
    \includegraphics[width=\columnwidth,clip=true,trim=0cm 0cm 0cm 0cm]{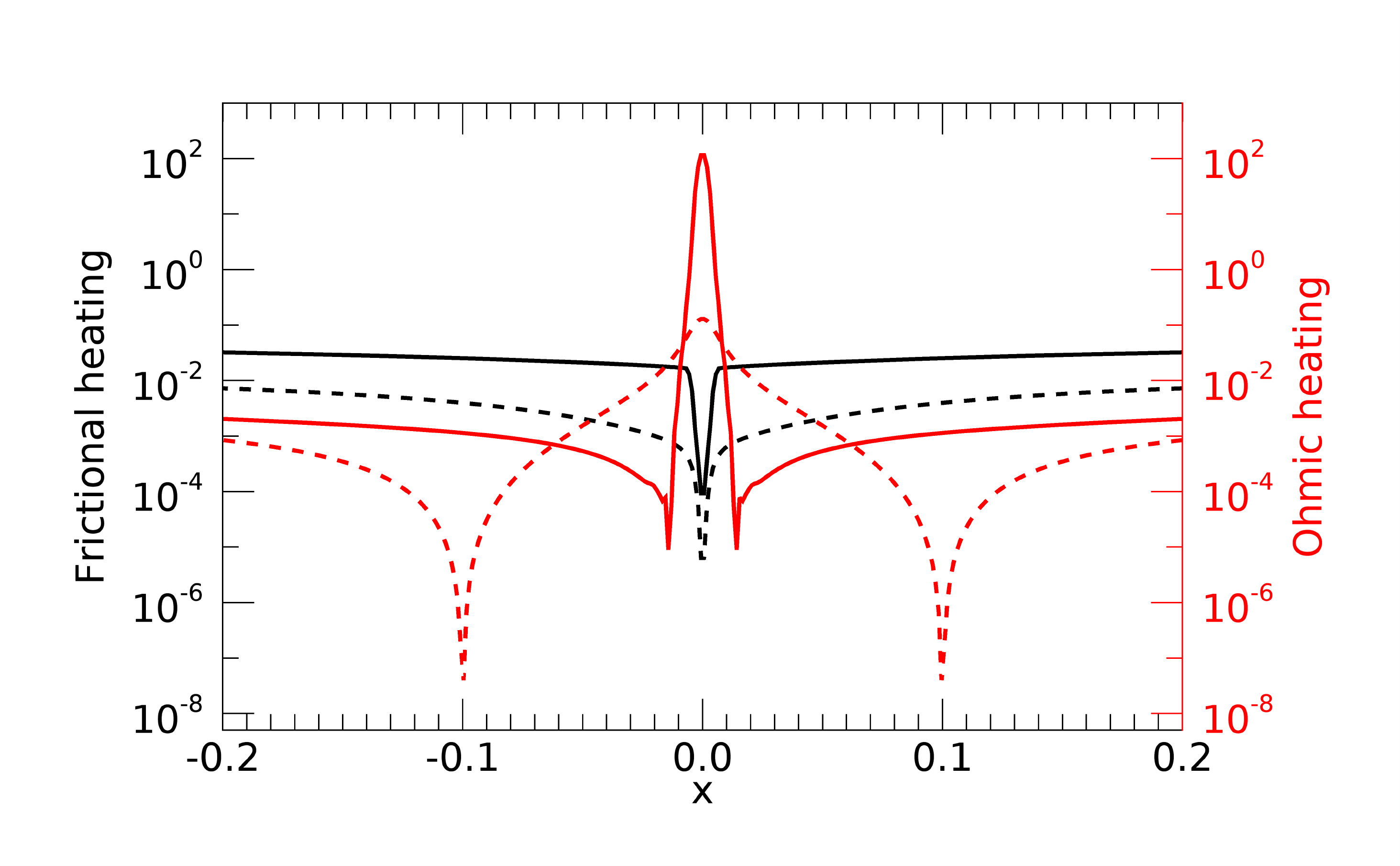}
    \caption{Frictional heating at $t = 1.62$ (black dashed line) and $t = 2.42$ (black solid line) along the $x-$axis in the PIP simulation, compared to the Ohmic heating (red dashed line for $t = 1.62$, red solid line for $t = 2.42$).}
    \label{fig:PIP_inflow_frictional_heating}
\end{figure}

A large amount of plasma builds up due to the attraction of the magnetic field and a current sheet is created by the magnetic field piling up. This corresponds to a strong increase in the plasma pressure which supports the current sheet and in the plasma temperature, as shown in the top and bottom right panels of Figure \ref{fig:PIP_inflow_properties}. The non-adiabatic spike in the plasma temperature produced by the Ohmic heating $\eta J^2$ (shown in Figure \ref{fig:PIP_inflow_frictional_heating}) has an effect on the neutral temperature, which increases due to the thermal coupling between the species. As the neutrals are not completely coupled to the ions they are expelled from the current sheet, as shown by the drop in the neutral density (bottom left panel of Figure \ref{fig:PIP_inflow_properties}).

Both fluids are also heated up in the inflow region through frictional heating, which non-dimensional definition \cite{2016A&A...591A.112H} is:
\begin{equation}
    (1/2)\alpha_c (T_n , T_p, v_D ) \rho_n \rho_p (\textbf{v}_n - \textbf{v}_p )^2 .
    \label{eq:frictional_heating}
\end{equation}
The frictional heating at $y = 0$ is shown in Figure \ref{fig:PIP_inflow_frictional_heating} compared to the Ohmic heating at $t = 1.62$ and $t = 2.42$. It increases with time during the first phases of reconnection, dropping only in the current sheet, where the Ohmic heating provides the major contribution by heating the plasma and increasing the plasma pressure and temperature.

\subsection{Current sheet and reconnection rate}
\label{sec:reconnection_rate}

Once the current sheet is generated, the MHD and PIP cases show very different reconnection processes. Laminar reconnection takes place in the MHD case, independent of the initial white noise perturbation without the onset of further instabilities. The long thin current sheet can be compared to the steady-state Sweet-Parker model. Both the length $\Delta_{\operatorname{MHD}}$ and the width $\delta_{\operatorname{MHD}}$ are estimated by taking the full width at $1/8$ of the maximum current density $J_z$, respectively along the $y-$axis and the $x-$axis. We choose this ratio as it accurately represents the termination of the reconnection region. The current sheet width and length are $\delta_{\operatorname{MHD}} \sim 0.02$ and $\Delta_{\operatorname{MHD}} \sim 1.49$. Using $\Delta_{\operatorname{MHD}}$ as characteristic length of the system and the maximum value of the Alfv\'en speed that occurs at the boundary of the current sheet, $v_A \sim 3.26$, it is possible to calculate the Lundquist number. We find that $S = \Delta_{\operatorname{MHD}} v_A / \eta \sim 9.7 \cdot 10^3$.

In case of Sweet-Parker-like reconnection, the current sheet aspect ratio scales as $S^{-1/2}$. From the value of $S$ we obtain that the expected aspect ratio for the MHD case is $ \delta / \Delta \sim 1.01 \cdot 10^{-2}$, an estimate comparable to the measure obtained by $\delta_{\operatorname{MHD}}/\Delta_{\operatorname{MHD}} \sim 1.57 \cdot 10^{-2}$. Having laminar reconnection in a long thin current sheet which does not develop instabilities nor break into smaller parts, we may suggest that the MHD case is subject to a reconnection process that is Sweet-Parker-like.

Unlike the MHD case, reconnection in the PIP case is nonlaminar. The presence of plasmoids breaks the current sheet into multiple thinner current sheets, which reconnect faster than the original structure. The dimensions of the current sheet at the time immediately before the generation of the first plasmoid ($t = 2.42$) are $\delta_{\operatorname{PIP}} \sim 0.01$ in width and $\Delta_{\operatorname{PIP}} \sim 0.46$ in length, estimated as $J_{z \operatorname{,max}}/8$ along the $x-$axis and the $y$-axis respectively. The onset of the plasmoid instability in a fully ionised plasma might take place below a critical aspect ratio \cite{2010PhPl...17f2104H} of $1/200 = 5 \cdot 10^{-3}$: the same result was obtained for a multi-fluid plasma \cite{2012ApJ...760..109L}. Our current sheet aspect ratio is $\sim 1.9 \cdot 10^{-2}$, about four times larger than the predicted aspect ratio implying further physics may be involved in the onset of plasmoid formation. This is investigated in Section \ref{sec:subcritical}.

In a partially ionised plasma three different Alfv\'en speeds can be identified \citep{1989ApJ...340..550Z}: a total Alfv\'en speed $v_{A,t}$ related to the total density $\rho_n + \rho_p$, an ion Alfv\'en speed $v_{A,p}$, based on the density of the charged particles, and an effective Alfv\'en speed $v_{A,e}$, based on the combined density of charged particles and neutrals that are coupled through collisions. The expression for  $v_{A,e}$ might be non-trivial, however it is possible to provide a close estimate for it from the plasma outflow velocity ($v_{\operatorname{out}} \sim v_{A,e}$). We chose to use the plasma velocity as the ionised fluid is the one directly accelerated by the reconnected magnetic field lines. The Alfv\'en speed is inversely proportional to the density: as $v_A$ increases at the decrease of density, the ion Alfv\'en speed is bigger than the total Alfv\'en speed. The Lundquist number calculated by using $v_{A,t} \sim 2.03$ is $S_t = v_{A,t} \Delta_{\operatorname{PIP}} / \eta \sim 1.75 \cdot 10^3$. If we consider the ion Alfv\'en speed only, which is $v_{A,p} \sim 23.29$, the Lundquist number becomes $S_p \sim 2.01 \cdot 10^4$, that is consistent with the threshold value $S = 4 \cdot 10^4$ for the onset of the tearing instability and plasmoid formation \citep{1986PhFl...29.1520B, 2009PhPl...16k2102B, 2012ApJ...760..109L, 2015PhPl...22j0706S}.

In presence of collisional coupling, reconnection scales with the Lundquist number associated to the effective Alfv\'en speed $v_{A,e}$. Estimating $v_{A,e} \sim 9.09$ from the maximum outflow velocity, we find $S_e \sim 7.8 \cdot 10^3$, which is below the threshold value for the onset of the tearing instability. The discrepancy suggests that effects due to partial ionisation might affect the dynamics of reconnection, allowing the formation of secondary plasmoids in the presence of a lower Lundquist number. The answer to this discrepancy between the models can be sought in the modifications due to the ion-neutral interaction.

The reconnection rate $M$ 
is defined by:
\begin{equation}
    M = \frac{\eta J_{\operatorname{max}}}{v_A B_{\operatorname{up}}},
    \label{eq:reconnection_rate}
\end{equation}
where $J_{\operatorname{max}}$ is the absolute maximum value of the current density inside the current sheet, $v_{A}$ is the initial maximum bulk Alfv\'en speed ($v_{A,t}$ in the PIP case) and $B_{\operatorname{up}}$ is the initial maximum value of $B_y$ in the inflow. The MHD mean reconnection rate is $M = 0.027 \pm 0.004$, and it appears to be higher than the Sweet-Parker rate by a factor of 2. The PIP reconnection rate, which average value is $0.08 \pm 0.01$, is approximately three times bigger than the MHD case, and displays very sharp variations during the merging. Such fluctuations correspond to the formation and expulsion of the secondary plasmoids that disrupt the current sheet.

\begin{figure}[htb]
    \centering
    \includegraphics[width=\columnwidth,clip=true,trim=0cm 0cm 0cm 0cm]{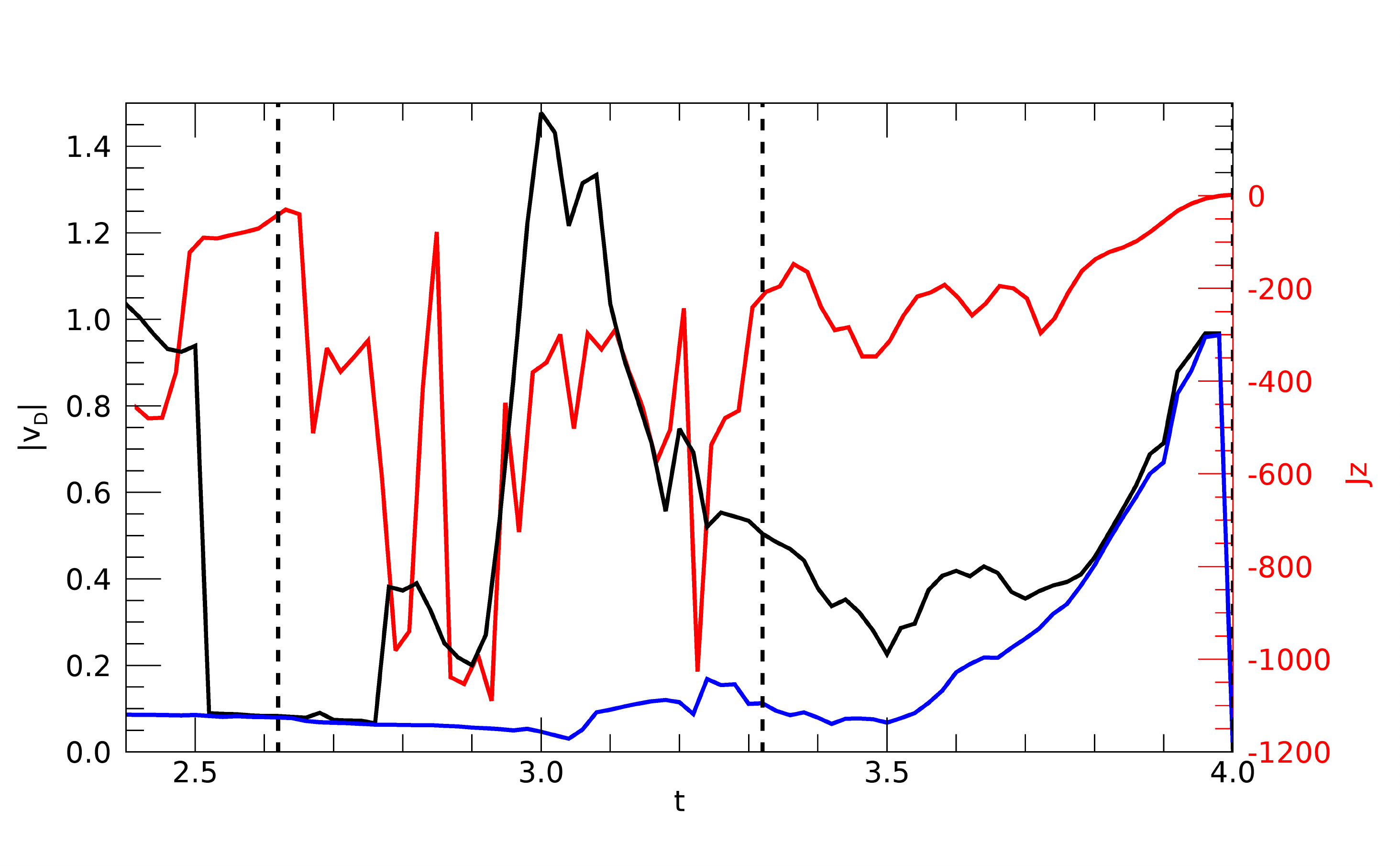}
    \caption{Temporal evolution of the drift velocity magnitude, compared to J$_z$ at the centre of the current sheet (in red) for the PIP case. The solid black lines indicate the maximum values of drift velocity and the solid blue lines indicate the median values. The vertical dashed lines indicate the times displayed in panels (b), (d) and (f) of Figure \ref{fig:current_evolution_frames}.}
    \label{fig:drift}
\end{figure}

Figure \ref{fig:drift} shows the temporal evolution of the maximum (black) and median (blue) drift velocity inside the current sheet during the reconnection phase at $t = [2.4, 4]$ for the PIP reference case. Both are compared to the evolution of current density at the centre of the current sheet ($x = 0, y = 0$), displayed in red. While the maximum drift velocity tends to oscillate quite drastically, especially in connection of the major fluctuation in the current density, the median value tends to be approximately constant with a value of $\sim 0.1$, increasing smoothly in the last phases of the coalescence after $t = 3.5$, as the magnitude of $J_z$ decreases. The peak in the median drift velocity is reached at the complete merging, where it reaches a value $|v_D | \sim 1$.

The huge increase in the absolute maximum drift corresponds to the neutrals being expelled from the current sheet in the $x$ direction during its collapse, and it takes place during the formation and expulsion of secondary plasmoids (for more details on secondary plasmoids see Section \ref{sec:secondary_plasmoids}). This can be seen from the peak in the maximum drift velocity, that is reached in the interval between the two central vertical dashed lines. These vertical lines represent the times identified in panels (b) and (d) of Figure \ref{fig:current_evolution_frames}. The drop in the maximum $|v_D|$ occurring approximately between $t = 2.5$ and $t = 2.8$ happens in correspondence of a relatively constant value of $J_z$. Such smooth variation of $J_z$ is linked to the formation and growth of the first secondary plasmoid ant the centre of the current sheet. More details about the investigation of the current density are presented in Section \ref{sec:alpha_c}.

\subsection{Shocks}
\label{sec:shocks}

\begin{figure}[htb]
    \centering
    \includegraphics[width=\columnwidth,clip=true,trim=0.7cm 0.5cm 0.7cm 0.7cm]{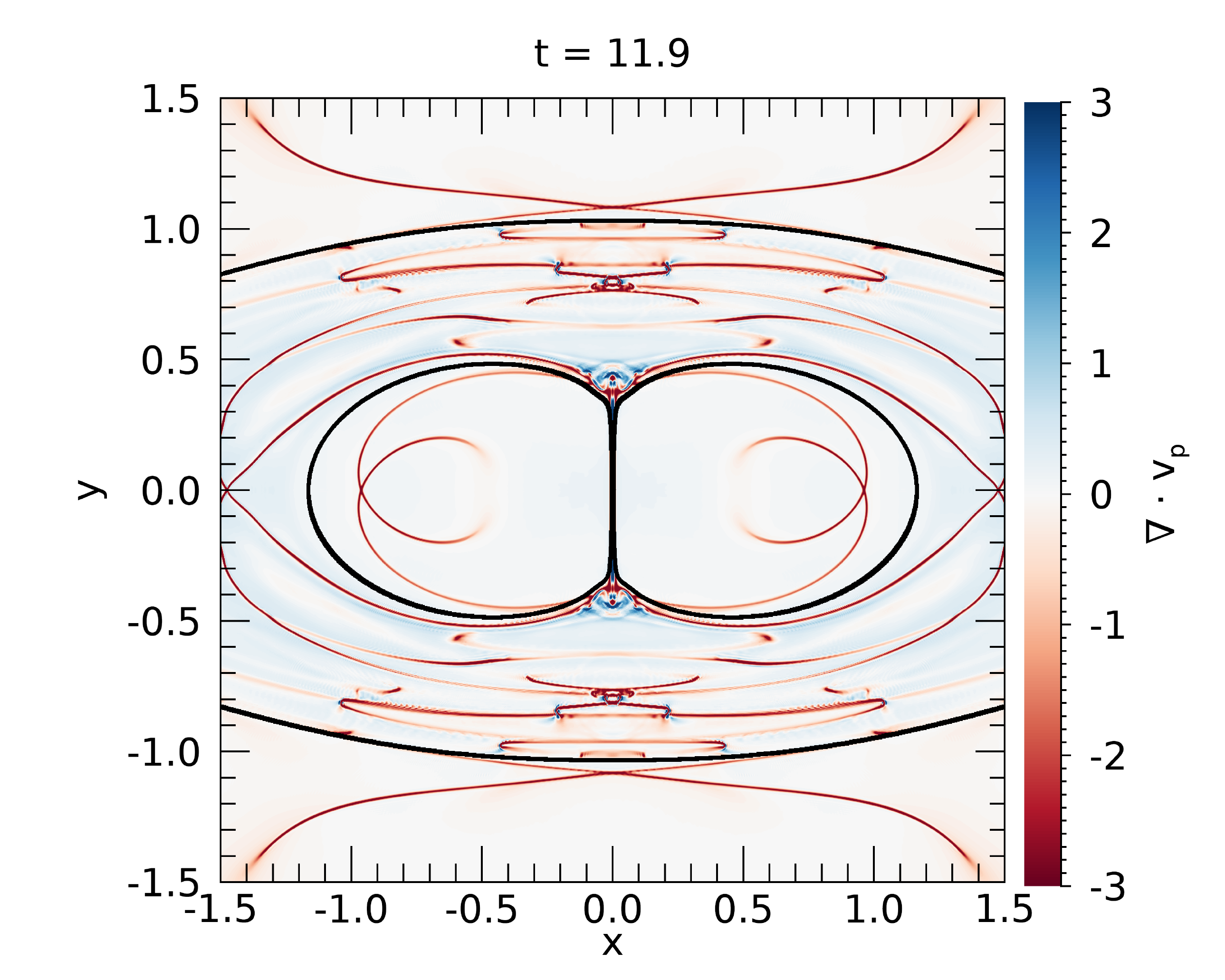}
    \caption{Divergence of the plasma velocity at $t = 11.9$ for the MHD simulation. The plot is saturated in order to enhance the structures associated to shocks. In black the inner separatrix shows the edges of the merging plasmoids, while the outer separatrix show the border of the final plasmoid that is forming.}
    \label{fig:shock_divv}
\end{figure}

During reconnection and in the final phase of the merging in the MHD case (panels (c) and (e) of Figure \ref{fig:current_evolution_frames}), there is evidence of shocks visible as thin elongated lines corresponding to both positive and negative peaks of the current density magnitude. The structures that can be distinguished in the current density are enhanced in the divergence of $v_p$, shown in Figure \ref{fig:shock_divv}, where they are identified as thin red lines.

The minimum in the divergence of the plasma velocity field identifies a region in which the flow is highly compressed, i.e. a shock. Across the shock the magnetic field components $B_y$ and $B_z$ drop, while plasma density and pressure rise steeply. The behaviour of magnetic field and pressure identifies this as a slow-shock. The presence of slow-mode shocks is expected as they are a common feature in reconnecting systems, being part of a huge variety of fine structures that can be identified when plasmoid dynamics takes place \citep{2011PhPl...18b2105Z}.

\begin{figure}[htb]
    \centering
    \includegraphics[width=\columnwidth,clip=true,trim=0cm 5cm 0cm 0cm]{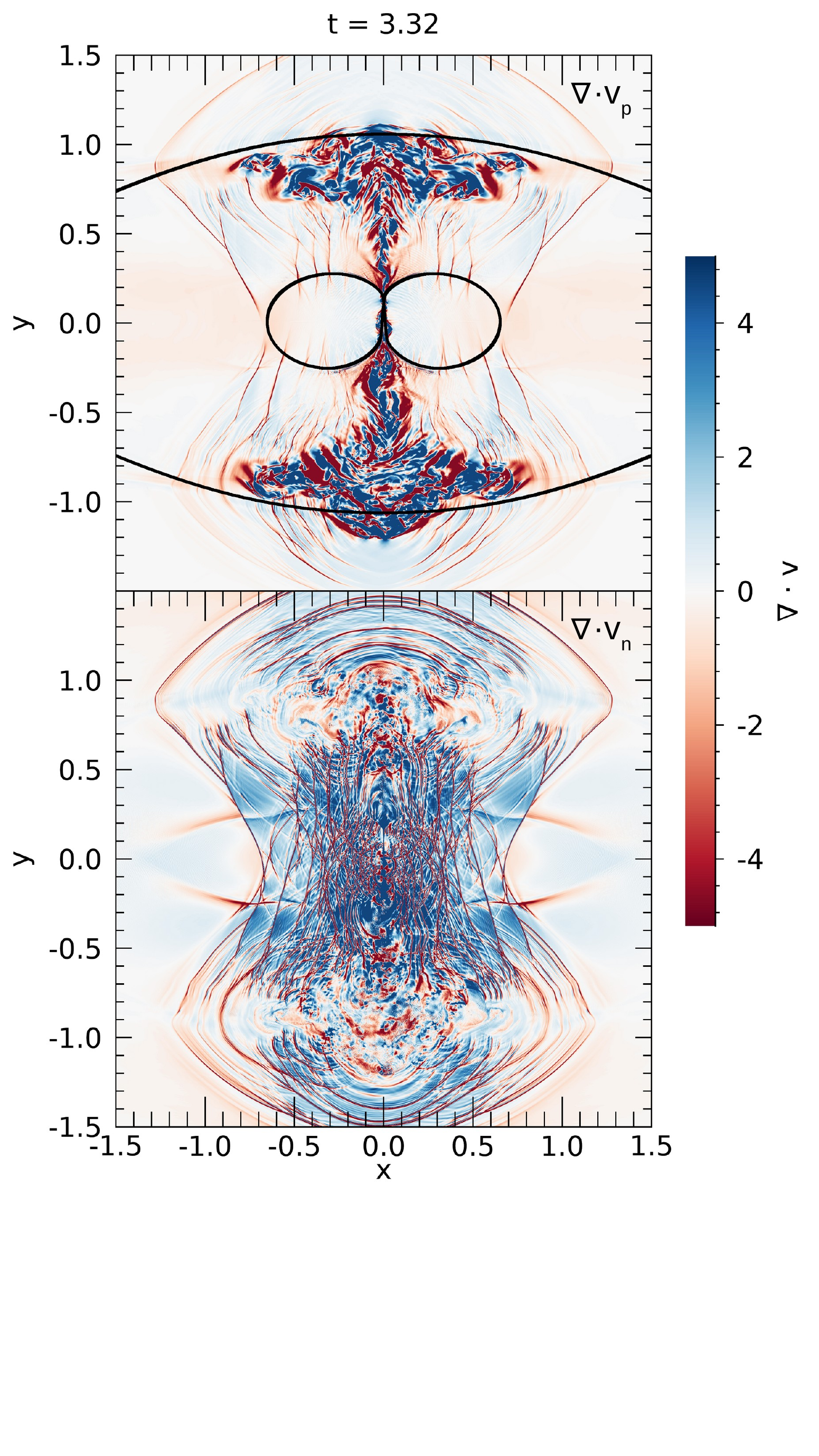}  
    \caption{Divergence of the plasma velocity (top) and the neutral velocity (bottom) at $t = 3.32$ for the PIP simulation. The plot is saturated in order to enhance the structures associated to shocks. In the top panel, the magnetic field lines separatrices are shown in black.}
    \label{fig:PIP_shock_divv}
\end{figure}

Comparing panels (c) and (d) of Figure \ref{fig:current_evolution_frames} the PIP case shows far fewer clear shock structures. Here we analyse the mechanisms suppressing slow-mode shocks in the PIP simulations. We examine the divergence of the plasma and neutral velocity fields (Figure \ref{fig:PIP_shock_divv}) at $t = 3.32$, the same time of panel (d) in Figure \ref{fig:current_evolution_frames}. We focus on the divergence of the velocity field as large values of $\nabla \cdot \mathbf{v}$ are a signature of shocks. Comparing it to the divergence of the velocity in the MHD case (Figure \ref{fig:shock_divv}) there are no structures that can be associated to slow-mode shocks. We present more information later on in Section \ref{sec:alpha_c}.

However, a wide range of structures appear in both $\nabla \cdot \mathbf{v}_n$ and $\nabla \cdot \mathbf{v}_p$, and they are particularly enhanced in the neutrals. In the plasma velocity divergence, the most prominent structure is associated to the neutral jet discussed in Section \ref{sec:neutral_jet}, but other structures cut the $x-$axis symmetrically at both sides of the reconnection region. These structures, which form in the neutrals and later couple to the plasma, are hydrodynamic shocks generated by the motion of the neutral species in the inflow and not slow-mode shocks as found in the MHD case.

During the merging, the neutrals are expelled and travel away from the reconnection region with a flow that is more dense in the direction perpendicular to the current sheet. This can be seen from panel (a) of Figure \ref{fig:jet_detail}. In their motion, the neutrals interact with the dense plasma flow (panel (b) of Figure \ref{fig:jet_detail}), and they are halted by the collisions. The compression of the neutral flow leads to the formation of multiple shock fronts that are perpendicular to the $x-$axis. The neutral shocks coupling to the plasma manifest as the lines in $\nabla \cdot \mathbf{v}_p$. The lines, whose front moves away from the current sheet centre, are present in the neutral $v_x$ component, pressure and density but do not display a counterpart in the plasma variables.

Other hydrodynamic shocks are visible along the $y-$axis. These shocks are formed by the material accelerated inside a neutral jet, which will be examined in Section \ref{sec:neutral_jet}.

\subsection{Secondary plasmoids}
\label{sec:secondary_plasmoids}

\begin{figure}[htb]
    \centering
    \includegraphics[width=\columnwidth,clip=true,trim=0.8cm 0cm 1cm 0cm]{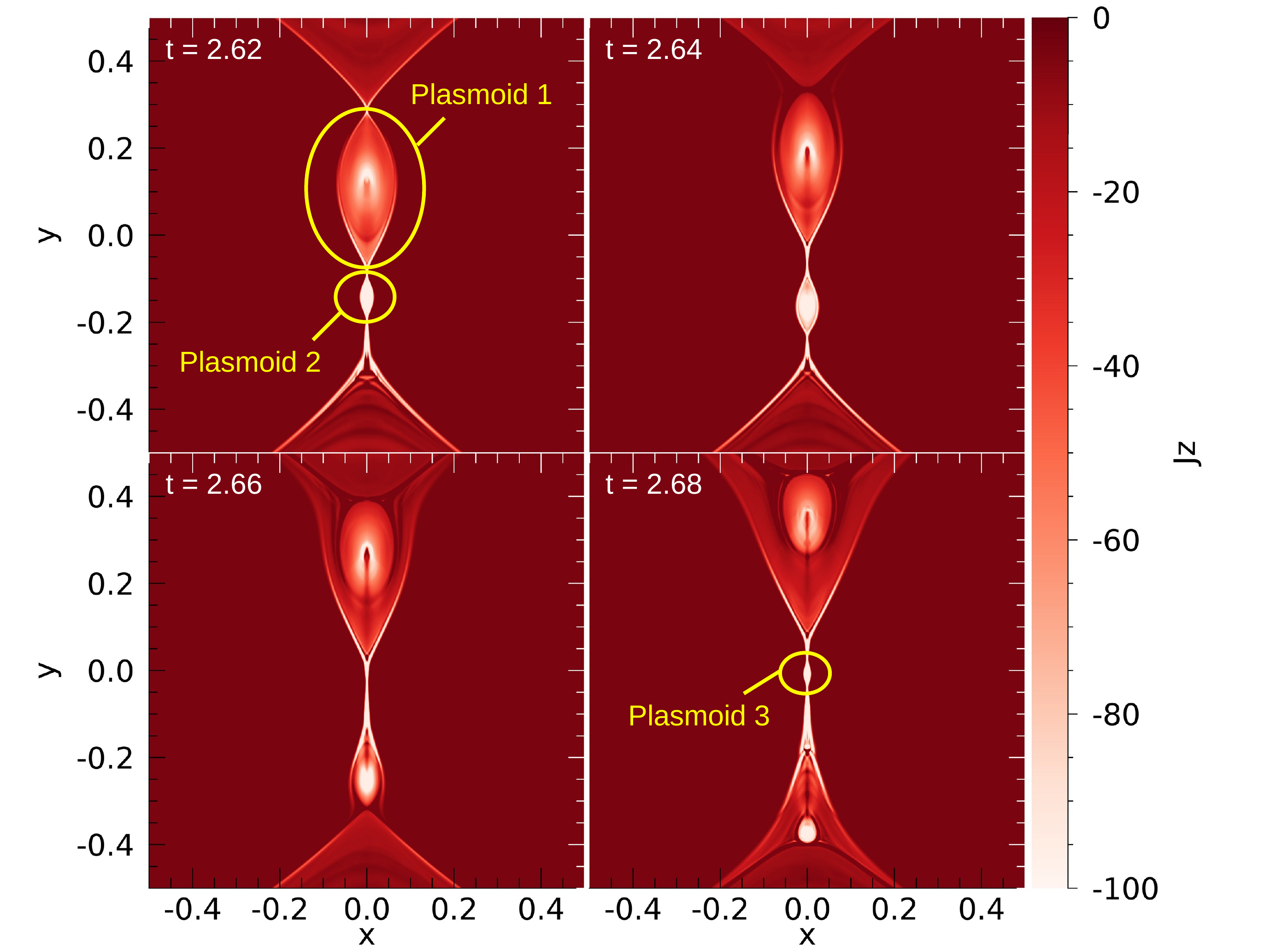}  
    \caption{Formation and expulsion of plasmoids from the central current sheet at $t = 2.62$ (top left), 2.64 (top right), 2.66 (bottom left) and 2.68 (bottom right).}
    \label{fig:plasmoids_timelapse}
\end{figure}

During coalescence in the PIP case the central current sheet is subject to the tearing instability, and secondary plasmoids are produced as evident in panel (b) of Figure \ref{fig:current_evolution_frames}. Figure \ref{fig:plasmoids_timelapse} shows three secondary plasmoids forming, moving along the current sheet and being expelled. The motion along the current sheet is triggered by the white noise perturbation that breaks the symmetry of the system.

\begin{figure}[htb]
    \centering
    \includegraphics[width=\columnwidth,clip=true,trim=0cm 0.5cm 0cm 1.7cm]{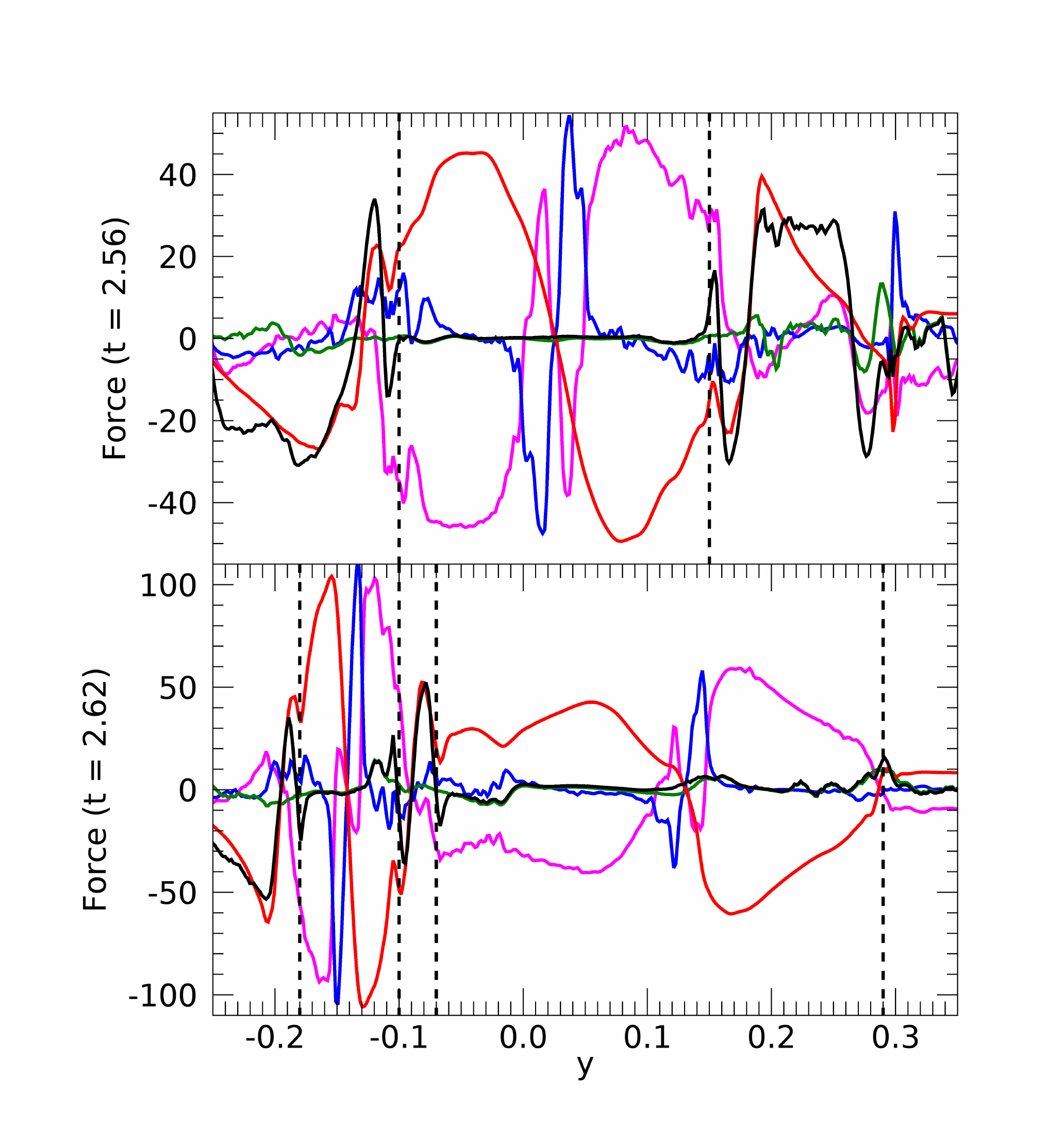}
    \caption{Force balance $\textbf{J} \times \textbf{B} - \nabla p$ (black solid line) calculated along the current sheet in the $y-$axis at $t = 2.56$ (top) and $t = 2.62$ (bottom). The black dashed lines indicate the edges of the secondary plasmoids 1 (top and bottom panels) and 2 (bottom panel, on the left). The force components that are represented are the $- \nabla p_p$ (blue), $- \nabla p_n$ (green), the magnetic pressure (magenta) and the magnetic tension (red).}
    \label{fig:plasmoids_force_balance}
\end{figure}

In the framework of plasmoid dynamics, it is interesting to investigate whether the secondary plasmoids have any characteristic in common with the initial plasmoids. We look at the force balance between the total pressure gradient and the Lorentz force ($ \textbf{J} \times \textbf{B} - \nabla p$), shown in Figure \ref{fig:plasmoids_force_balance} along the $y-$axis at $t = 2.56$ and $t = 2.62$.
The vertical dashed lines are representative of the edges of the secondary plasmoids. At $t = 2.56$, plasmoid 1 can be identified at $y = [-0.1, 0.15]$ and moves to the right at $t= 2.62$, while plasmoid 2 forms at $y = [-0.18,-0.1]$ at the later time. The force components cancel each other at the plasmoids location, while the current sheet around is still out of balance. Inside the plasmoids, the major contributions to the total force are provided by the magnetic pressure $B^2 /2$ (magenta) and the $y$ component of the magnetic tension $(\textbf{B} \cdot \nabla) \cdot \textbf{B}$ (red), while the gradient of the neutral pressure (green) is negligible across the whole region. This suggests that the secondary plasmoids are in an almost force-free condition other than a small region at their centre where the plasma pressure ($- \nabla p_p$ is shown in blue in Figure \ref{fig:plasmoids_force_balance}) becomes significant.

\begin{figure}[htb]
    \centering
    \includegraphics[width=\columnwidth,clip=true,trim=0cm 0cm 0cm 0cm]{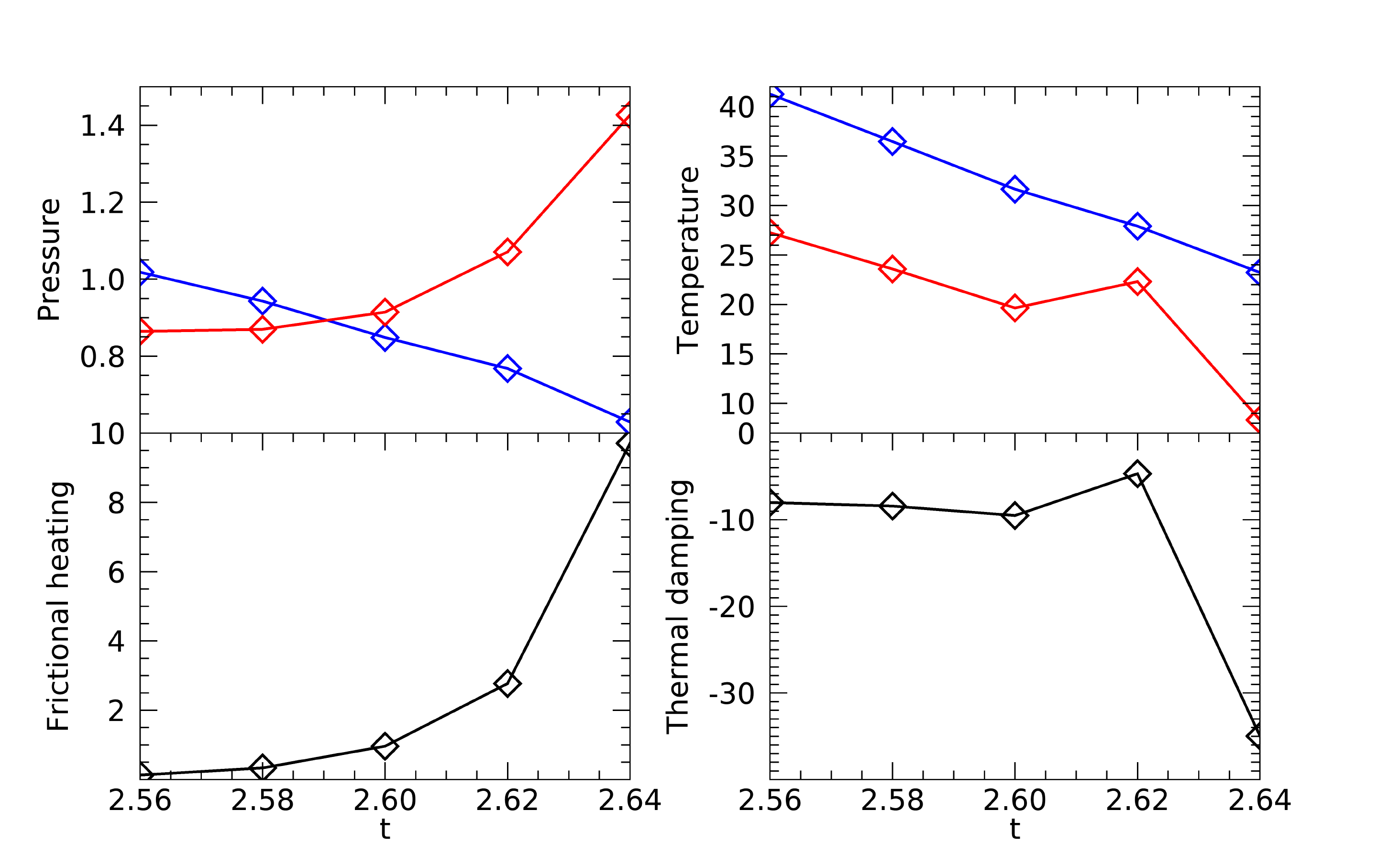}
    \caption{Evolution of pressure, temperature, frictional heating and thermal damping at the centre of secondary plasmoid 1 in the time interval $t = [2.56, 2.64]$. In the top panels plasma properties are displayed in blue while neutral properties are shown in red.}
    \label{fig:secondary_plasmoids_properties}
\end{figure}

The thermal coupling between ions and neutrals, whose terms are displayed on the right hand side of equations (\ref{eq:neutral_energy_2}) and (\ref{plasma_energy_equation}), however, contributes to change the plasma pressure gradient with time. The effect of the thermal coupling is shown in Figure \ref{fig:secondary_plasmoids_properties}, where the plasma pressure, the plasma temperature and the two terms associated with the coupling with the neutrals (frictional heating and thermal damping) are displayed at the centre of the secondary plasmoid 1 in the time interval $t = [2.56, 2.64]$. The thermal damping term, whose non-dimensional definition is $(3/2 \gamma) \alpha_c (T_n , T_p, v_D )\rho_n \rho_p (T_n - T_p)$, drives the thermal equilibrium. When negative, the thermal damping indicates that energy is transferred from the hotter plasma to the neutrals. The plasma pressure decreases, together with the plasma temperature, under the effect of the thermal damping that reaches more negative values in time, a trend that is associated with energy passing from the plasma to the neutrals. The trend shown by the thermal damping reflects the neutral temperature, which tends to the plasma temperature until $t = 2.62$: after this time the plasmoid begins moving faster to the end of the current sheet as shown in Figure \ref{fig:plasmoids_timelapse}. The frictional heating is also seen to increase remarkably with respect to time, as a result of the combined effect of the neutral pressure increasing from the inflow and the plasmoid motion. The coupling with the neutrals acts on $\nabla p_p$ by sharpening its peak at the plasmoid centre, but the gradient continues to be a relevant contribution to the total force. For this reason, the secondary plasmoids do not become completely force-free before they are expelled from the current sheet.

\begin{figure}[htb]
    \centering
    \includegraphics[width=\columnwidth,clip=true,trim=0cm 6.5cm 0cm 0cm]{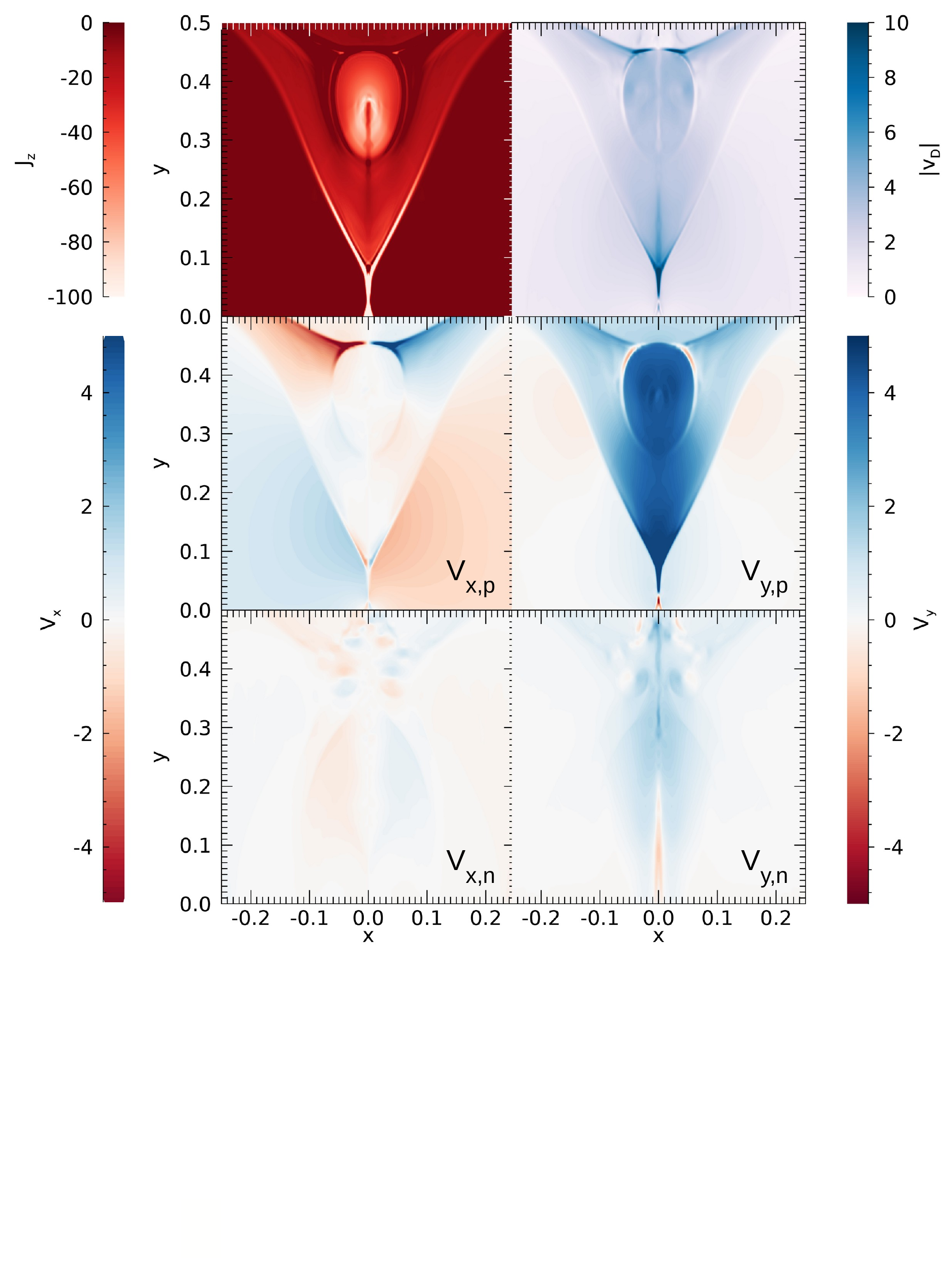}
    \caption{Detail of the secondary plasmoid leaving the current sheet at $t = 2.68$, seen in the current density $J_z$ (top left) and drift velocity magnitude (top right). The plasma velocity components are displayed in the central panels ($v_{x,p}$ on the left, $v_{y,p}$ on the right), while the neutral velocity components are shown in the bottom panels ($v_{x,n}$ on the left, $v_{y,n}$ on the right).}
    \label{fig:secondary_plasmoids_contour}
\end{figure}

The detail of one of the secondary plasmoids reconnecting at one end of the current sheet is shown in Figure \ref{fig:secondary_plasmoids_contour}, where $J_z$, the drift velocity magnitude, and the plasma and neutral velocity components are displayed. The plasma flow is faster than the neutrals, and this leads to a non-negligible drift velocity between the two fluids. Looking at both the current density and the drift velocity magnitude, a thin elongated vertical structure is observed from the current sheet to the centre of the secondary plasmoid. This structure, visible in red in the bottom right panel of Figure \ref{fig:secondary_plasmoids_contour} (neutral $v_{y}$), is a jet in the neutral flow, which extends in the direction opposite to the plasmoid motion, going back to the current sheet. This jet, accelerated by the neutral pressure gradient inside the plasmoid, is present only in correspondence of secondary plasmoids: there is no similar structure forming in the bigger coalescing plasmoids. The absence of this feature might depend on the fact that the bigger plasmoids are initially in a force-free condition and the pressure gradients are too small to expel the neutrals through a jet.

\subsection{Extended neutral reconnection jet}
\label{sec:neutral_jet}

A prominent feature developing in the PIP simulation is the formation of a jet-like structure that extends asymmetrically along the $y-$axis during coalescence. This large structure must be distinguished from the small-scale neutral jets discussed in Section \ref{sec:secondary_plasmoids}. In standard reconnection models magnetic energy can be released to form a plasma jet, a feature that is also found in many observations.

\begin{figure}[htb]
    \centering
    \includegraphics[width=\columnwidth,clip=true,trim=0cm 0cm 14cm 0cm]{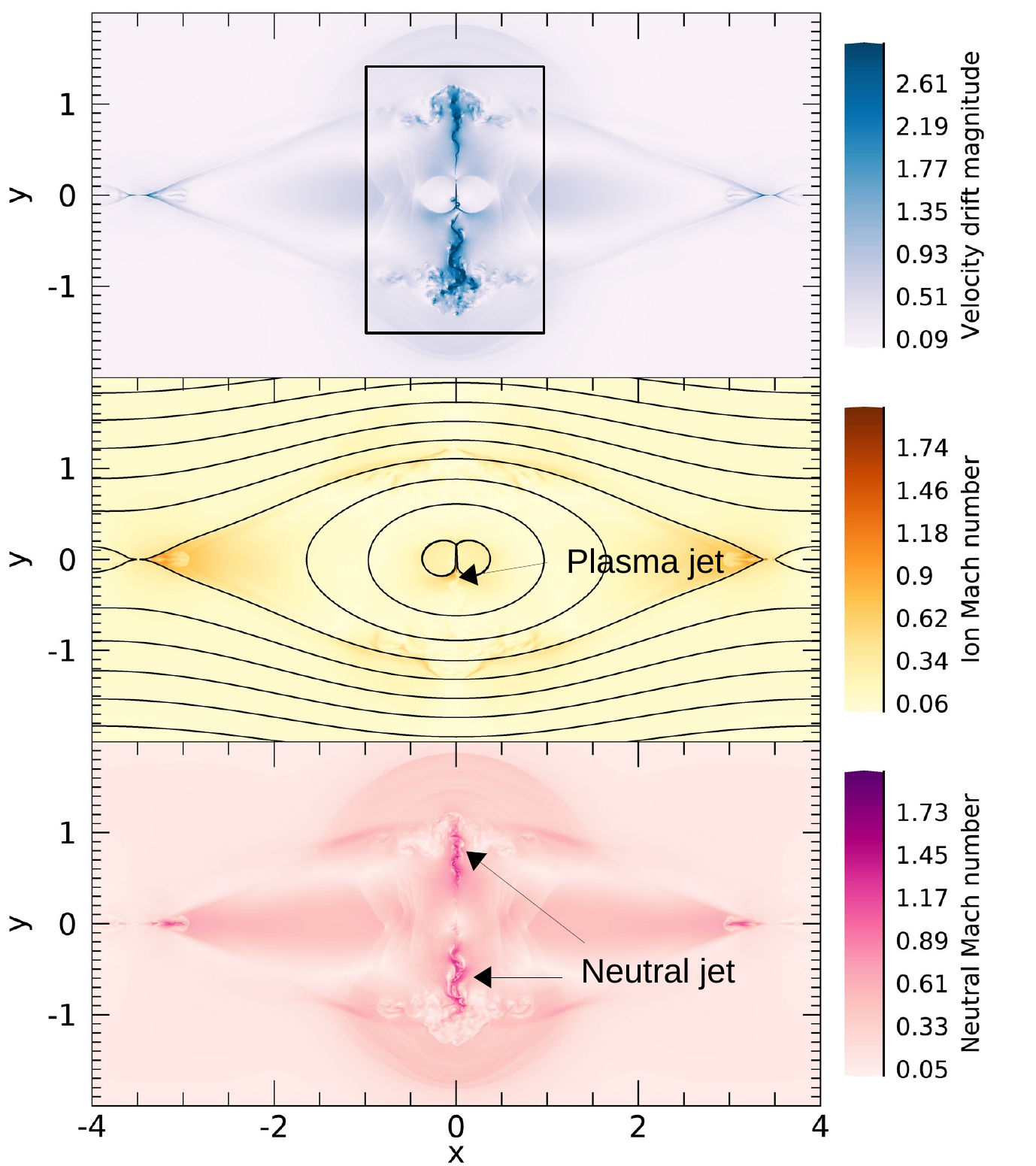}
    \caption{Top: drift velocity magnitude $|v_D|$ between neutral and charged fluids. Centre: ion Mach number $v_p / c_{s,p}$. The magnetic field lines are shown in black. Bottom: neutral Mach number $v_n / c_{s,n}$. All the plots display the quantities distribution at $t = 3.48$. The box in the top panel shows the domain selection displayed in Figure \ref{fig:jet_detail}.}
    \label{fig:jet_vel_profile}
\end{figure}

\begin{figure*}[htb]
    \centering
    \includegraphics[width=0.95\textwidth,clip=true,trim=3cm 0cm 3cm 0cm]{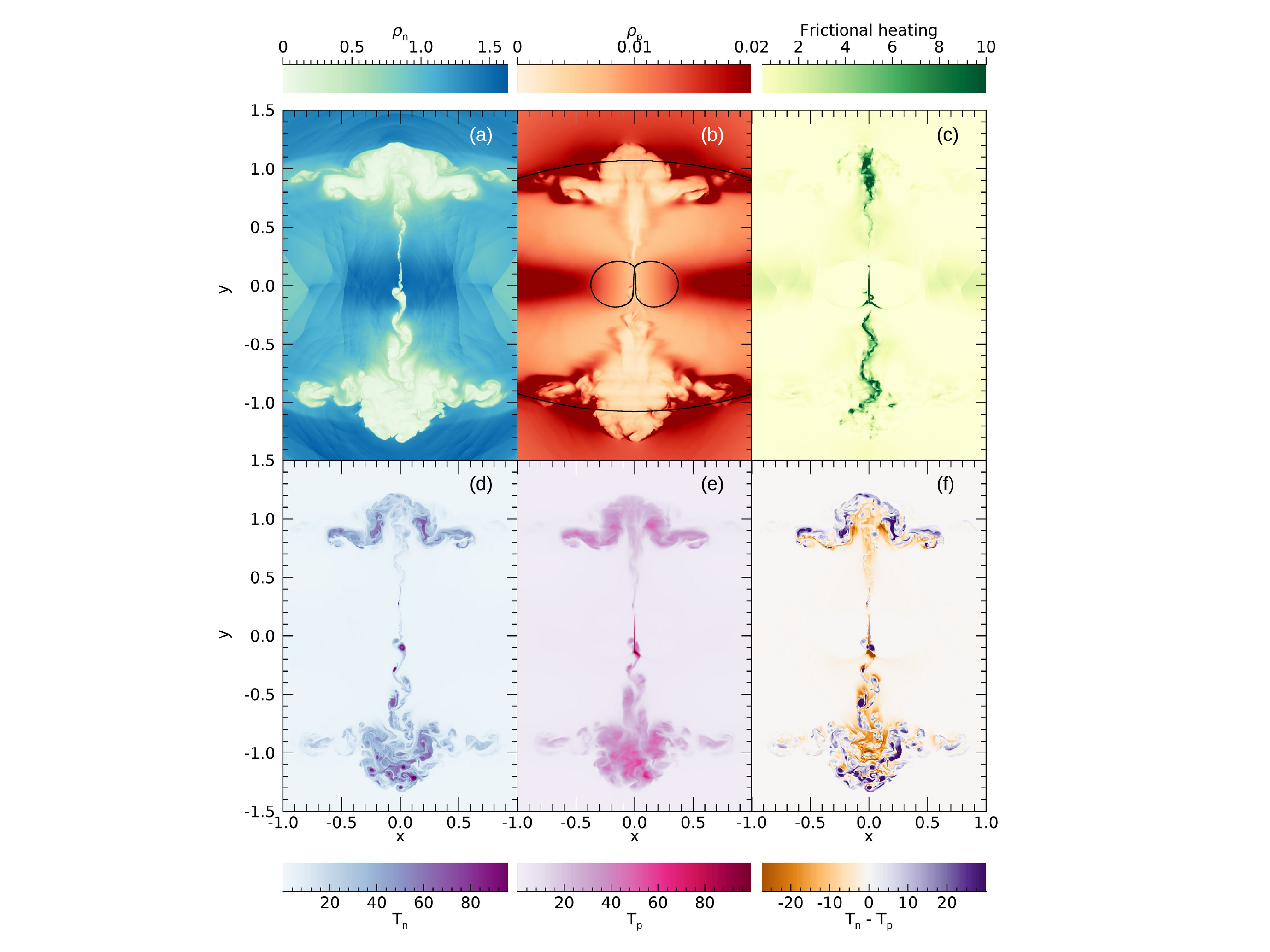}
    \caption{Profiles of neutral (panel \textit{a}) and ion density (panel \textit{b}), frictional heating (panel \textit{c}), neutral (panel \textit{d}) and plasma temperature (panel \textit{e}) and difference between neutral and ion temperatures (panel \textit{f}) are displayed at $t = 3.48$. In panel \textit{b}, the separatrices are shown in black. The selected region is shown in Figure \ref{fig:jet_vel_profile}.}
    \label{fig:jet_detail}
\end{figure*}

From Figure \ref{fig:jet_vel_profile}, however, we can clearly see that the neutral jet is significantly longer than the plasma jet. The ion velocity increases to supersonic values along the reconnection region, but the enhancement is localised near the centre of the domain. The velocity of the extended neutral jet is supersonic, and the neutral Mach number reaches values of $\sim 1.6$. The larger drift velocity in Figure \ref{fig:jet_vel_profile} shows that the species are significantly decoupled in the jet. 

A more detailed picture of the interaction of the two fluids along the jet structure can be provided by looking at the physical properties in the smaller region where the jet develops. Such region is identified in the top panel of Figure \ref{fig:jet_vel_profile}. Plots of neutral and plasma densities and temperatures, frictional heating and temperature difference between the two species are shown in Figure \ref{fig:jet_detail}.

The decoupling of neutrals and plasma along the jet is favoured by the very low density of both species, as shown in panels (a) and (b) of Figure \ref{fig:jet_detail}. The two species reach similar peaks in temperature, as shown in panels (d) and (e) of Figure \ref{fig:jet_detail}, but the heating distribution is different for each fluid. The species are thermally decoupled, as shown by the difference between the neutral and ion temperatures in panel (f) of Figure \ref{fig:jet_detail}. During its evolution, the jet appears to be very turbulent. There is presence of many coherent vortices mostly concentrated at the jet truncation that are particularly evident in panels (c) and (f) of Figure \ref{fig:jet_detail}. Along the jet the ion temperature is the highest and reaches its maximum in correspondence of the current sheet, while the neutral temperature is higher than the ion temperature at the centre of the vortices. Neutrals and ions are however heated up in the current sheet and along the jet by the thermal energy. The thermal energy is released through the frictional heating, defined in Equation (\ref{eq:frictional_heating}), which is associated to collisions between the two fluids and it is shown in panel (c) of Figure \ref{fig:jet_detail}.

The velocity difference at the interfaces between the jet and the environment leads to the onset of shear flow instabilities. The sinusoidal shape of the jet is characteristic of the Kelvin-Helmholtz instability (KHI), a classical shear flow instability that tears apart vorticity sheets at the surface of separation of the two fluids \citep{1981STIA...8217950D,DRAZIN2015343, 2018ApJ...864L..10H}. In order to confirm whether the system is KH unstable, we compare the neutral jet to the simple Bickley jet, a steady two-dimensional laminar jet which is unstable to the sinusoidal-mode of the KHI \citep{1981STIA...8217950D}. Under the action of the KHI, the Bickley jet develops a sinusoidal structure at a preferred wavelength of $\sim 6.3$ times the characteristic flow half width.

The instability wavelength $\lambda$ can be calculated at $t = 3.16$, where the instability is taking place from $y = 0$ and propagating downward along the jet. Measuring the wavelength as the distance between two vortices on the same side of the jet, $\lambda \sim 0.115$. The average half width of the jet, calculated as the distance between the peaks of maximum and minimum vorticity, is $\delta \sim 0.017$. We find an aspect ratio $\lambda / \delta$ of 6.6, which is similar to the value predicted for the Bickley jet undergoing KHI. We can conclude that the jet is subject to the KHI.

For our jet, the KHI is seen to evolve to a turbulent state. At the termination point, in this location shocks are generated and can be seen as weak structures in the neutral and drift velocities in Figure \ref{fig:jet_vel_profile} (top and bottom panels). These structures are the hydrodynamic shocks discussed in Section \ref{sec:shocks}.

\section{Parameter survey}
\label{sec:parameters}

We investigate the changes in the coalescence process due to the diffusivity ($\eta$), the collisional coupling ($\alpha_c$), the ion fraction ($\xi_p$) and the plasma $\beta$. In the following section we present a survey over these four key parameters of our physical system. The simulations are identified by numbers, and the respective physical  parameters and the spatial resolution are listed in Table \ref{tab:parameters}.

\begin{table*}
\caption{List of the simulation parameters. \label{tab:parameters}}
\begin{ruledtabular}
\begin{tabular}{cccccccccc}
 Nr. & Type & $\eta$ & $\alpha_c$ & $\xi_p$ & $\beta$ &  Nr. $x$ grid points & Nr. $y$ grid points & $\Delta x$ & $\Delta y$\\
\hline
 1 & MHD & 0.0005 & $\infty$\footnotemark[1] & 1\footnotemark[1] & 0.1 & 2062 & 3086 & $1.95 \cdot 10^{-3}$ & $2.6 \cdot 10^{-3}$ \\ 
 2 & PIP & 0.0005 & 100 & 0.01 & 0.1 & 6478 & 4862 & $1.2 \cdot 10^{-3}$ & $1.6 \cdot 10^{-3}$ \\ 
 \hline
 3 & PIP & 0.0015 & 100 & 0.01 & 0.1 & 2062 & 3086 & $1.95 \cdot 10^{-3}$ & $2.6 \cdot 10^{-3}$ \\
 4 & PIP & 0.005 & 100 & 0.01 & 0.1 & 1038 & 1550 & $3.9 \cdot 10^{-3}$ & $5.2 \cdot 10^{-3}$ \\
 5 & PIP & 0.015 & 100 & 0.01 & 0.1 & 1038 & 1550 & $3.9 \cdot 10^{-3}$ & $5.2 \cdot 10^{-3}$ \\
 6 & PIP & 0.05 & 100 & 0.01 & 0.1 & 1038 & 1550 & $3.9 \cdot 10^{-3}$ & $5.2 \cdot 10^{-3}$ \\
 7 & PIP & 0.15 & 100 & 0.01 & 0.1 & 1038 & 1550 & $3.9 \cdot 10^{-3}$ & $5.2 \cdot 10^{-3}$ \\
 8 & PIP & 0.5 & 100 & 0.01 & 0.1 & 1038 & 1550 & $3.9 \cdot 10^{-3}$ & $5.2 \cdot 10^{-3}$ \\
 \hline
 9 & MHD & 0.0015 & 0\footnotemark[1] & 0.01\footnotemark[1] & 0.1 & 2062 & 3086 & $1.95 \cdot 10^{-3}$ & $2.6 \cdot 10^{-3}$ \\
 10 & PIP & 0.0015 & 1 & 0.01 & 0.1 & 2062 & 3086 & $1.95 \cdot 10^{-3}$ & $2.6 \cdot 10^{-3}$ \\
 11 & PIP & 0.0015 & 10 & 0.01 & 0.1 & 2062 & 3086 & $1.95 \cdot 10^{-3}$ & $2.6 \cdot 10^{-3}$ \\
 12 & PIP & 0.0015 & 1000 & 0.01 & 0.1 & 2062 & 3086 & $1.95 \cdot 10^{-3}$ & $2.6 \cdot 10^{-3}$ \\
 13 & PIP & 0.0015 & 3000 & 0.01 & 0.1 & 2062 & 3086 & $1.95 \cdot 10^{-3}$ & $2.6 \cdot 10^{-3}$ \\
 14 & MHD & 0.0015 & $\infty$\footnotemark[1] & 1\footnotemark[1] & 0.1 & 2062 & 3086 & $1.95 \cdot 10^{-3}$ & $2.6 \cdot 10^{-3}$ \\
 \hline
 15 & PIP & 0.0015 & 100 & 0.5 & 0.1 & 2062 & 3086 & $1.95 \cdot 10^{-3}$ & $2.6 \cdot 10^{-3}$ \\
 16 & PIP & 0.0015 & 100 & 0.1 & 0.1 & 2062 & 3086 & $1.95 \cdot 10^{-3}$ & $2.6 \cdot 10^{-3}$ \\
 17 & PIP & 0.0015 & 100 & 0.001 & 0.1 & 2062 & 3086 & $1.95 \cdot 10^{-3}$ & $2.6 \cdot 10^{-3}$ \\
 \hline
 18 & PIP & 0.0015 & 100 & 0.01 & 1 & 1038 & 1550 & $3.9 \cdot 10^{-3}$ & $5.2 \cdot 10^{-3}$ \\
 19 & PIP & 0.0015 & 100 & 0.01 & 0.01 & 1038 & 1550 & $3.9 \cdot 10^{-3}$ & $5.2 \cdot 10^{-3}$
\end{tabular}
\end{ruledtabular}
\footnotetext[1]{These data are the effective values of the two-fluid parameters $\alpha_c$ and $\xi_p$ for the single-fluid cases, which are chosen as limits for the PIP simulations.}
\end{table*}

The PIP simulations in this Section have the same resolution as the MHD cases or an even lower resolution ($\Delta x = 3.9 \cdot 10^{-3}$, $\Delta y = 5.2 \cdot 10^{-3}$). Due to the parameter variation changing the size of the central current sheet, it was possible to use a lower resolution without losing the possibility to resolve the current sheet.

\subsection{Variation of resistivity}
\label{sec:resistivity}

We begin the investigation of partial ionisation effects on the coalescence instability by considering the role of varying the resistivity in PIP simulations. The seven cases that are examined in this Section ($\eta =$ 0.0005, 0.0015, 0.005, 0.015, 0.05, 0.15 and 0.5) are listed in Table \ref{tab:parameters} with the numbers from 2 to 8. 
\begin{figure}[htb]
    \centering
    \includegraphics[width=\columnwidth,clip=true,trim=3cm 1.2cm 1cm 2cm]{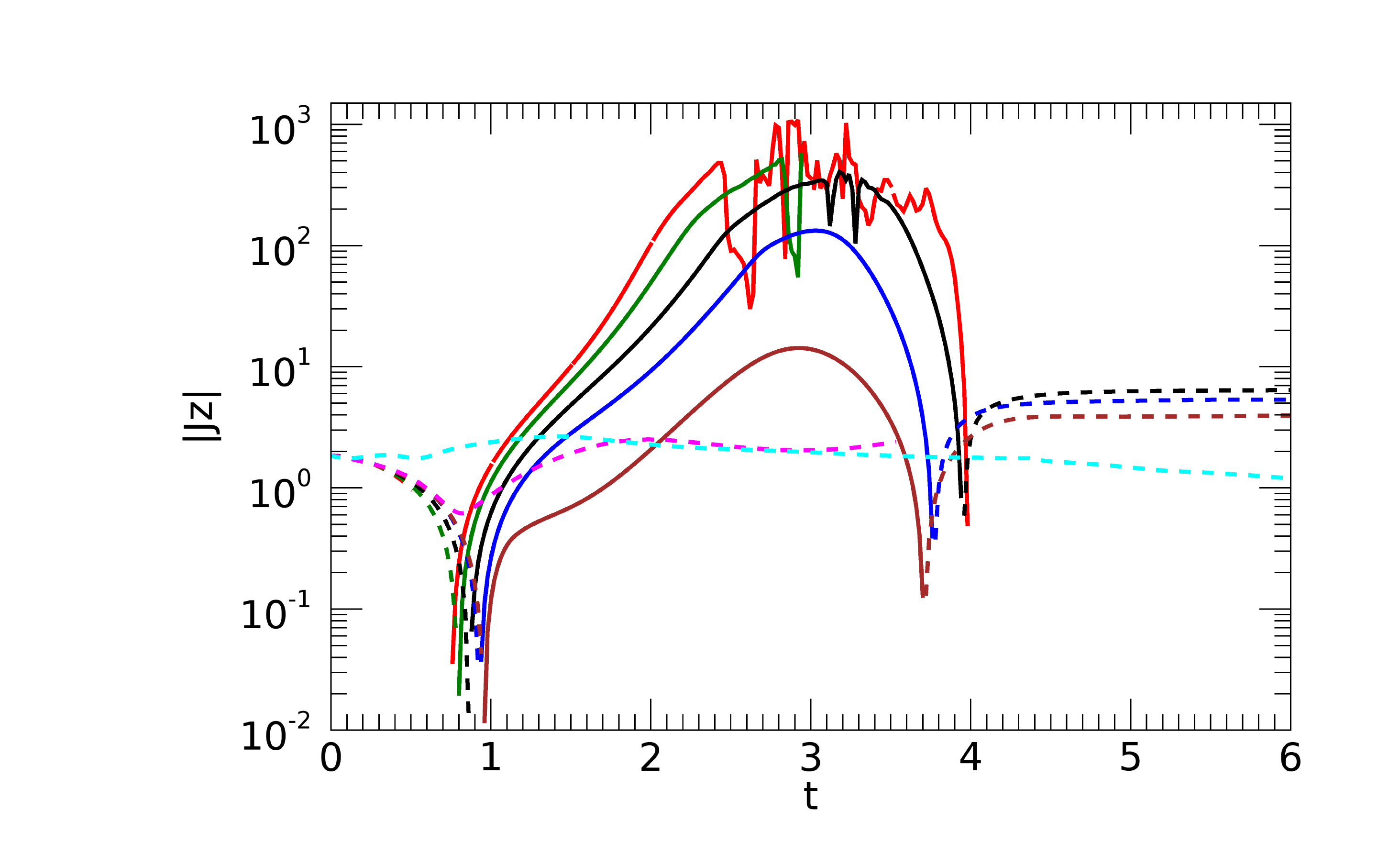}
    \caption{Evolution over time of the current density $J_z $ at the central point of the current sheet formed during coalescence $(x=0, y=0)$ at the variation of the initial resistivity. The cases displayed are for $\eta = 0.0005$ (red), $\eta=0.0015$ (green), $\eta=0.005$ (black), $\eta=0.015$ (blue), $\eta=0.05$ (brown), $\eta=0.15$ (magenta) and $\eta = 0.5$ (cyan). The solid lines indicate the negative part of the curves, the dashed lines the positive part.}
    \label{fig:res_current}
\end{figure}
The magnitude of $J_z$ at the centre of the current sheet ($x=0$, $y=0$) is displayed in Figure \ref{fig:res_current}, and it is seen to decrease as result of the increasing diffusion. The chosen location allows us to identify the beginning and the end of reconnection with the formation of the final plasmoid.

For $\eta = 0.005$ (in black in Figure \ref{fig:res_current}) reconnection happens in the central current sheet with the formation of some secondary plasmoids, although fewer with respect to the less diffusive cases ($\eta = 0.0005$, examined in Section \ref{sec:simulations}, and $\eta = 0.0015$). The secondary plasmoids formation and expulsion are identified by a sudden drop (when the plasmoid is formed) followed by a drastic increase back to previous values of $|J_z|$ as soon as it moves away from the centre of the current sheet. In the simulation with $\eta = 0.015$ the current is highly diffused. During the plasmoid merger there is no observed formation of secondary plasmoids. This is the first case in which the diffusion is high enough to prevent the formation of secondary plasmoids during coalescence in a PIP case. A similar result is obtained for the case with $\eta = 0.05$. The cases having the highest resistivity ($\eta = 0.15, 0.5$) do not develop the coalescence instability, as the initial plasmoids are quickly diffused. As $\eta$ is increased, the Lundquist number $S$ decreases: when $S$ becomes sufficiently low, the tearing instability stops taking place in the central current sheet, thus explaining the lack of secondary plasmoids.

The variation of resistivity also plays a role in changing the timescale of coalescence. For higher $\eta$, coalescence starts at a later time in each simulation. This is shown in Figure \ref{fig:res_current} by the position in time of the first drops in $J_z$. Such effect depends on how efficiently the current sheet is generated in the phase where the two plasmoids are approaching, as for smaller $\eta$ it is easier to build up current.

\subsection{Variation of collisional frequency}
\label{sec:alpha_c}

Here we investigate the effects of the collisional coupling between ions and neutrals. We compare the simulations listed in Table \ref{tab:parameters} with the numbers 3, 9, 10, 11, 12, 13 and 14. An MHD case with total density equal to $\rho_p = 1$ (simulation 14) is taken as the limit case $\alpha_c \rightarrow \infty$. A second MHD case (simulation 9) having initial density equal to $\rho_p = 0.01$ is considered as the limit for $\alpha_c = 0$ (i.e. no collisions), as this is the same plasma density set for the PIP cases. For $\alpha_c=\infty$ the neutral and plasma species are entirely coupled hence the system behaves like a single fluid MHD model, with the density and pressure being the bulk (ion + neutral) values. For $\alpha_c=0$, the species are completely decoupled, i.e., the plasma evolves independently from the neutrals and can be considered to be a single fluid MHD system with the density/pressure based on the plasma values only. An equivalent MHD simulation can be performed by changing the initial plasma beta. While the magnetic field strength is unchanged, the variation of plasma beta modifies the pressure. To maintain the same initial temperature of the calculation, we use a lower density (the same as the plasma density of our two-fluid calculations). Therefore, the difference in the plasma density between the two limit MHD cases result in an effective difference in the plasma $\beta$ of the two simulations.

\begin{figure}[htb]
    \centering
    \includegraphics[width=\columnwidth,clip=true,trim=0.7cm 1cm 1cm 2cm]{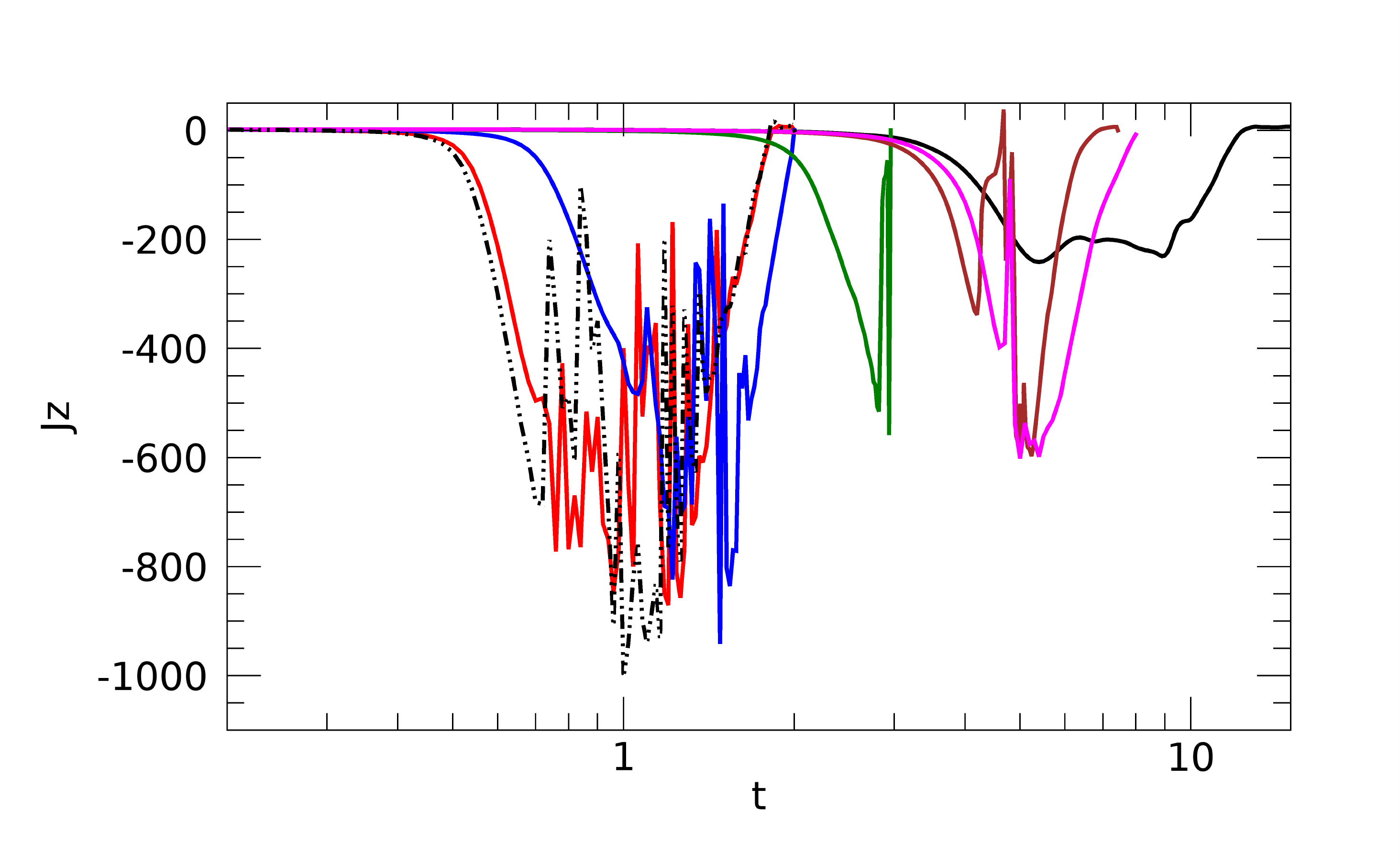}
    \caption{Evolution over time of the current density $J_z $ at the central point of the current sheet formed during coalescence $(x=0, y=0)$ at the variation of the initial collisional coupling. The cases displayed are the MHD limit for $\alpha=0$ (black, dotted-dashed line), the PIP simulations with $\alpha=1$ (red), $\alpha=10$ (blue), $\alpha = 100$ (green), $\alpha=1000$ (brown) and $\alpha=3000$ (magenta), and the MHD limit for $\alpha \rightarrow \infty$ (black, solid line).}
    \label{fig:current_different_alpha}
\end{figure}

In Figure \ref{fig:current_different_alpha} we display the variation of current density $J_z$ at $x,y = 0$ with respect to time. The beginning of reconnection is identified with the first minimum occurring in the current density, when the current sheet is compressed the most by the two plasmoids. Decreasing $\alpha_c$ the timescale for the plasmoid coalescence becomes considerably shorter. The timescale does not vary linearly, but shows the presence of two accumulation points, corresponding to the two limits identified by the MHD simulations. At lower $\alpha_c$ ($\alpha_c = 1, 10$), the timescale tends to approach the one for $\alpha_c \rightarrow 0$, while at higher $\alpha_c$ ($\alpha_c = 1000, 3000$) coalescence takes place in a time interval similar to the one obtained for $\alpha_c \rightarrow \infty$.

The more ions and neutrals are coupled, the later reconnection starts. The slowing down of the first phase prior to the onset of reconnection can be associated to the damping effect that neutrals have on the ions in the inflow, which increases at the increase of $\alpha_c$ as the ions (which fraction is much smaller than the neutral fraction) interact with a higher number of neutrals. This result is consistent with previous simulations \cite{2008A&A...486..569S}. We can understand this result by comparing the ion Alfv\'en time $\tau_{A,p}$ with both the ion and the neutral collisional times. This leads to the identification of two values for $\alpha_c$ that define when the species couple with each other through collisions. For $\alpha_c \sim 20$, the plasma-neutral collisional time $\tau_{\operatorname{col,pn}} = (\alpha_c \rho_n)^{-1}$ becomes smaller than $\tau_{A,p}$ and the ions couple with the neutrals. The coupling of the neutrals on the ions, that happens when the neutral-plasma collisional time $(\alpha_c \rho_p)^{-1} $ becomes smaller than $ \tau_{A,p}$, takes place for $\alpha_c \sim 2000$.

\begin{figure}[htb]
    \centering
    \includegraphics[width=\columnwidth,clip=true,trim=0.2cm 0cm 2cm 0.5cm]{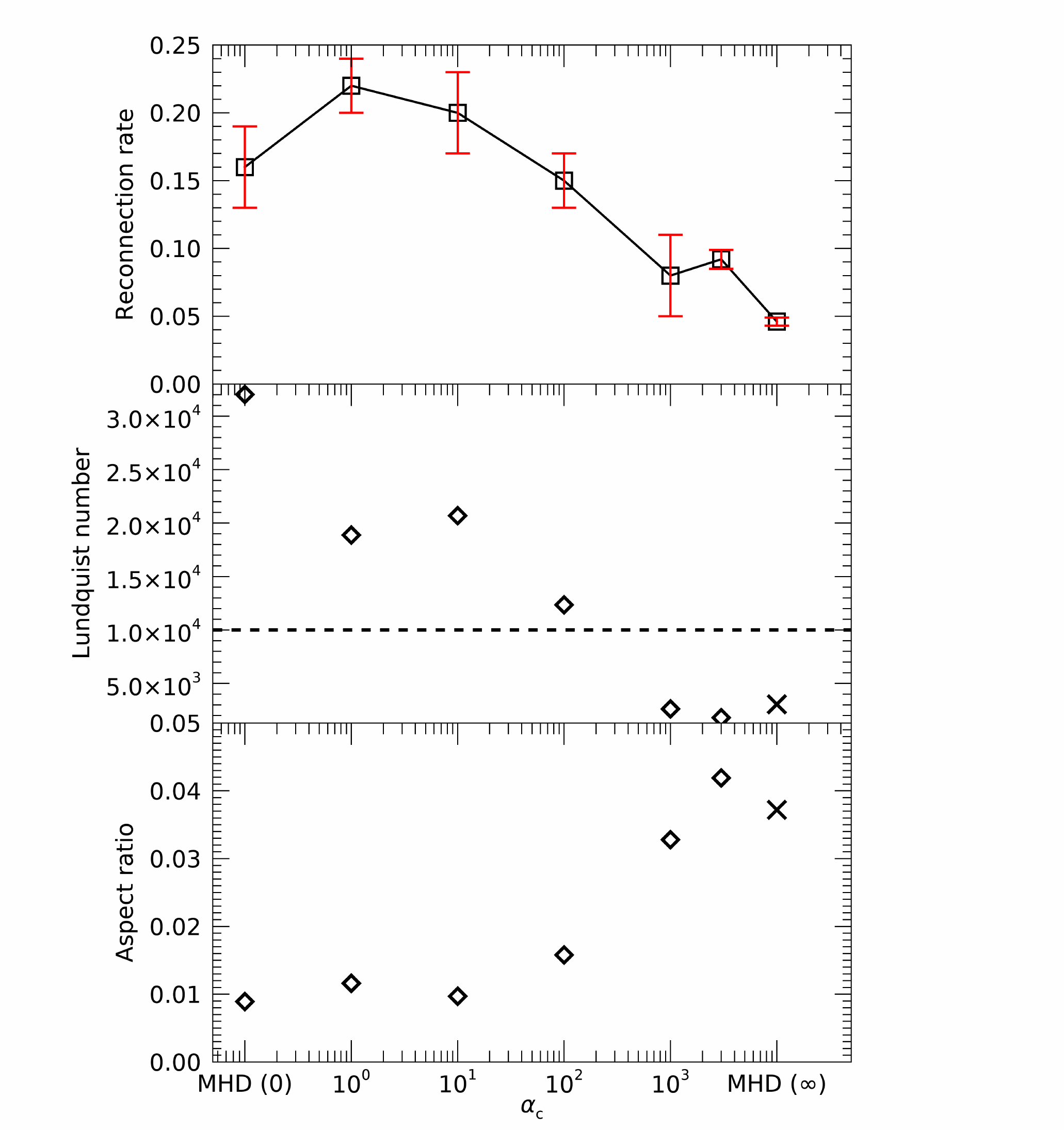}
    \caption{Top: average reconnection rate as a function of $\alpha_c$. The error bar is associated to the standard deviation on each measure. Centre: Lundquist number $S$ with respect to $\alpha_c$ calculated using the effective Alfv\'en speed $v_{A,e}$. The horizontal dashed line indicates the threshold limit of $S = 10^4$ for the onset of the tearing instability. Bottom: current sheet aspect ratio $\delta / \Delta$ as a function of $\alpha_c$. In both central and bottom panels, diamonds are associated to simulations with secondary plasmoids, crosses are associated to cases without secondary plasmoids.}
    \label{fig:alpha_properties}
\end{figure}

As $\alpha_c$ increases, reconnection takes a longer time before the plasmoids are completely merged. This results in a decrease of the average reconnection rate as $\alpha_c$ increases, as shown in the top panel of Figure \ref{fig:alpha_properties}, where the error bar is given by the standard deviation. The reduced reconnection rate can be explained by the variation of the effective Alfv\'en speed, as introduced for the two-fluid case in Section \ref{sec:reconnection_rate}. If the collisional frequency is increased, the effective density of the magnetic fluid increases as more neutrals interact with the ions and consequently $v_{A,e}$ decreases. The variation of the Lundquist number is shown in the central panel of Figure \ref{fig:alpha_properties}, as calculated from $v_{A,e}$.

After reconnection begins, the PIP cases (Figure \ref{fig:current_different_alpha}) show a strong fluctuation of the current towards less negative values, behaviour that is not present for a full coupling (MHD case for $\alpha_c \rightarrow \infty$). This fluctuation is a consequence of plasmoid formation in the current sheet. The growth and expulsion of plasmoids contributes initially to slow down the reconnection by saturating the negative current in a blob between the merging plasmoids, and then to leave the current sheet unstable, leading to the formation of further plasmoids. For the simulations with secondary plasmoids, $S$ is calculated at the time before the formation of the first plasmoid in the central current sheet, while in case of simulations without secondary plasmoid $S$ is evaluated at the first minimum of $J_z$ at the centre of the current sheet. The value of the Lundquist number decreases at the increase of the collisional coupling, following the variation of both the effective Alfv\'en speed and the characteristic length of the current sheet. $S$ is more sensitive to the variation of $v_{A,e}$, as such factor varies over an interval covering orders of magnitude, while $\Delta$, being about the same size of the plasmoids, displays a lesser variation in length across the cases. As shown in Figure \ref{fig:alpha_properties}, $S$ appears to be above the threshold of $10^4$ for all the cases showing formation of secondary plasmoids, with the exception of the case with $\alpha_c =1000$ and 3000. The effects that might lead to the sub-critical plasmoid formation we have found here will be discussed in Section \ref{sec:subcritical}.

\begin{figure*}[htb]
    \centering
    \includegraphics[width=\textwidth,clip=true,trim=0cm 6cm 0cm 0cm]{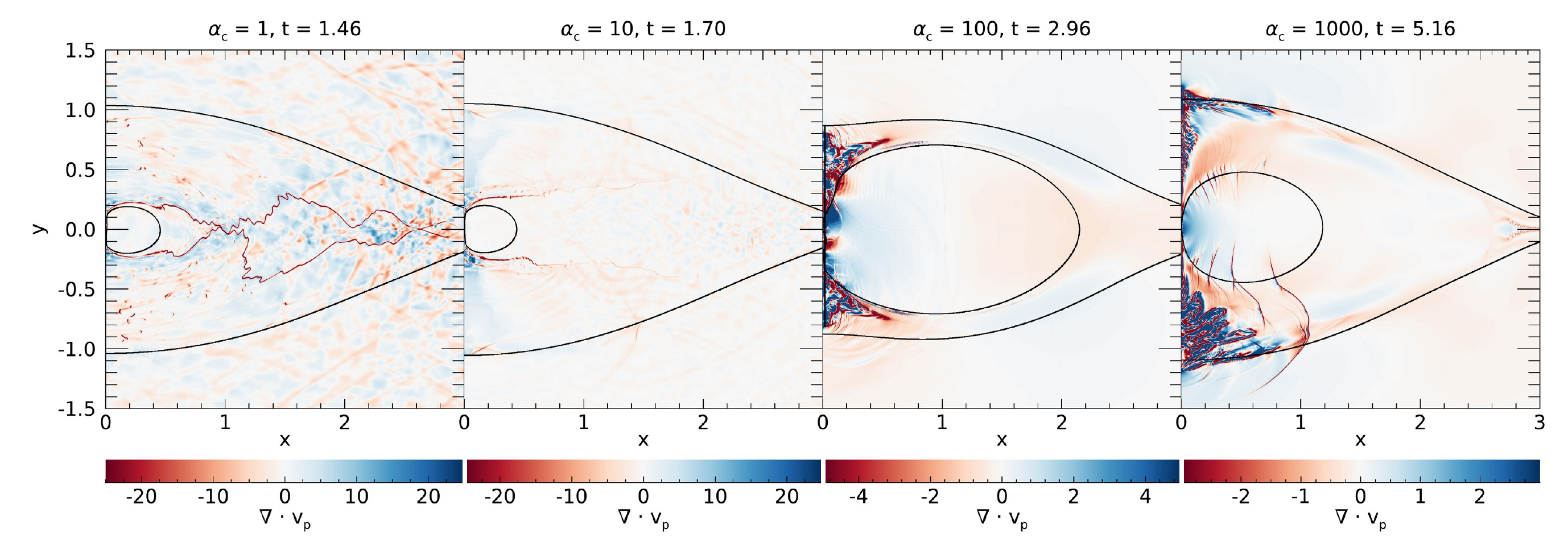}
    \caption{Divergence of the plasma velocity for the PIP cases at $\alpha = 1, 10, 100$ and 1000. The plots are saturated in order to enhance the structures associated to slow-mode shocks. The separatrices are displayed as black contour lines.}
    \label{fig:alpha_shocks}
\end{figure*}

Looking at the results of Section \ref{sec:shocks}, slow-mode shocks are apparently absent in the PIP case with $\alpha_c = 100$. However, slow-mode shocks can be generate in two-fluid environments \cite{refId0}. Therefore, we want to investigate whether slow-mode shocks are produced at a higher or lower collisional coupling that approach the MHD cases, and whether they form but are dissipated at a later time of coalescence. Slow-mode shocks are indeed generated in the two-fluid simulations, and their propagation is clearly visible particularly at low $\alpha_c$, as shown in the first two panels (simulations 10 and 11) of Figure \ref{fig:alpha_shocks}. At the increase of collisional coupling ($\alpha_c \ge 10$) they are damped and disappear as a consequence of two-fluid effects. At higher collisional frequencies ($\alpha_c = 100, 1000$ displayed in the last two panels of Figure \ref{fig:alpha_shocks}), the turbulent motion set by the neutral jet and the propagation of hydrodynamic shocks disrupt the slow-mode shock front in proximity of the reconnection region. The presence of hydrodynamic shocks, absent for $\alpha_c \le 1$, increases with the coupling between the two species around the inflow region, as the two species that move in opposite directions are subject to an increasing interaction. For $\alpha_c = 100$ the slow-mode shocks can't be detected. Due to the better coupling, in the case with initial $\alpha_c = 1000$ the hydrodynamic shocks show a similar behaviour as the slow-mode shocks: their front moves along the magnetic field lines, as shown by their position with respect to the separatrices in Figure \ref{fig:alpha_shocks}.

\subsection{Variation of ion fraction}

In this section we compare five cases with a different ion fraction $\xi_p$ in the range $10^{-3} - 5 \cdot 10^{-1}$, correspondent to the the numbers 3, 14, 15, 16 and 17 listed in Table \ref{tab:parameters}.

\begin{figure}[htb]
    \centering
    \includegraphics[width=\columnwidth,clip=true,trim=0.7cm 1cm 1cm 2cm]{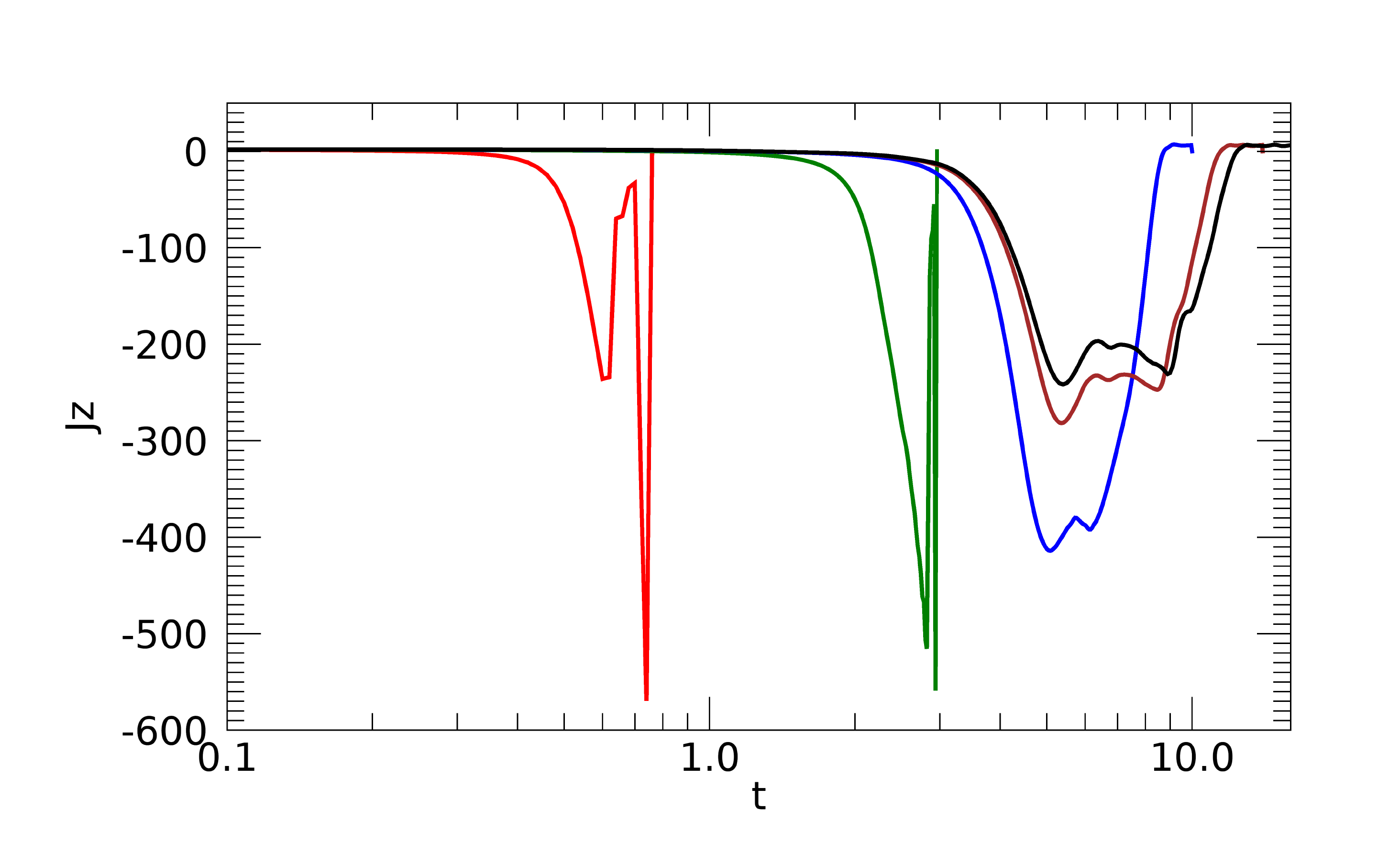}
    \caption{Evolution over time of the current density $J_z $ at the central point of the current sheet formed during coalescence $(x=0, y=0)$ at the variation of the initial ion fraction. The cases displayed are for the MHD case ($\xi_p = 1$, black) $\xi_p = 0.5$ (brown), $\xi_p = 0.1$ (blue), $\xi_p = 0.01$ (green) and $\xi_p = 0.001$ (red).}
    \label{fig:ion_current}
\end{figure}

The variation in time of $J_z$ at the centre of the current sheet is displayed in Figure \ref{fig:ion_current}, with the speed in the process of coalescence drastically increased at the decrease of $\xi_p$. Such behaviour, which shows a variation in the timescale of both ideal phase (when the plasmoids attract each other and the current sheet is formed) and reconnection phase, might be explained by similar arguments to those presented in Section \ref{sec:alpha_c}.

At the variation of the ion fraction, the initial ion Alfv\'en timescale $\tau_{A,p}$ and the ion collisional timescale increase with $\xi_p$, while the neutral collisional timescales decreases. We compare $\tau_{A,p}$ with the ion and the neutral collisional times to find the values of $\xi_p$ for which the two species couple with each other. In our range, the ion collisional time $(\alpha_c \rho_n)^{-1}$ is always smaller than $\tau_{A,p}$ with the only exception of the MHD case (simulation 14) which is taken into account as the limit value for $\xi_p = 1$. Therefore the ions are always coupled to the neutrals in all the PIP simulations in this Section. The value of $\xi_p$ below which the ion dynamics becomes fast enough to decouple from the neutrals is $\sim 4 \cdot 10^{-4}$. On the other side, the neutrals coupling on the ions takes place for $\xi_p = 0.075$. In the PIP cases with the highest ion fractions ($\xi_p = 0.1,0.5$), in which the magnetic forces are felt by a significant portion of the fluid, the coalescence develops in a similar way as in the MHD case.

\begin{figure}[htb]
    \centering
    \includegraphics[width=\columnwidth,clip=true,trim=0.2cm 0cm 2cm 0.5cm]{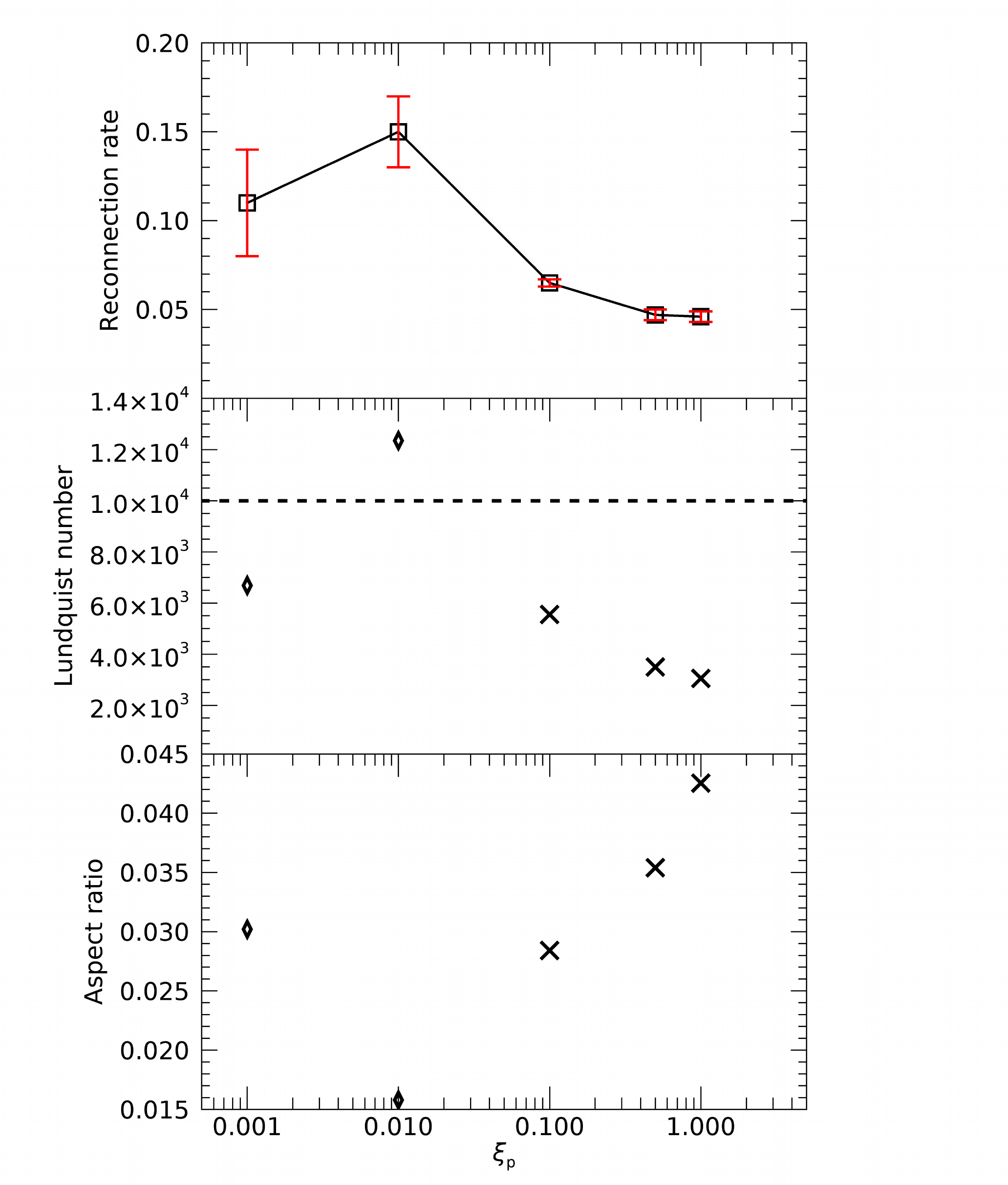}
    \caption{Top: average reconnection rate as a function of $\xi_p$. The error bar is associated to the standard deviation on each measure. Centre: Lundquist number $S$ with respect of $\xi_p$ calculated using $v_{A,e}$. The horizontal dashed line indicates the threshold limit of $S = 10^4$ for the onset of the tearing instability. Diamonds are associated to simulations with secondary plasmoids, crosses are associated with cases without secondary plasmoids. Bottom left: current sheet aspect ratio $\delta / \Delta$ as a function of $\xi_p$. }
    \label{fig:ion_properties}
\end{figure}

The reconnection rate, shown in the top panel of Figure \ref{fig:ion_properties}, decreases as the ion fraction increases, following the variation in the coalescence timescale. At the increase of $\xi_p$ the effective Alfv\'en speed $v_{A,e}$ decreases, following the increase in the plasma density. This affects the Lundquist number (central panel of Figure \ref{fig:ion_properties}), which in turn extends the timescale of the reconnection phase. Sub-critical plasmoid formation, discussed in Section \ref{sec:subcritical}, is observed for the case at the lowest ion fraction ($\xi_p = 0.001$), with $S$ below the threshold limit of $10^4$ at the onset of the tearing instability.

\subsection{Variation of plasma beta}

\begin{figure}[htb]
    \centering
    \includegraphics[width=\columnwidth,clip=true,trim=0.7cm 1cm 1cm 2cm]{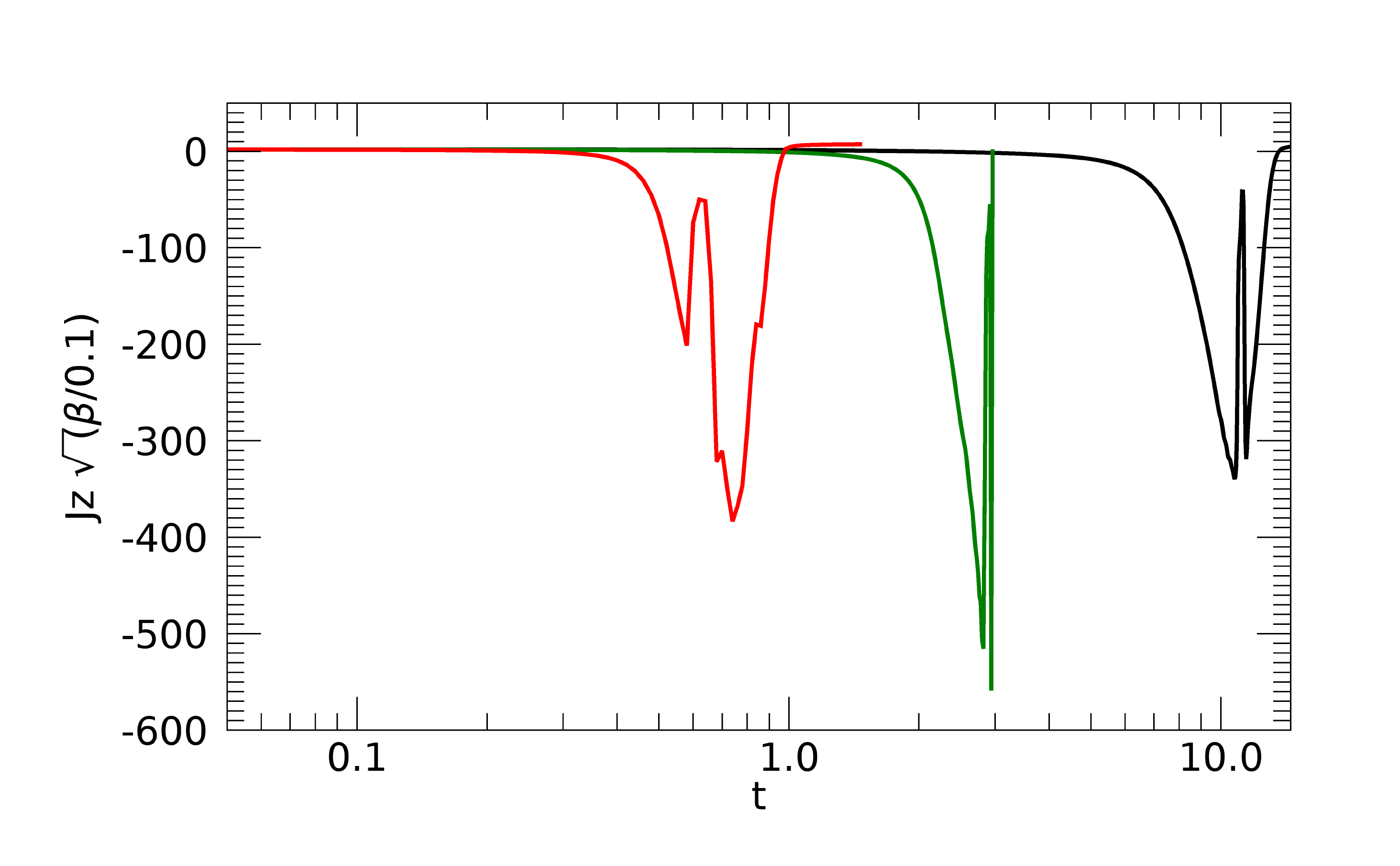}
    \caption{Evolution  over  time  of  the  current  density $J_z$ at the central point of the current sheet formed during coalescence  ($x =  0, y =  0$)  at  the  variation  of  the  initial plasma $\beta$. The cases displayed are for $\beta = 0.01$ (red), $\beta = 0.1$ (green) and $\beta = 1$ (black). For better comparison the current density $J_z$ was normalized with respect to the case with plasma $\beta = 0.1$ by multiplying all curves by a factor $\sqrt{\beta / 0.1}$.}
    \label{fig:beta_current}
\end{figure}

In the following section, three cases at different plasma $\beta$ (simulations 3, 18 and 19 in Table \ref{tab:parameters}) are investigated. The speed of coalescence is greatly affected by the change of $\beta$, and increases at the decrease of this parameter as shown in Figure \ref{fig:beta_current}. This is explained by the fact that at smaller $\beta$ the magnetic field get stronger with respect to the plasma pressure: the Alfv\'en speed increases (as its numerator increases), and so does the Lundquist number.

The consequences of plasma $\beta$ variations are not directly associated to a two-fluid effect, as the different dynamics depends uniquely on the variation of \textbf{B} and can be purely reduced to an MHD effect. However it provides important context for how changing different parameters in the two-fluid case, which effectively alter $v_A$, can increase the merger rate.

The presence of sub-critical plasmoid formation is observed for the case with the largest plasma $\beta$ ($\beta = 1$), where the formation of a single secondary plasmoid takes place for a Lundquist number smaller than $10^4$. The physics behind this case is discussed in Section \ref{sec:subcritical}.

\subsection{Sub-critical plasmoid formation}
\label{sec:subcritical}

\begin{figure}[h!]
    \centering
    \includegraphics[width=\columnwidth,clip=true,trim=0cm 0cm 2.5cm 0cm]{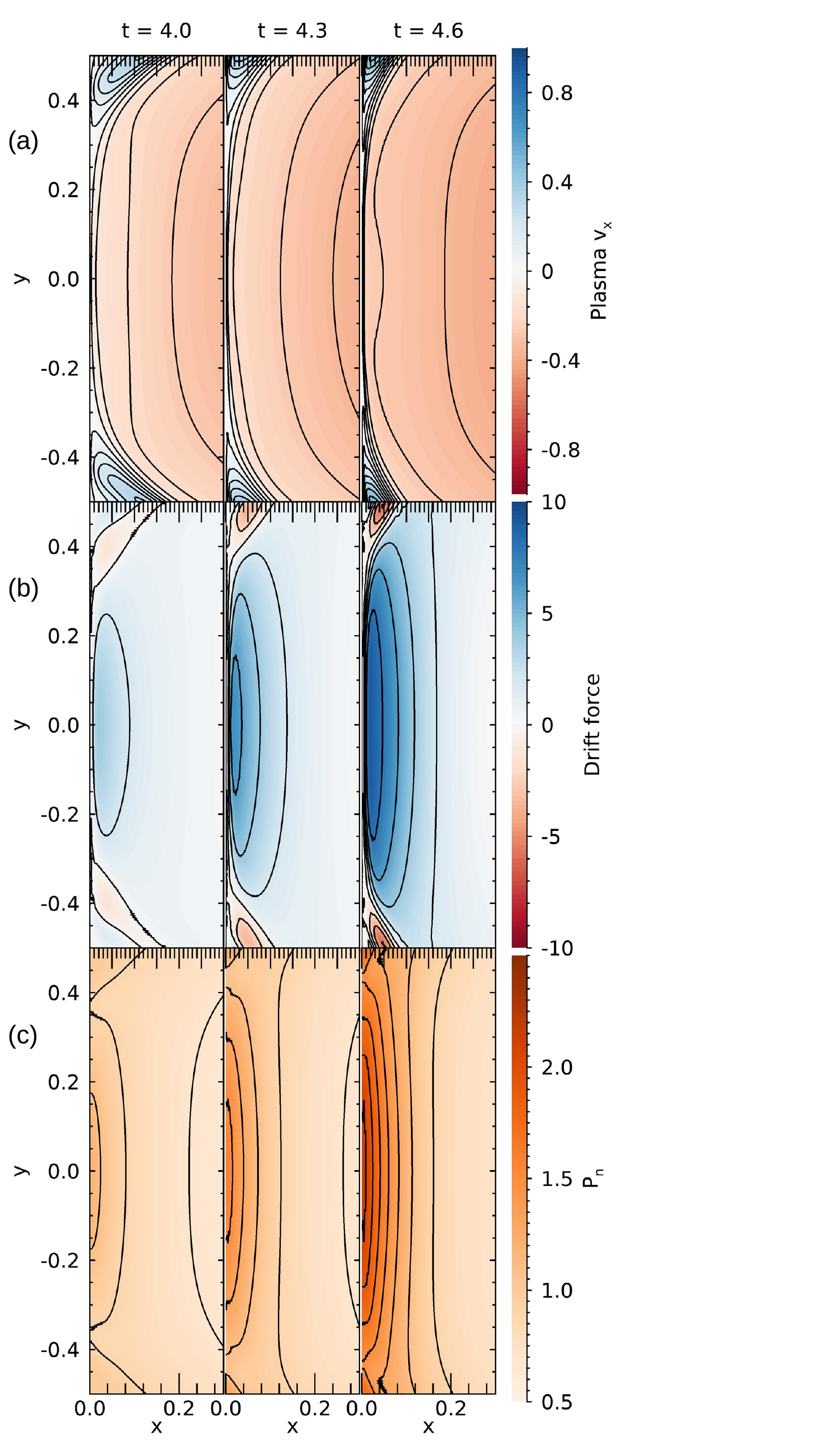}
    \caption{Evolution of $v_{p,x}$ (row $a$), $x$ component of the drift force (row $b$) and $p_n$ (row $c$) in a PIP simulation with an initial strong coupling between the two species ($\alpha_c = 3000$). In black the level lines of the variables are displayed.}
    \label{fig:alpha_3000_timelapse}
\end{figure}

Several PIP simulations presented in the previous sections (identified in Table \ref{tab:parameters} by number 2, 4, 12, 13, 17 and 18) show the signs of secondary plasmoids formation in the central current sheet below the Lundquist number threshold $S= 10^4$. The values of the critical $S$ found in these cases are consistent with the results obtained by recent studies of magnetic reconnection in a multi-fluid partially ionised plasma \cite{2018ApJ...868..144N}, already discussed in Section \ref{sec:intro}. One might expect that the lower critical Lundquist number in our simulations could depend on the initial setup, and in particular the low ion fraction. In terms of the initial setup, a comparison with previous works is difficult: many studies\cite{2012ApJ...760..109L,2013PhPl...20f1202L} are performed with a static current sheet setup, while our simulations present a driven reconnection, which leads to a different current sheet structure and dynamics. A more detailed evaluation of the role played by the initial perturbation and the random white noise in equation (\ref{velocity_perturbation}) on the onset of the tearing instability must be investigated in future developments of this research. Several studies already found that the role of the amplitude of perturbation noises \cite{2016PhPl...23c2111C,2016PhPl...23j0702C,2017ApJ...850..142C,2017ApJ...849...75H} is major in determining the critical Lundquist number and current sheet aspect ratio in a range of initial configurations. However, tests performed on the variation of the white noise perturbations on our simulations proved that the secondary plasmoids are always generated at the same time in every simulation that has the same initial set of parameters.

In this Section, we often refer to the effective Lundquist number as the critical Lundquist number for simulations displaying secondary plasmoid formation. The critical Lundquist number for the onset of the tearing instability was not isolated in previous studies\cite{2010PhPl...17f2104H,2018ApJ...868..144N}, but it was considered to fall in an interval of values whose limits are set by the absence (lower boundary) and presence (upper boundary) of plasmoid formation. Such intervals had been determined by varying two main parameters of the current sheet: characteristic length and resistivity. In our study, this interval is determined by changing the characteristic length scale of the current sheet. We do not change $L$ directly, but it varies in time as the system evolves. The time interval between two outputs is small enough that the current sheet length does not display a large variation: this reduces to have Lundquist numbers that are very close to each other before and after the first secondary plasmoid appear, which we identify by the formation of an $O-$point in the current sheet magnetic field. For such reason, we choose to identify the critical Lundquist number as a single value rather than specifying an interval.

We might expect the sub-critical plasmoid formation to be triggered in all the PIP cases because of the small ion fraction. However, the critical Lundquist number obtained for the MHD case at lower ion density (matching that of the plasma density of the PIP simulations), shown in Figure \ref{fig:alpha_properties}, is well above the threshold of $10^4$ (limit case for $\alpha_c = 0$, listed with number 9 in Table 1). This proves that the change in the plasma density itself, along with the variation of the plasma $\beta$ of the plasma that this change implies, does not affect the system so that it develops sub-critical plasmoids. We also see that for the MHD calculation that is equivalent to $\alpha_c =\infty$ the current sheet is stable for $S = 9.7 \cdot 10^3$, again implying a critical $ S > 10^4$. At the same time, PIP cases characterised by an $S$ and an aspect ratio $\delta / \Delta$ incredibly similar to this MHD calculation (see cases 12 and 13, where $\alpha_c$ is respectively $10^3$ and $3 \cdot 10^3$, in Figure \ref{fig:alpha_properties}), show sub-critical plasmoid formation. The inclusion of the coupling between the ion and neutral fluids (effectively looking at systems between those two MHD simulations) allows plasmoids to form below $S = 10^4$, even without changing the plasma $\beta$, the initial conditions or the perturbation magnitude. Therefore, such sub-critical plasmoid formation does not depend on the characteristics of the plasma itself, but it might be ascribed to the combined result of two-fluid effects.

From previous studies of the onset of the tearing instability in partially ionised plasmas \cite{2020arXiv200603957P}, the critical aspect ratio for the initiation of a growth rate independent of the Lundquist number and neutral to ionic density was derived for three regimes (coupled by collisions, intermediate and uncoupled). In the intermediate regime, where ions and neutral are partly coupled, the critical aspect ratio for a generic equilibrium configuration is:
\begin{equation}
    \frac{\delta}{\Delta} \sim S_p ^{-1/3} (\nu_{pn} \tau_{A,p})^{-1/6},
    \label{eq:critical_aspect_ratio}
\end{equation}
where the collisional frequency $\nu_{pn}$ is equal to $\alpha_c \rho_n$ and $\tau_{A,p} = \Delta / v_{A,p}$ is the ion Alfv\'en time scale. In general, this critical aspect ratio is bigger than both the threshold $S^{-1/3}$ obtained from classical arguments that involve the Lundquist number and  the value of $1/200$ found in several works \cite{2010PhPl...17f2104H,2012ApJ...760..109L}.

Comparing our current sheet aspect ratios at the time of the onset of the tearing instability in the PIP simulations with the critical aspect ratio, we find that in most of the cases the aspect ratio is still smaller than the critical value obtained by equation (\ref{eq:critical_aspect_ratio}). Therefore, the sub-critical plasmoid formation can't be explained uniquely by applying such condition for the onset of the tearing instability.

Focusing on a physical explanation for the sub-critical plasmoid formation, we take a close look at what happens in the current sheet before the onset of tearing instability. We examine in detail the PIP case correspondent to simulation 13 in Table \ref{tab:parameters} and evaluate how the two-fluid parameters vary up to the time immediately before the formation of a central plasmoid, which signature appears in the current density and magnetic field at $t = 4.7$. This PIP case is characterised by an initial strong coupling between the two species as the initial $\alpha_c (0)$ is set equal to 3000, and the Lundquist number before the onset of the tearing instability is $S = 3 \cdot 10^3$.

The plasma velocity along $x$ is shown in row $(a)$ of Figure \ref{fig:alpha_3000_timelapse}. As enhanced by the contour lines, before the formation of the secondary plasmoid the plasma in the inflow slows down around the centre of the current sheet where a stagnation point in the flow exists, forming a pinching flow that promotes reconnection in two points above and below the current sheet central point.

In order to understand how the plasma motion is slowed down, we evaluate the force balance in the inflow region. All the force contributions are in equilibrium with each other, with the exception of the drift force, stated on the right hand side of Equations (\ref{eq:force_neutral}) and (\ref{eq:force_plasma}), which $x$ component for the plasma is shown in row $(b)$ of Figure \ref{fig:alpha_3000_timelapse}. The drift force magnitude appears to be quite large, however this does not mean that the drift velocity between the two species is also big in the inflow region. The drift force has a strong dependency on the collisional term $\alpha_c$, which is itself dependent on both the drift velocity and the temperatures of the two fluids. Because of the strong dependence on the $\alpha_c$ term, when the species are weakly coupled (and consequently the drift velocities are large), the drift force is small.

The drift force acts by slowing down the flow as it approaches the current sheet: as its sign depends on the term $(\textbf{v}_n - \textbf{v}_p)$, its direction suggests that the plasma is slowed down by the interaction with the neutrals, whose velocity in the inflow is slower than the plasma. The neutrals are naturally slower as they move towards the centre of the current sheet as they are being pulled in by the plasma motions. However, as shown in panel $(c)$, the neutral pressure increases at the centre slowing the inflow motions and increasing the drag. The neutral pressure increases under the effect of the Ohmic heating $\eta J^2$ and of the adiabatic compression of the neutrals in the inflow. As already discussed for the reference PIP case in Section \ref{sec:inflow}, the Ohmic heating plays a major role in heating the plasma inside the current sheet (see Figure \ref{fig:PIP_inflow_frictional_heating}), which results in the increase of the plasma pressure and temperature. As the two species are thermally coupled, the Ohmic heating acts by indirectly heating the neutrals, driving their pressure to increase. The non-adiabatic contribution from the Ohmic heating is responsible for the increase of the plasma temperature inside the current sheet, which affect the neutral temperature. The heated neutrals expand in the inflow region, leading to an adiabatic compression directed out of the current sheet, which contributes to increase the neutral pressure.

The sub-critical plasmoid formation can therefore be triggered by the two-fluid interaction between plasma and neutrals. The PIP cases at lower collisional coupling and higher ion fraction, however, develop the formation of critical plasmoids following the linear condition for the tearing instability before this non-linear mechanism becomes important. Therefore the two-fluid  interaction plays a lesser role in the formation of plasmoids when the effect of neutrals is weak, i.e., at smaller $\alpha_c$ and higher $\xi_p$.

\section{Discussion}
\label{sec:discussion}

Among the processes promoting the development of fast magnetic reconnection, the coalescence instability can play an important role by driving the interaction of plasmoids (and their subsequent reconnection) on dynamic timescales. We have investigated how plasmoid coalescence behaves in a partially ionised plasma, a situation reflected in a range of solar atmospheric layers and in particular the solar chromosphere. Through the comparison of a fully ionised case (MHD) and a partially ionised case (PIP) we find that partial ionisation noticeably shortens the coalescence timescale and creates new dynamics, producing neutral jets and secondary plasmoids, suppressing slow-mode shocks while promoting hydrodynamic shocks, and leading to sub-critical plasmoid formation.

Observations of chromospheric anemone jets performed with the Ca II H filtergram of the Solar Optical Telescope onboard Hinode\cite{2012ApJ...759...33S} showed the presence of recurrent plasmoids expulsion with a size of about a few hundred km from the jets. Assuming an approximate diameter $\varnothing_{\operatorname{MAX}} \sim 500$ km our characteristic length is $\sim 100$ km, as in our system the initial plasmoid length is equal to $4 L$. Knowing that the sound speed is about 10 km s$^{-1}$, we identify a time scale of $\sim 10$ s. Taking the appropriate ion-neutral collisional cross section for elastic scattering and charge exchange \cite{2013A&A...554A..22V} and the characteristic temperature, density and ion fraction \cite{2015ApJ...799...79N,1981ApJS...45..635V,1986A&A...154..231P,Khomenko2008} of the chromosphere the ion-neutral collisional frequency \cite{2005A&A...442.1091L,2012ASPC..463..281K} is between $10^3$ s$^{-1}$ and $10^6$ s$^{-1}$. In non-dimensional form, to compare with our simulations, an equivalent $\alpha_c$ to the one observed for the Hinode plasmoids is in range $10^4 - 10^7$.

A prediction on the behaviour at these higher $\alpha_c$ can be made by studying the simulations presented in Section \ref{alpha_c}, whose trend is shown in Figures \ref{fig:current_different_alpha} and \ref{fig:alpha_properties}. When $\alpha_c$ increases, the evolution tends to the one of the upper limiting MHD case ($\alpha_c \rightarrow \infty$): such case can be considered as an accumulation point. In the evolution of the current density, all the simulations at the chromospheric $\alpha_c$ would reach the beginning of reconnection in a time interval between the current minimum of the PIP case with $\alpha_c = 3 \cdot 10^3$ and the one of the MHD case, so it is sensible to say that the coalescence evolution and time scale can be predicted. These values suggest that the coalescence between the biggest observed plasmoids would take place in a regime that is almost MHD. At the specific values for ion fraction, plasma $\beta$ and resistivity used in this study, faster coalescence would become relevant for plasmoids with a diameter of $\sim 3$ km. Such length scale can be found by comparing the highest collisional frequency of our simulations with the dimensional chromospheric collisional frequency. Taking 1 km as a characteristic length and a sound speed of 10 km s$^{-1}$, we obtain a time scale $\tau = 0.1 s$. This leads to a collisional frequency of $\sim 10^4$ s$^{-1}$, consistent with the collisional frequency observed in the chromosphere and with our case at a non-dimensional $\alpha_c = 3 \cdot 10^3$.

This length scale is however dependent on the parameters of the medium studied. In many regions of the solar chromosphere the resistivity is often smaller, which results in the formation of more plasmoid dynamics, and both $\xi_p$ (equal to 0.01 in the $\alpha_c$ section of the parameter survey) and the plasma $\beta$ (set equal to 0.1) can be lower in the chromosphere, leading to an enhancement of two-fluid effects. At lower $\xi_p$ and $\beta$, for the range of $\alpha_c$ considered in this study, PIP effects that include a faster coalescence and the sub-critical plasmoid formation would become important for larger plasmoids and are potentially observable on scales that are currently resolved. The 3 km plasmoid length scale must therefore be considered as a lower limit for PIP effects to become observable.

Partial ionisation plays a role in changing the way many effects, such as the Hall diffusion, develop and act on magnetic reconnection. The Hall effect results from the different velocities of electrons and ions\cite{2012ApJ...747...87K}, is dependent on the ion fraction and is seen to play a major role in both highly ionised and weakly ionised plasmas\cite{2008MNRAS.385.2269P}. In a partially ionised environment, the physical scales over which the Hall diffusion becomes important drastically change from the ones found for fully ionised plasmas \cite{2008MNRAS.385.2269P}. Several works\cite{1986A&A...154..231P,Khomenko2008,2012ApJ...747...87K} show that in magnetic flux tubes, of which plasmoids are to be considered 2D sections, the Hall diffusion is less important than the contribution from ion-neutral effects at the chromospheric heights. As such we viewed that including the Hall effect to be beyond the scope of this paper.

A very important effect in evidence in our results is the sub-critical plasmoid formation discussed in Section \ref{sec:subcritical}. Many studies on the onset of the tearing instability in partially ionised plasmas often focused on linear stability criteria \cite{2014ApJ...780L..19P,2015ApJ...801..145T,2015PASJ...67...96S,2018PhPl...25c2113P,2020arXiv200603957P}, while in our study plasmoid formation is also promoted by a nonlinear effect linked to the two-fluid collisional and thermal coupling. The role played by the coupling between ions and neutrals in determining the dynamics of plasmoid formation has already been acknowledged in a previous study \cite{2015PASJ...67...96S} looking at the onset of the tearing mode at several levels down to the kinetic scale. In that study the authors only use a linear criterion for the onset of the tearing instability, while in this work we find that nonlinear effects are able to play a major role in plasmoid formation. Including the nonlinear physics may result in new physical parameter regimes that are able to tear down to kinetic scales.

The physics that leads to sub-critical plasmoid formation in our simulations is expected to be largely affected by the non-equilibrium ionisation-recombination processes. These are currently not included in our model, as the interaction between the two species is provided uniquely by elastic collisions and charge-exchange, but future developments of this research strictly require their inclusion. As pointed out from previous studies of magnetic reconnection in a multi-fluid partially ionised plasma at low plasma $\beta$ \cite{2012ApJ...760..109L,2013PhPl...20f1202L,2015ApJ...805..134M,2018ApJ...852...95N,2018PhPl...25d2903N,2018ApJ...868..144N}, the non-equilibrium ionization-recombination effect leads to a strong ionisation of the material in the reconnection region. In a recent paper \cite{2018ApJ...852...95N} magnetic reconnection has been examined through 2.5D simulations in weakly ionised plasmas with an initial $\xi_p$ ($\xi_p = 10^{-4}$) lower than our reference cases. Their results suggest that, for a plasma $\beta$ larger than 1 and weak reconnection magnetic fields, the non-equilibrium ionization-recombination effect is responsible for a strong ionisation of the neutral fluid in the reconnection region and a faster reconnection rate occurring before the onset of plasmoid instabilities. These are consistent with a previous paper \cite{2013PhPl...20f1202L}, where an increase by an order of magnitude was recorded for the ionization degree within the current sheet during the reconnection process. The strong ionisation is responsible for a bigger interaction between the neutral fluid and the plasma, which will be better coupled both in the inflow and outflow region. These very same effects are to be expected in the case discussed in Section \ref{sec:subcritical}: the drastic increase in temperature due to the Ohmic heating in the reconnection region would promote the ionisation of the neutral fluid, that in the absence of such process is forced to expand outwards, halting the plasma inflow. In case of a small plasma $\beta$ smaller than 1 like in our study, however, plasmoid instability is still the main process promoting fast magnetic reconnection\cite{2018ApJ...852...95N}.

The ionisation/recombination effects are largely affected by the action of the ionisation potential. When collisional ionisation takes place, the work done against the ionisation potential to ionise the atom removes energy from the electrons and acts as a cooling term in the plasma \cite{2020arXiv201006303S}. As the recombination process is associated with photons being released, this overall effect can be modelled as a radiative loss. Researches investigating the role of radiative cooling in magnetic reconnection \cite{1995ApJ...449..777D,2011PhPl...18d2105U} proved that such process, linked to the collisional ionisation, thins the reconnection layer by decreasing the plasma pressure and density inside the current sheet. Therefore, the inclusion of radiative losses speeds up reconnection to rates that are bigger than the ones found in models without radiation, and might lead to timescales and outflows that are consistent with those found in spicules and chromospheric jets \cite{2017ApJ...842..117A}. In a recent study \cite{2017ApJ...842..117A} it was found that a strong recombination process in the reconnection region, combined with Alfv\'enic outflows, can lead to a fast reconnection rate independent of Lundquist number. While the decoupling of neutrals and plasma has been recorded in the inflow region, these findings show that the two fluids are well coupled in the outflows, which is opposite to what we find for our intermediate coupled case (see Section \ref{sec:neutral_jet}). We therefore expect that including the radiative losses by adding a ionisation potential would lead to a better coupling of the two species around the reconnection region, as well as lower temperatures and a faster reconnection rate.

Radiation is not only important in terms of the radiative losses from the ionisation/recombination processes in the chromosphere. Beyond the action of the ionisation potential, effects of ionisation and excitation could be triggered by an external radiation field. This is a fundamental factor in the chromosphere, where the ionisation degree is largely determined by the incident external radiation \cite{2007A&A...473..625L,2014ApJ...784...30G,2017ApJ...847...36M,2019SoPh..294..165R}.

In a recent study \cite{2018ApJ...868..144N} it was demonstrated that the plasmoid cascading process, for which the current sheet breaks into smaller sections following the formation of multiple plasmoids, is terminated in the MHD scale. The progressive reduction of secondary plasmoids in the PIP cases as an MHD-like regime is approached is an aspect that has been already marginally observed in our simulations, especially in the parameter survey linked to the variation of the collisional coupling (Section \ref{sec:alpha_c}). Multiple plasmoids are seen to form in the simulations having an initial lower $\alpha_c$, while at higher collisional coupling only two and one secondary plasmoids are produced respectively for $\alpha_c = 1000$ and 3000. Moreover, as discussed in Section \ref{sec:subcritical}, the secondary plasmoids formed at higher $\alpha_c$ are generated by the pinching action of the neutrals in the inflow, rather than the onset of instabilities in the current sheet. Therefore, we expect the plasmoid cascading process to be interrupted in these cases approaching the MHD regime. A more detailed study needs to be performed on these secondary plasmoids number and characteristics, however we can already confirm a good agreement with the trend showed by previous studies \cite{2018ApJ...868..144N}.

\acknowledgments

The authors are grateful to Dr. N. Nakamura for the inspiration of his original work on this problem that lead to this study. AH and BS are supported by STFC Research Grant No. ST/R000891/1. AH is also supported by his STFC Ernest Rutherford Fellowship grant number ST/L00397X/1.

\section*{Data availability}

The data that support the findings of this study are available from the corresponding author upon reasonable request. The (P\underline{I}P) code is available at the following url: \url{https://github.com/AstroSnow/PIP}.

\nocite{*}
\bibliography{biblio}

\begin{thebibliography}{88}%
\makeatletter
\providecommand \@ifxundefined [1]{%
 \@ifx{#1\undefined}
}%
\providecommand \@ifnum [1]{%
 \ifnum #1\expandafter \@firstoftwo
 \else \expandafter \@secondoftwo
 \fi
}%
\providecommand \@ifx [1]{%
 \ifx #1\expandafter \@firstoftwo
 \else \expandafter \@secondoftwo
 \fi
}%
\providecommand \natexlab [1]{#1}%
\providecommand \enquote  [1]{``#1''}%
\providecommand \bibnamefont  [1]{#1}%
\providecommand \bibfnamefont [1]{#1}%
\providecommand \citenamefont [1]{#1}%
\providecommand \href@noop [0]{\@secondoftwo}%
\providecommand \href [0]{\begingroup \@sanitize@url \@href}%
\providecommand \@href[1]{\@@startlink{#1}\@@href}%
\providecommand \@@href[1]{\endgroup#1\@@endlink}%
\providecommand \@sanitize@url [0]{\catcode `\\12\catcode `\$12\catcode
  `\&12\catcode `\#12\catcode `\^12\catcode `\_12\catcode `\%12\relax}%
\providecommand \@@startlink[1]{}%
\providecommand \@@endlink[0]{}%
\providecommand \url  [0]{\begingroup\@sanitize@url \@url }%
\providecommand \@url [1]{\endgroup\@href {#1}{\urlprefix }}%
\providecommand \urlprefix  [0]{URL }%
\providecommand \Eprint [0]{\href }%
\providecommand \doibase [0]{http://dx.doi.org/}%
\providecommand \selectlanguage [0]{\@gobble}%
\providecommand \bibinfo  [0]{\@secondoftwo}%
\providecommand \bibfield  [0]{\@secondoftwo}%
\providecommand \translation [1]{[#1]}%
\providecommand \BibitemOpen [0]{}%
\providecommand \bibitemStop [0]{}%
\providecommand \bibitemNoStop [0]{.\EOS\space}%
\providecommand \EOS [0]{\spacefactor3000\relax}%
\providecommand \BibitemShut  [1]{\csname bibitem#1\endcsname}%
\let\auto@bib@innerbib\@empty
\bibitem [{\citenamefont {{Priest}}\ and\ \citenamefont
  {{Forbes}}(2000)}]{2000mrmt.conf.....P}%
  \BibitemOpen
  \bibfield  {author} {\bibinfo {author} {\bibfnamefont {E.}~\bibnamefont
  {{Priest}}}\ and\ \bibinfo {author} {\bibfnamefont {T.}~\bibnamefont
  {{Forbes}}},\ }\href@noop {} {\emph {\bibinfo {title} {Magnetic reconnection
  : MHD theory and applications / Eric Priest}}}\ (\bibinfo {year}
  {2000})\BibitemShut {NoStop}%
\bibitem [{\citenamefont {{Biskamp}}(2000)}]{2000mrp..book.....B}%
  \BibitemOpen
  \bibfield  {author} {\bibinfo {author} {\bibfnamefont {D.}~\bibnamefont
  {{Biskamp}}},\ }\href@noop {} {\emph {\bibinfo {title} {Magnetic reconnection
  in plasmas, Cambridge, UK: Cambridge University Press, 2000 xiv, 387
  p.~Cambridge monographs on plasma physics, vol.~3, ISBN 0521582881}}}\
  (\bibinfo {year} {2000})\BibitemShut {NoStop}%
\bibitem [{\citenamefont {{Parker}}(1957)}]{1957JGR....62..509P}%
  \BibitemOpen
  \bibfield  {author} {\bibinfo {author} {\bibfnamefont {E.~N.}\ \bibnamefont
  {{Parker}}},\ }\bibfield  {title} {\enquote {\bibinfo {title} {{Sweet's
  Mechanism for Merging Magnetic Fields in Conducting Fluids}},}\ }\href
  {\doibase 10.1029/JZ062i004p00509} {\bibfield  {journal} {\bibinfo  {journal}
  {\jgr}\ }\textbf {\bibinfo {volume} {62}},\ \bibinfo {pages} {509--520}
  (\bibinfo {year} {1957})}\BibitemShut {NoStop}%
\bibitem [{\citenamefont {{Sweet}}(1958)}]{1958IAUS....6..123S}%
  \BibitemOpen
  \bibfield  {author} {\bibinfo {author} {\bibfnamefont {P.~A.}\ \bibnamefont
  {{Sweet}}},\ }\bibfield  {title} {\enquote {\bibinfo {title} {{The Neutral
  Point Theory of Solar Flares}},}\ }in\ \href@noop {} {\emph {\bibinfo
  {booktitle} {Electromagnetic Phenomena in Cosmical Physics}}},\ \bibinfo
  {series} {IAU Symposium}, Vol.~\bibinfo {volume} {6},\ \bibinfo {editor}
  {edited by\ \bibinfo {editor} {\bibfnamefont {B.}~\bibnamefont {{Lehnert}}}}\
  (\bibinfo {year} {1958})\ p.\ \bibinfo {pages} {123}\BibitemShut {NoStop}%
\bibitem [{\citenamefont {{Shibata}}\ \emph {et~al.}(2007)\citenamefont
  {{Shibata}}, \citenamefont {{Nakamura}}, \citenamefont {{Matsumoto}},
  \citenamefont {{Otsuji}}, \citenamefont {{Okamoto}}, \citenamefont
  {{Nishizuka}}, \citenamefont {{Kawate}}, \citenamefont {{Watanabe}},
  \citenamefont {{Nagata}}, \citenamefont {{UeNo}}, \citenamefont {{Kitai}},
  \citenamefont {{Nozawa}}, \citenamefont {{Tsuneta}}, \citenamefont
  {{Suematsu}}, \citenamefont {{Ichimoto}}, \citenamefont {{Shimizu}},
  \citenamefont {{Katsukawa}}, \citenamefont {{Tarbell}}, \citenamefont
  {{Berger}}, \citenamefont {{Lites}}, \citenamefont {{Shine}},\ and\
  \citenamefont {{Title}}}]{2007Sci...318.1591S}%
  \BibitemOpen
  \bibfield  {author} {\bibinfo {author} {\bibfnamefont {K.}~\bibnamefont
  {{Shibata}}}, \bibinfo {author} {\bibfnamefont {T.}~\bibnamefont
  {{Nakamura}}}, \bibinfo {author} {\bibfnamefont {T.}~\bibnamefont
  {{Matsumoto}}}, \bibinfo {author} {\bibfnamefont {K.}~\bibnamefont
  {{Otsuji}}}, \bibinfo {author} {\bibfnamefont {T.~J.}\ \bibnamefont
  {{Okamoto}}}, \bibinfo {author} {\bibfnamefont {N.}~\bibnamefont
  {{Nishizuka}}}, \bibinfo {author} {\bibfnamefont {T.}~\bibnamefont
  {{Kawate}}}, \bibinfo {author} {\bibfnamefont {H.}~\bibnamefont
  {{Watanabe}}}, \bibinfo {author} {\bibfnamefont {S.}~\bibnamefont
  {{Nagata}}}, \bibinfo {author} {\bibfnamefont {S.}~\bibnamefont {{UeNo}}},
  \bibinfo {author} {\bibfnamefont {R.}~\bibnamefont {{Kitai}}}, \bibinfo
  {author} {\bibfnamefont {S.}~\bibnamefont {{Nozawa}}}, \bibinfo {author}
  {\bibfnamefont {S.}~\bibnamefont {{Tsuneta}}}, \bibinfo {author}
  {\bibfnamefont {Y.}~\bibnamefont {{Suematsu}}}, \bibinfo {author}
  {\bibfnamefont {K.}~\bibnamefont {{Ichimoto}}}, \bibinfo {author}
  {\bibfnamefont {T.}~\bibnamefont {{Shimizu}}}, \bibinfo {author}
  {\bibfnamefont {Y.}~\bibnamefont {{Katsukawa}}}, \bibinfo {author}
  {\bibfnamefont {T.~D.}\ \bibnamefont {{Tarbell}}}, \bibinfo {author}
  {\bibfnamefont {T.~E.}\ \bibnamefont {{Berger}}}, \bibinfo {author}
  {\bibfnamefont {B.~W.}\ \bibnamefont {{Lites}}}, \bibinfo {author}
  {\bibfnamefont {R.~A.}\ \bibnamefont {{Shine}}}, \ and\ \bibinfo {author}
  {\bibfnamefont {A.~M.}\ \bibnamefont {{Title}}},\ }\bibfield  {title}
  {\enquote {\bibinfo {title} {{Chromospheric Anemone Jets as Evidence of
  Ubiquitous Reconnection}},}\ }\href {\doibase 10.1126/science.1146708}
  {\bibfield  {journal} {\bibinfo  {journal} {Science}\ }\textbf {\bibinfo
  {volume} {318}},\ \bibinfo {pages} {1591} (\bibinfo {year} {2007})},\ \Eprint
  {http://arxiv.org/abs/0810.3974} {arXiv:0810.3974 [astro-ph]} \BibitemShut
  {NoStop}%
\bibitem [{\citenamefont {{Nishizuka}}\ \emph {et~al.}(2011)\citenamefont
  {{Nishizuka}}, \citenamefont {{Nakamura}}, \citenamefont {{Kawate}},
  \citenamefont {{Singh}},\ and\ \citenamefont
  {{Shibata}}}]{2011ApJ...731...43N}%
  \BibitemOpen
  \bibfield  {author} {\bibinfo {author} {\bibfnamefont {N.}~\bibnamefont
  {{Nishizuka}}}, \bibinfo {author} {\bibfnamefont {T.}~\bibnamefont
  {{Nakamura}}}, \bibinfo {author} {\bibfnamefont {T.}~\bibnamefont
  {{Kawate}}}, \bibinfo {author} {\bibfnamefont {K.~A.~P.}\ \bibnamefont
  {{Singh}}}, \ and\ \bibinfo {author} {\bibfnamefont {K.}~\bibnamefont
  {{Shibata}}},\ }\bibfield  {title} {\enquote {\bibinfo {title} {{Statistical
  Study of Chromospheric Anemone Jets Observed with Hinode/SOT}},}\ }\href
  {\doibase 10.1088/0004-637X/731/1/43} {\bibfield  {journal} {\bibinfo
  {journal} {\apj}\ }\textbf {\bibinfo {volume} {731}},\ \bibinfo {eid} {43}
  (\bibinfo {year} {2011})}\BibitemShut {NoStop}%
\bibitem [{\citenamefont {{Singh}}\ \emph {et~al.}(2011)\citenamefont
  {{Singh}}, \citenamefont {{Shibata}}, \citenamefont {{Nishizuka}},\ and\
  \citenamefont {{Isobe}}}]{2011PhPl...18k1210S}%
  \BibitemOpen
  \bibfield  {author} {\bibinfo {author} {\bibfnamefont {K.~A.~P.}\
  \bibnamefont {{Singh}}}, \bibinfo {author} {\bibfnamefont {K.}~\bibnamefont
  {{Shibata}}}, \bibinfo {author} {\bibfnamefont {N.}~\bibnamefont
  {{Nishizuka}}}, \ and\ \bibinfo {author} {\bibfnamefont {H.}~\bibnamefont
  {{Isobe}}},\ }\bibfield  {title} {\enquote {\bibinfo {title} {{Chromospheric
  anemone jets and magnetic reconnection in partially ionized solar
  atmosphere}},}\ }\href {\doibase 10.1063/1.3655444} {\bibfield  {journal}
  {\bibinfo  {journal} {Physics of Plasmas}\ }\textbf {\bibinfo {volume}
  {18}},\ \bibinfo {pages} {111210--111210} (\bibinfo {year}
  {2011})}\BibitemShut {NoStop}%
\bibitem [{\citenamefont {{Singh}}\ \emph {et~al.}(2012)\citenamefont
  {{Singh}}, \citenamefont {{Isobe}}, \citenamefont {{Nishizuka}},
  \citenamefont {{Nishida}},\ and\ \citenamefont
  {{Shibata}}}]{2012ApJ...759...33S}%
  \BibitemOpen
  \bibfield  {author} {\bibinfo {author} {\bibfnamefont {K.~A.~P.}\
  \bibnamefont {{Singh}}}, \bibinfo {author} {\bibfnamefont {H.}~\bibnamefont
  {{Isobe}}}, \bibinfo {author} {\bibfnamefont {N.}~\bibnamefont
  {{Nishizuka}}}, \bibinfo {author} {\bibfnamefont {K.}~\bibnamefont
  {{Nishida}}}, \ and\ \bibinfo {author} {\bibfnamefont {K.}~\bibnamefont
  {{Shibata}}},\ }\bibfield  {title} {\enquote {\bibinfo {title} {{Multiple
  Plasma Ejections and Intermittent Nature of Magnetic Reconnection in Solar
  Chromospheric Anemone Jets}},}\ }\href {\doibase 10.1088/0004-637X/759/1/33}
  {\bibfield  {journal} {\bibinfo  {journal} {\apj}\ }\textbf {\bibinfo
  {volume} {759}},\ \bibinfo {eid} {33} (\bibinfo {year} {2012})}\BibitemShut
  {NoStop}%
\bibitem [{\citenamefont {{Guo}}\ \emph {et~al.}(2020)\citenamefont {{Guo}},
  \citenamefont {{De Pontieu}}, \citenamefont {{Huang}}, \citenamefont
  {{Peter}},\ and\ \citenamefont {{Bhattacharjee}}}]{2020arXiv200911475G}%
  \BibitemOpen
  \bibfield  {author} {\bibinfo {author} {\bibfnamefont {L.~J.}\ \bibnamefont
  {{Guo}}}, \bibinfo {author} {\bibfnamefont {B.}~\bibnamefont {{De Pontieu}}},
  \bibinfo {author} {\bibfnamefont {Y.~M.}\ \bibnamefont {{Huang}}}, \bibinfo
  {author} {\bibfnamefont {H.}~\bibnamefont {{Peter}}}, \ and\ \bibinfo
  {author} {\bibfnamefont {A.}~\bibnamefont {{Bhattacharjee}}},\ }\bibfield
  {title} {\enquote {\bibinfo {title} {{Observations and modeling of the onset
  of fast reconnection in the solar transition region}},}\ }\href@noop {}
  {\bibfield  {journal} {\bibinfo  {journal} {arXiv e-prints}\ ,\ \bibinfo
  {eid} {arXiv:2009.11475}} (\bibinfo {year} {2020})},\ \Eprint
  {http://arxiv.org/abs/2009.11475} {arXiv:2009.11475 [astro-ph.SR]}
  \BibitemShut {NoStop}%
\bibitem [{\citenamefont {{Furth}}, \citenamefont {{Killeen}},\ and\
  \citenamefont {{Rosenbluth}}(1963)}]{1963PhFl....6..459F}%
  \BibitemOpen
  \bibfield  {author} {\bibinfo {author} {\bibfnamefont {H.~P.}\ \bibnamefont
  {{Furth}}}, \bibinfo {author} {\bibfnamefont {J.}~\bibnamefont {{Killeen}}},
  \ and\ \bibinfo {author} {\bibfnamefont {M.~N.}\ \bibnamefont
  {{Rosenbluth}}},\ }\bibfield  {title} {\enquote {\bibinfo {title}
  {{Finite-Resistivity Instabilities of a Sheet Pinch}},}\ }\href {\doibase
  10.1063/1.1706761} {\bibfield  {journal} {\bibinfo  {journal} {Phys. Fluids}\
  }\textbf {\bibinfo {volume} {6}},\ \bibinfo {pages} {459--484} (\bibinfo
  {year} {1963})}\BibitemShut {NoStop}%
\bibitem [{\citenamefont {{Tanuma}}\ \emph {et~al.}(2001)\citenamefont
  {{Tanuma}}, \citenamefont {{Yokoyama}}, \citenamefont {{Kudoh}},\ and\
  \citenamefont {{Shibata}}}]{2001ApJ...551..312T}%
  \BibitemOpen
  \bibfield  {author} {\bibinfo {author} {\bibfnamefont {S.}~\bibnamefont
  {{Tanuma}}}, \bibinfo {author} {\bibfnamefont {T.}~\bibnamefont
  {{Yokoyama}}}, \bibinfo {author} {\bibfnamefont {T.}~\bibnamefont {{Kudoh}}},
  \ and\ \bibinfo {author} {\bibfnamefont {K.}~\bibnamefont {{Shibata}}},\
  }\bibfield  {title} {\enquote {\bibinfo {title} {{Two-dimensional
  Magnetohydrodynamic Numerical Simulations of Magnetic Reconnection Triggered
  by a Supernova Shock in the Interstellar Medium: Generation of X-Ray Gas in
  the Galaxy}},}\ }\href {\doibase 10.1086/320058} {\bibfield  {journal}
  {\bibinfo  {journal} {\apj}\ }\textbf {\bibinfo {volume} {551}},\ \bibinfo
  {pages} {312--332} (\bibinfo {year} {2001})},\ \Eprint
  {http://arxiv.org/abs/astro-ph/0009088} {arXiv:astro-ph/0009088 [astro-ph]}
  \BibitemShut {NoStop}%
\bibitem [{\citenamefont {{Shibata}}\ and\ \citenamefont
  {{Tanuma}}(2001)}]{2001EP&S...53..473S}%
  \BibitemOpen
  \bibfield  {author} {\bibinfo {author} {\bibfnamefont {K.}~\bibnamefont
  {{Shibata}}}\ and\ \bibinfo {author} {\bibfnamefont {S.}~\bibnamefont
  {{Tanuma}}},\ }\bibfield  {title} {\enquote {\bibinfo {title}
  {{Plasmoid-induced-reconnection and fractal reconnection}},}\ }\href
  {\doibase 10.1186/BF03353258} {\bibfield  {journal} {\bibinfo  {journal}
  {Earth, Planets, and Space}\ }\textbf {\bibinfo {volume} {53}},\ \bibinfo
  {pages} {473--482} (\bibinfo {year} {2001})},\ \Eprint
  {http://arxiv.org/abs/astro-ph/0101008} {arXiv:astro-ph/0101008 [astro-ph]}
  \BibitemShut {NoStop}%
\bibitem [{\citenamefont {{Samtaney}}\ \emph {et~al.}(2009)\citenamefont
  {{Samtaney}}, \citenamefont {{Loureiro}}, \citenamefont {{Uzdensky}},
  \citenamefont {{Schekochihin}},\ and\ \citenamefont
  {{Cowley}}}]{2009PhRvL.103j5004S}%
  \BibitemOpen
  \bibfield  {author} {\bibinfo {author} {\bibfnamefont {R.}~\bibnamefont
  {{Samtaney}}}, \bibinfo {author} {\bibfnamefont {N.~F.}\ \bibnamefont
  {{Loureiro}}}, \bibinfo {author} {\bibfnamefont {D.~A.}\ \bibnamefont
  {{Uzdensky}}}, \bibinfo {author} {\bibfnamefont {A.~A.}\ \bibnamefont
  {{Schekochihin}}}, \ and\ \bibinfo {author} {\bibfnamefont {S.~C.}\
  \bibnamefont {{Cowley}}},\ }\bibfield  {title} {\enquote {\bibinfo {title}
  {{Formation of Plasmoid Chains in Magnetic Reconnection}},}\ }\href {\doibase
  10.1103/PhysRevLett.103.105004} {\bibfield  {journal} {\bibinfo  {journal}
  {\prl}\ }\textbf {\bibinfo {volume} {103}},\ \bibinfo {eid} {105004}
  (\bibinfo {year} {2009})},\ \Eprint {http://arxiv.org/abs/0903.0542}
  {arXiv:0903.0542 [astro-ph.SR]} \BibitemShut {NoStop}%
\bibitem [{\citenamefont {{Loureiro}}, \citenamefont {{Schekochihin}},\ and\
  \citenamefont {{Cowley}}(2007)}]{2007PhPl...14j0703L}%
  \BibitemOpen
  \bibfield  {author} {\bibinfo {author} {\bibfnamefont {N.~F.}\ \bibnamefont
  {{Loureiro}}}, \bibinfo {author} {\bibfnamefont {A.~A.}\ \bibnamefont
  {{Schekochihin}}}, \ and\ \bibinfo {author} {\bibfnamefont {S.~C.}\
  \bibnamefont {{Cowley}}},\ }\bibfield  {title} {\enquote {\bibinfo {title}
  {{Instability of current sheets and formation of plasmoid chains}},}\ }\href
  {\doibase 10.1063/1.2783986} {\bibfield  {journal} {\bibinfo  {journal}
  {Phys. Plasmas}\ }\textbf {\bibinfo {volume} {14}},\ \bibinfo {pages}
  {100703--100703} (\bibinfo {year} {2007})},\ \Eprint
  {http://arxiv.org/abs/astro-ph/0703631} {astro-ph/0703631} \BibitemShut
  {NoStop}%
\bibitem [{\citenamefont {{Loureiro}}\ \emph {et~al.}(2012)\citenamefont
  {{Loureiro}}, \citenamefont {{Samtaney}}, \citenamefont {{Schekochihin}},\
  and\ \citenamefont {{Uzdensky}}}]{2012PhPl...19d2303L}%
  \BibitemOpen
  \bibfield  {author} {\bibinfo {author} {\bibfnamefont {N.~F.}\ \bibnamefont
  {{Loureiro}}}, \bibinfo {author} {\bibfnamefont {R.}~\bibnamefont
  {{Samtaney}}}, \bibinfo {author} {\bibfnamefont {A.~A.}\ \bibnamefont
  {{Schekochihin}}}, \ and\ \bibinfo {author} {\bibfnamefont {D.~A.}\
  \bibnamefont {{Uzdensky}}},\ }\bibfield  {title} {\enquote {\bibinfo {title}
  {{Magnetic reconnection and stochastic plasmoid chains in
  high-Lundquist-number plasmas}},}\ }\href {\doibase 10.1063/1.3703318}
  {\bibfield  {journal} {\bibinfo  {journal} {Physics of Plasmas}\ }\textbf
  {\bibinfo {volume} {19}},\ \bibinfo {pages} {042303--042303} (\bibinfo {year}
  {2012})},\ \Eprint {http://arxiv.org/abs/1108.4040} {arXiv:1108.4040
  [astro-ph.SR]} \BibitemShut {NoStop}%
\bibitem [{\citenamefont {{Loureiro}}\ and\ \citenamefont
  {{Uzdensky}}(2016)}]{2016PPCF...58a4021L}%
  \BibitemOpen
  \bibfield  {author} {\bibinfo {author} {\bibfnamefont {N.~F.}\ \bibnamefont
  {{Loureiro}}}\ and\ \bibinfo {author} {\bibfnamefont {D.~A.}\ \bibnamefont
  {{Uzdensky}}},\ }\bibfield  {title} {\enquote {\bibinfo {title} {{Magnetic
  reconnection: from the Sweet-Parker model to stochastic plasmoid chains}},}\
  }\href {\doibase 10.1088/0741-3335/58/1/014021} {\bibfield  {journal}
  {\bibinfo  {journal} {Plasma Physics and Controlled Fusion}\ }\textbf
  {\bibinfo {volume} {58}},\ \bibinfo {eid} {014021} (\bibinfo {year}
  {2016})},\ \Eprint {http://arxiv.org/abs/1507.07756} {arXiv:1507.07756
  [physics.plasm-ph]} \BibitemShut {NoStop}%
\bibitem [{\citenamefont {{Zweibel}}(1989)}]{1989ApJ...340..550Z}%
  \BibitemOpen
  \bibfield  {author} {\bibinfo {author} {\bibfnamefont {E.~G.}\ \bibnamefont
  {{Zweibel}}},\ }\bibfield  {title} {\enquote {\bibinfo {title} {{Magnetic
  Reconnection in Partially Ionized Gases}},}\ }\href {\doibase 10.1086/167416}
  {\bibfield  {journal} {\bibinfo  {journal} {The Astrophysical Journal}\
  }\textbf {\bibinfo {volume} {340}},\ \bibinfo {pages} {550} (\bibinfo {year}
  {1989})}\BibitemShut {NoStop}%
\bibitem [{\citenamefont {{Ni}}\ \emph {et~al.}(2015)\citenamefont {{Ni}},
  \citenamefont {{Kliem}}, \citenamefont {{Lin}},\ and\ \citenamefont
  {{Wu}}}]{2015ApJ...799...79N}%
  \BibitemOpen
  \bibfield  {author} {\bibinfo {author} {\bibfnamefont {L.}~\bibnamefont
  {{Ni}}}, \bibinfo {author} {\bibfnamefont {B.}~\bibnamefont {{Kliem}}},
  \bibinfo {author} {\bibfnamefont {J.}~\bibnamefont {{Lin}}}, \ and\ \bibinfo
  {author} {\bibfnamefont {N.}~\bibnamefont {{Wu}}},\ }\bibfield  {title}
  {\enquote {\bibinfo {title} {{Fast Magnetic Reconnection in the Solar
  Chromosphere Mediated by the Plasmoid Instability}},}\ }\href {\doibase
  10.1088/0004-637X/799/1/79} {\bibfield  {journal} {\bibinfo  {journal}
  {\apj}\ }\textbf {\bibinfo {volume} {799}},\ \bibinfo {eid} {79} (\bibinfo
  {year} {2015})},\ \Eprint {http://arxiv.org/abs/1509.06895} {arXiv:1509.06895
  [astro-ph.SR]} \BibitemShut {NoStop}%
\bibitem [{\citenamefont {{Park}}, \citenamefont {{Monticello}},\ and\
  \citenamefont {{White}}(1984)}]{1984PhFl...27..137P}%
  \BibitemOpen
  \bibfield  {author} {\bibinfo {author} {\bibfnamefont {W.}~\bibnamefont
  {{Park}}}, \bibinfo {author} {\bibfnamefont {D.~A.}\ \bibnamefont
  {{Monticello}}}, \ and\ \bibinfo {author} {\bibfnamefont {R.~B.}\
  \bibnamefont {{White}}},\ }\bibfield  {title} {\enquote {\bibinfo {title}
  {{Reconnection rates of magnetic fields including the effects of
  viscosity}},}\ }\href {\doibase 10.1063/1.864502} {\bibfield  {journal}
  {\bibinfo  {journal} {Physics of Fluids}\ }\textbf {\bibinfo {volume} {27}},\
  \bibinfo {pages} {137--149} (\bibinfo {year} {1984})}\BibitemShut {NoStop}%
\bibitem [{\citenamefont {{Steinolfson}}\ and\ \citenamefont {{van
  Hoven}}(1984)}]{1984PhFl...27.1207S}%
  \BibitemOpen
  \bibfield  {author} {\bibinfo {author} {\bibfnamefont {R.~S.}\ \bibnamefont
  {{Steinolfson}}}\ and\ \bibinfo {author} {\bibfnamefont {G.}~\bibnamefont
  {{van Hoven}}},\ }\bibfield  {title} {\enquote {\bibinfo {title} {{Nonlinear
  evolution of the resistive tearing mode}},}\ }\href {\doibase
  10.1063/1.864728} {\bibfield  {journal} {\bibinfo  {journal} {Physics of
  Fluids}\ }\textbf {\bibinfo {volume} {27}},\ \bibinfo {pages} {1207--1214}
  (\bibinfo {year} {1984})}\BibitemShut {NoStop}%
\bibitem [{\citenamefont {{Biskamp}}(1986)}]{1986PhFl...29.1520B}%
  \BibitemOpen
  \bibfield  {author} {\bibinfo {author} {\bibfnamefont {D.}~\bibnamefont
  {{Biskamp}}},\ }\bibfield  {title} {\enquote {\bibinfo {title} {{Magnetic
  reconnection via current sheets}},}\ }\href {\doibase 10.1063/1.865670}
  {\bibfield  {journal} {\bibinfo  {journal} {Physics of Fluids}\ }\textbf
  {\bibinfo {volume} {29}},\ \bibinfo {pages} {1520--1531} (\bibinfo {year}
  {1986})}\BibitemShut {NoStop}%
\bibitem [{\citenamefont {{Lee}}\ and\ \citenamefont
  {{Fu}}(1986)}]{1986JGR....91.6807L}%
  \BibitemOpen
  \bibfield  {author} {\bibinfo {author} {\bibfnamefont {L.~C.}\ \bibnamefont
  {{Lee}}}\ and\ \bibinfo {author} {\bibfnamefont {Z.~F.}\ \bibnamefont
  {{Fu}}},\ }\bibfield  {title} {\enquote {\bibinfo {title} {{Multiple X line
  reconnection 1. A criterion for the transition from a single X line to a
  multiple X line reconnection}},}\ }\href {\doibase 10.1029/JA091iA06p06807}
  {\bibfield  {journal} {\bibinfo  {journal} {\jgr}\ }\textbf {\bibinfo
  {volume} {91}},\ \bibinfo {pages} {6807--6815} (\bibinfo {year}
  {1986})}\BibitemShut {NoStop}%
\bibitem [{\citenamefont {{Jin}}\ and\ \citenamefont
  {{Ip}}(1991)}]{1991PhFlB...3.1927J}%
  \BibitemOpen
  \bibfield  {author} {\bibinfo {author} {\bibfnamefont {S.~P.}\ \bibnamefont
  {{Jin}}}\ and\ \bibinfo {author} {\bibfnamefont {W.~H.}\ \bibnamefont
  {{Ip}}},\ }\bibfield  {title} {\enquote {\bibinfo {title} {{Two-dimensional
  compressible magnetohydrodynamic simulation of the driven reconnection
  process}},}\ }\href {\doibase 10.1063/1.859661} {\bibfield  {journal}
  {\bibinfo  {journal} {Physics of Fluids B}\ }\textbf {\bibinfo {volume}
  {3}},\ \bibinfo {pages} {1927--1936} (\bibinfo {year} {1991})}\BibitemShut
  {NoStop}%
\bibitem [{\citenamefont {{Ugai}}(1995)}]{1995PhPl....2..388U}%
  \BibitemOpen
  \bibfield  {author} {\bibinfo {author} {\bibfnamefont {M.}~\bibnamefont
  {{Ugai}}},\ }\bibfield  {title} {\enquote {\bibinfo {title} {{Computer
  studies on powerful magnetic energy conversion by the spontaneous fast
  reconnection mechanism}},}\ }\href {\doibase 10.1063/1.870965} {\bibfield
  {journal} {\bibinfo  {journal} {Physics of Plasmas}\ }\textbf {\bibinfo
  {volume} {2}},\ \bibinfo {pages} {388--397} (\bibinfo {year}
  {1995})}\BibitemShut {NoStop}%
\bibitem [{\citenamefont {{Loureiro}}\ \emph {et~al.}(2005)\citenamefont
  {{Loureiro}}, \citenamefont {{Cowley}}, \citenamefont {{Dorland}},
  \citenamefont {{Haines}},\ and\ \citenamefont
  {{Schekochihin}}}]{2005PhRvL..95w5003L}%
  \BibitemOpen
  \bibfield  {author} {\bibinfo {author} {\bibfnamefont {N.~F.}\ \bibnamefont
  {{Loureiro}}}, \bibinfo {author} {\bibfnamefont {S.~C.}\ \bibnamefont
  {{Cowley}}}, \bibinfo {author} {\bibfnamefont {W.~D.}\ \bibnamefont
  {{Dorland}}}, \bibinfo {author} {\bibfnamefont {M.~G.}\ \bibnamefont
  {{Haines}}}, \ and\ \bibinfo {author} {\bibfnamefont {A.~A.}\ \bibnamefont
  {{Schekochihin}}},\ }\bibfield  {title} {\enquote {\bibinfo {title} {{X-Point
  Collapse and Saturation in the Nonlinear Tearing Mode Reconnection}},}\
  }\href {\doibase 10.1103/PhysRevLett.95.235003} {\bibfield  {journal}
  {\bibinfo  {journal} {\prl}\ }\textbf {\bibinfo {volume} {95}},\ \bibinfo
  {eid} {235003} (\bibinfo {year} {2005})},\ \Eprint
  {http://arxiv.org/abs/physics/0507206} {arXiv:physics/0507206
  [physics.plasm-ph]} \BibitemShut {NoStop}%
\bibitem [{\citenamefont {{Takasao}}\ \emph {et~al.}(2012)\citenamefont
  {{Takasao}}, \citenamefont {{Asai}}, \citenamefont {{Isobe}},\ and\
  \citenamefont {{Shibata}}}]{2012ApJ...745L...6T}%
  \BibitemOpen
  \bibfield  {author} {\bibinfo {author} {\bibfnamefont {S.}~\bibnamefont
  {{Takasao}}}, \bibinfo {author} {\bibfnamefont {A.}~\bibnamefont {{Asai}}},
  \bibinfo {author} {\bibfnamefont {H.}~\bibnamefont {{Isobe}}}, \ and\
  \bibinfo {author} {\bibfnamefont {K.}~\bibnamefont {{Shibata}}},\ }\bibfield
  {title} {\enquote {\bibinfo {title} {{Simultaneous Observation of
  Reconnection Inflow and Outflow Associated with the 2010 August 18 Solar
  Flare}},}\ }\href {\doibase 10.1088/2041-8205/745/1/L6} {\bibfield  {journal}
  {\bibinfo  {journal} {\apjl}\ }\textbf {\bibinfo {volume} {745}},\ \bibinfo
  {eid} {L6} (\bibinfo {year} {2012})},\ \Eprint
  {http://arxiv.org/abs/1112.1398} {arXiv:1112.1398 [astro-ph.SR]} \BibitemShut
  {NoStop}%
\bibitem [{\citenamefont {{Bhattacharjee}}\ \emph {et~al.}(2009)\citenamefont
  {{Bhattacharjee}}, \citenamefont {{Huang}}, \citenamefont {{Yang}},\ and\
  \citenamefont {{Rogers}}}]{2009PhPl...16k2102B}%
  \BibitemOpen
  \bibfield  {author} {\bibinfo {author} {\bibfnamefont {A.}~\bibnamefont
  {{Bhattacharjee}}}, \bibinfo {author} {\bibfnamefont {Y.-M.}\ \bibnamefont
  {{Huang}}}, \bibinfo {author} {\bibfnamefont {H.}~\bibnamefont {{Yang}}}, \
  and\ \bibinfo {author} {\bibfnamefont {B.}~\bibnamefont {{Rogers}}},\
  }\bibfield  {title} {\enquote {\bibinfo {title} {{Fast reconnection in
  high-Lundquist-number plasmas due to the plasmoid Instability}},}\ }\href
  {\doibase 10.1063/1.3264103} {\bibfield  {journal} {\bibinfo  {journal}
  {Phys. Plasmas}\ }\textbf {\bibinfo {volume} {16}},\ \bibinfo {eid} {112102}
  (\bibinfo {year} {2009})},\ \Eprint {http://arxiv.org/abs/0906.5599}
  {arXiv:0906.5599 [physics.plasm-ph]} \BibitemShut {NoStop}%
\bibitem [{\citenamefont {{Cassak}}, \citenamefont {{Shay}},\ and\
  \citenamefont {{Drake}}(2009)}]{2009PhPl...16l0702C}%
  \BibitemOpen
  \bibfield  {author} {\bibinfo {author} {\bibfnamefont {P.~A.}\ \bibnamefont
  {{Cassak}}}, \bibinfo {author} {\bibfnamefont {M.~A.}\ \bibnamefont
  {{Shay}}}, \ and\ \bibinfo {author} {\bibfnamefont {J.~F.}\ \bibnamefont
  {{Drake}}},\ }\bibfield  {title} {\enquote {\bibinfo {title} {{Scaling of
  Sweet-Parker reconnection with secondary islands}},}\ }\href {\doibase
  10.1063/1.3274462} {\bibfield  {journal} {\bibinfo  {journal} {Physics of
  Plasmas}\ }\textbf {\bibinfo {volume} {16}},\ \bibinfo {eid} {120702}
  (\bibinfo {year} {2009})}\BibitemShut {NoStop}%
\bibitem [{\citenamefont {{Huang}}\ and\ \citenamefont
  {{Bhattacharjee}}(2010)}]{2010PhPl...17f2104H}%
  \BibitemOpen
  \bibfield  {author} {\bibinfo {author} {\bibfnamefont {Y.-M.}\ \bibnamefont
  {{Huang}}}\ and\ \bibinfo {author} {\bibfnamefont {A.}~\bibnamefont
  {{Bhattacharjee}}},\ }\bibfield  {title} {\enquote {\bibinfo {title}
  {{Scaling laws of resistive magnetohydrodynamic reconnection in the
  high-Lundquist-number, plasmoid-unstable regime}},}\ }\href {\doibase
  10.1063/1.3420208} {\bibfield  {journal} {\bibinfo  {journal} {Physics of
  Plasmas}\ }\textbf {\bibinfo {volume} {17}},\ \bibinfo {pages}
  {062104--062104} (\bibinfo {year} {2010})},\ \Eprint
  {http://arxiv.org/abs/1003.5951} {arXiv:1003.5951 [physics.plasm-ph]}
  \BibitemShut {NoStop}%
\bibitem [{\citenamefont {{Ni}}\ \emph {et~al.}(2010)\citenamefont {{Ni}},
  \citenamefont {{Germaschewski}}, \citenamefont {{Huang}}, \citenamefont
  {{Sullivan}}, \citenamefont {{Yang}},\ and\ \citenamefont
  {{Bhattacharjee}}}]{2010PhPl...17e2109N}%
  \BibitemOpen
  \bibfield  {author} {\bibinfo {author} {\bibfnamefont {L.}~\bibnamefont
  {{Ni}}}, \bibinfo {author} {\bibfnamefont {K.}~\bibnamefont
  {{Germaschewski}}}, \bibinfo {author} {\bibfnamefont {Y.-M.}\ \bibnamefont
  {{Huang}}}, \bibinfo {author} {\bibfnamefont {B.~P.}\ \bibnamefont
  {{Sullivan}}}, \bibinfo {author} {\bibfnamefont {H.}~\bibnamefont {{Yang}}},
  \ and\ \bibinfo {author} {\bibfnamefont {A.}~\bibnamefont
  {{Bhattacharjee}}},\ }\bibfield  {title} {\enquote {\bibinfo {title} {{Linear
  plasmoid instability of thin current sheets with shear flow}},}\ }\href
  {\doibase 10.1063/1.3428553} {\bibfield  {journal} {\bibinfo  {journal}
  {Physics of Plasmas}\ }\textbf {\bibinfo {volume} {17}},\ \bibinfo {eid}
  {052109} (\bibinfo {year} {2010})}\BibitemShut {NoStop}%
\bibitem [{\citenamefont {{Ni}}\ \emph {et~al.}(2012)\citenamefont {{Ni}},
  \citenamefont {{Ziegler}}, \citenamefont {{Huang}}, \citenamefont {{Lin}},\
  and\ \citenamefont {{Mei}}}]{2012PhPl...19g2902N}%
  \BibitemOpen
  \bibfield  {author} {\bibinfo {author} {\bibfnamefont {L.}~\bibnamefont
  {{Ni}}}, \bibinfo {author} {\bibfnamefont {U.}~\bibnamefont {{Ziegler}}},
  \bibinfo {author} {\bibfnamefont {Y.-M.}\ \bibnamefont {{Huang}}}, \bibinfo
  {author} {\bibfnamefont {J.}~\bibnamefont {{Lin}}}, \ and\ \bibinfo {author}
  {\bibfnamefont {Z.}~\bibnamefont {{Mei}}},\ }\bibfield  {title} {\enquote
  {\bibinfo {title} {{Effects of plasma {\ensuremath{\beta}} on the plasmoid
  instability}},}\ }\href {\doibase 10.1063/1.4736993} {\bibfield  {journal}
  {\bibinfo  {journal} {Physics of Plasmas}\ }\textbf {\bibinfo {volume}
  {19}},\ \bibinfo {eid} {072902} (\bibinfo {year} {2012})}\BibitemShut
  {NoStop}%
\bibitem [{\citenamefont {{Ni}}, \citenamefont {{Lin}},\ and\ \citenamefont
  {{Murphy}}(2013)}]{2013PhPl...20f1206N}%
  \BibitemOpen
  \bibfield  {author} {\bibinfo {author} {\bibfnamefont {L.}~\bibnamefont
  {{Ni}}}, \bibinfo {author} {\bibfnamefont {J.}~\bibnamefont {{Lin}}}, \ and\
  \bibinfo {author} {\bibfnamefont {N.~A.}\ \bibnamefont {{Murphy}}},\
  }\bibfield  {title} {\enquote {\bibinfo {title} {{Effects of the non-uniform
  initial environment and the guide field on the plasmoid instability}},}\
  }\href {\doibase 10.1063/1.4811144} {\bibfield  {journal} {\bibinfo
  {journal} {Physics of Plasmas}\ }\textbf {\bibinfo {volume} {20}},\ \bibinfo
  {eid} {061206} (\bibinfo {year} {2013})},\ \Eprint
  {http://arxiv.org/abs/1307.1963} {arXiv:1307.1963 [physics.space-ph]}
  \BibitemShut {NoStop}%
\bibitem [{\citenamefont {{Pucci}}\ and\ \citenamefont
  {{Velli}}(2014)}]{2014ApJ...780L..19P}%
  \BibitemOpen
  \bibfield  {author} {\bibinfo {author} {\bibfnamefont {F.}~\bibnamefont
  {{Pucci}}}\ and\ \bibinfo {author} {\bibfnamefont {M.}~\bibnamefont
  {{Velli}}},\ }\bibfield  {title} {\enquote {\bibinfo {title} {{Reconnection
  of Quasi-singular Current Sheets: The ``Ideal'' Tearing Mode}},}\ }\href
  {\doibase 10.1088/2041-8205/780/2/L19} {\bibfield  {journal} {\bibinfo
  {journal} {\apjl}\ }\textbf {\bibinfo {volume} {780}},\ \bibinfo {eid} {L19}
  (\bibinfo {year} {2014})}\BibitemShut {NoStop}%
\bibitem [{\citenamefont {{Tenerani}}\ \emph
  {et~al.}(2015{\natexlab{a}})\citenamefont {{Tenerani}}, \citenamefont
  {{Rappazzo}}, \citenamefont {{Velli}},\ and\ \citenamefont
  {{Pucci}}}]{2015ApJ...801..145T}%
  \BibitemOpen
  \bibfield  {author} {\bibinfo {author} {\bibfnamefont {A.}~\bibnamefont
  {{Tenerani}}}, \bibinfo {author} {\bibfnamefont {A.~F.}\ \bibnamefont
  {{Rappazzo}}}, \bibinfo {author} {\bibfnamefont {M.}~\bibnamefont {{Velli}}},
  \ and\ \bibinfo {author} {\bibfnamefont {F.}~\bibnamefont {{Pucci}}},\
  }\bibfield  {title} {\enquote {\bibinfo {title} {{The Tearing Mode
  Instability of Thin Current Sheets: the Transition to Fast Reconnection in
  the Presence of Viscosity}},}\ }\href {\doibase 10.1088/0004-637X/801/2/145}
  {\bibfield  {journal} {\bibinfo  {journal} {\apj}\ }\textbf {\bibinfo
  {volume} {801}},\ \bibinfo {eid} {145} (\bibinfo {year}
  {2015}{\natexlab{a}})},\ \Eprint {http://arxiv.org/abs/1412.0047}
  {arXiv:1412.0047 [physics.plasm-ph]} \BibitemShut {NoStop}%
\bibitem [{\citenamefont {{Tenerani}}\ \emph
  {et~al.}(2015{\natexlab{b}})\citenamefont {{Tenerani}}, \citenamefont
  {{Velli}}, \citenamefont {{Rappazzo}},\ and\ \citenamefont
  {{Pucci}}}]{2015ApJ...813L..32T}%
  \BibitemOpen
  \bibfield  {author} {\bibinfo {author} {\bibfnamefont {A.}~\bibnamefont
  {{Tenerani}}}, \bibinfo {author} {\bibfnamefont {M.}~\bibnamefont {{Velli}}},
  \bibinfo {author} {\bibfnamefont {A.~F.}\ \bibnamefont {{Rappazzo}}}, \ and\
  \bibinfo {author} {\bibfnamefont {F.}~\bibnamefont {{Pucci}}},\ }\bibfield
  {title} {\enquote {\bibinfo {title} {{Magnetic Reconnection: Recursive
  Current Sheet Collapse Triggered by
  {\textquotedblleft}Ideal{\textquotedblright} Tearing}},}\ }\href {\doibase
  10.1088/2041-8205/813/2/L32} {\bibfield  {journal} {\bibinfo  {journal}
  {\apjl}\ }\textbf {\bibinfo {volume} {813}},\ \bibinfo {eid} {L32} (\bibinfo
  {year} {2015}{\natexlab{b}})},\ \Eprint {http://arxiv.org/abs/1506.08921}
  {arXiv:1506.08921 [physics.plasm-ph]} \BibitemShut {NoStop}%
\bibitem [{\citenamefont {{Pucci}}\ \emph {et~al.}(2018)\citenamefont
  {{Pucci}}, \citenamefont {{Velli}}, \citenamefont {{Tenerani}},\ and\
  \citenamefont {{Del Sarto}}}]{2018PhPl...25c2113P}%
  \BibitemOpen
  \bibfield  {author} {\bibinfo {author} {\bibfnamefont {F.}~\bibnamefont
  {{Pucci}}}, \bibinfo {author} {\bibfnamefont {M.}~\bibnamefont {{Velli}}},
  \bibinfo {author} {\bibfnamefont {A.}~\bibnamefont {{Tenerani}}}, \ and\
  \bibinfo {author} {\bibfnamefont {D.}~\bibnamefont {{Del Sarto}}},\
  }\bibfield  {title} {\enquote {\bibinfo {title} {{Onset of fast ``ideal''
  tearing in thin current sheets: Dependence on the equilibrium current
  profile}},}\ }\href {\doibase 10.1063/1.5022988} {\bibfield  {journal}
  {\bibinfo  {journal} {Physics of Plasmas}\ }\textbf {\bibinfo {volume}
  {25}},\ \bibinfo {eid} {032113} (\bibinfo {year} {2018})},\ \Eprint
  {http://arxiv.org/abs/1801.08412} {arXiv:1801.08412 [physics.plasm-ph]}
  \BibitemShut {NoStop}%
\bibitem [{\citenamefont {{Huang}}, \citenamefont {{Comisso}},\ and\
  \citenamefont {{Bhattacharjee}}(2017)}]{2017ApJ...849...75H}%
  \BibitemOpen
  \bibfield  {author} {\bibinfo {author} {\bibfnamefont {Y.~M.}\ \bibnamefont
  {{Huang}}}, \bibinfo {author} {\bibfnamefont {L.}~\bibnamefont {{Comisso}}},
  \ and\ \bibinfo {author} {\bibfnamefont {A.}~\bibnamefont
  {{Bhattacharjee}}},\ }\bibfield  {title} {\enquote {\bibinfo {title}
  {{Plasmoid Instability in Evolving Current Sheets and Onset of Fast
  Reconnection}},}\ }\href {\doibase 10.3847/1538-4357/aa906d} {\bibfield
  {journal} {\bibinfo  {journal} {\apj}\ }\textbf {\bibinfo {volume} {849}},\
  \bibinfo {eid} {75} (\bibinfo {year} {2017})},\ \Eprint
  {http://arxiv.org/abs/1707.01863} {arXiv:1707.01863 [physics.plasm-ph]}
  \BibitemShut {NoStop}%
\bibitem [{\citenamefont {{Ni}}\ and\ \citenamefont
  {{Lukin}}(2018)}]{2018ApJ...868..144N}%
  \BibitemOpen
  \bibfield  {author} {\bibinfo {author} {\bibfnamefont {L.}~\bibnamefont
  {{Ni}}}\ and\ \bibinfo {author} {\bibfnamefont {V.~S.}\ \bibnamefont
  {{Lukin}}},\ }\bibfield  {title} {\enquote {\bibinfo {title} {{Onset of
  Secondary Instabilities and Plasma Heating during Magnetic Reconnection in
  Strongly Magnetized Regions of the Low Solar Atmosphere}},}\ }\href {\doibase
  10.3847/1538-4357/aaeb97} {\bibfield  {journal} {\bibinfo  {journal} {\apj}\
  }\textbf {\bibinfo {volume} {868}},\ \bibinfo {eid} {144} (\bibinfo {year}
  {2018})},\ \Eprint {http://arxiv.org/abs/1810.09874} {arXiv:1810.09874
  [astro-ph.SR]} \BibitemShut {NoStop}%
\bibitem [{\citenamefont {{Comisso}}\ and\ \citenamefont
  {{Grasso}}(2016)}]{2016PhPl...23c2111C}%
  \BibitemOpen
  \bibfield  {author} {\bibinfo {author} {\bibfnamefont {L.}~\bibnamefont
  {{Comisso}}}\ and\ \bibinfo {author} {\bibfnamefont {D.}~\bibnamefont
  {{Grasso}}},\ }\bibfield  {title} {\enquote {\bibinfo {title}
  {{Visco-resistive plasmoid instability}},}\ }\href {\doibase
  10.1063/1.4942940} {\bibfield  {journal} {\bibinfo  {journal} {Physics of
  Plasmas}\ }\textbf {\bibinfo {volume} {23}},\ \bibinfo {eid} {032111}
  (\bibinfo {year} {2016})},\ \Eprint {http://arxiv.org/abs/1603.00090}
  {arXiv:1603.00090 [physics.plasm-ph]} \BibitemShut {NoStop}%
\bibitem [{\citenamefont {{Comisso}}\ \emph {et~al.}(2016)\citenamefont
  {{Comisso}}, \citenamefont {{Lingam}}, \citenamefont {{Huang}},\ and\
  \citenamefont {{Bhattacharjee}}}]{2016PhPl...23j0702C}%
  \BibitemOpen
  \bibfield  {author} {\bibinfo {author} {\bibfnamefont {L.}~\bibnamefont
  {{Comisso}}}, \bibinfo {author} {\bibfnamefont {M.}~\bibnamefont {{Lingam}}},
  \bibinfo {author} {\bibfnamefont {Y.~M.}\ \bibnamefont {{Huang}}}, \ and\
  \bibinfo {author} {\bibfnamefont {A.}~\bibnamefont {{Bhattacharjee}}},\
  }\bibfield  {title} {\enquote {\bibinfo {title} {{General theory of the
  plasmoid instability}},}\ }\href {\doibase 10.1063/1.4964481} {\bibfield
  {journal} {\bibinfo  {journal} {Physics of Plasmas}\ }\textbf {\bibinfo
  {volume} {23}},\ \bibinfo {eid} {100702} (\bibinfo {year} {2016})},\ \Eprint
  {http://arxiv.org/abs/1608.04692} {arXiv:1608.04692 [physics.plasm-ph]}
  \BibitemShut {NoStop}%
\bibitem [{\citenamefont {{Comisso}}\ \emph {et~al.}(2017)\citenamefont
  {{Comisso}}, \citenamefont {{Lingam}}, \citenamefont {{Huang}},\ and\
  \citenamefont {{Bhattacharjee}}}]{2017ApJ...850..142C}%
  \BibitemOpen
  \bibfield  {author} {\bibinfo {author} {\bibfnamefont {L.}~\bibnamefont
  {{Comisso}}}, \bibinfo {author} {\bibfnamefont {M.}~\bibnamefont {{Lingam}}},
  \bibinfo {author} {\bibfnamefont {Y.~M.}\ \bibnamefont {{Huang}}}, \ and\
  \bibinfo {author} {\bibfnamefont {A.}~\bibnamefont {{Bhattacharjee}}},\
  }\bibfield  {title} {\enquote {\bibinfo {title} {{Plasmoid Instability in
  Forming Current Sheets}},}\ }\href {\doibase 10.3847/1538-4357/aa9789}
  {\bibfield  {journal} {\bibinfo  {journal} {\apj}\ }\textbf {\bibinfo
  {volume} {850}},\ \bibinfo {eid} {142} (\bibinfo {year} {2017})},\ \Eprint
  {http://arxiv.org/abs/1707.01862} {arXiv:1707.01862 [astro-ph.HE]}
  \BibitemShut {NoStop}%
\bibitem [{\citenamefont {{Leake}}\ \emph {et~al.}(2012)\citenamefont
  {{Leake}}, \citenamefont {{Lukin}}, \citenamefont {{Linton}},\ and\
  \citenamefont {{Meier}}}]{2012ApJ...760..109L}%
  \BibitemOpen
  \bibfield  {author} {\bibinfo {author} {\bibfnamefont {J.~E.}\ \bibnamefont
  {{Leake}}}, \bibinfo {author} {\bibfnamefont {V.~S.}\ \bibnamefont
  {{Lukin}}}, \bibinfo {author} {\bibfnamefont {M.~G.}\ \bibnamefont
  {{Linton}}}, \ and\ \bibinfo {author} {\bibfnamefont {E.~T.}\ \bibnamefont
  {{Meier}}},\ }\bibfield  {title} {\enquote {\bibinfo {title} {{Multi-fluid
  Simulations of Chromospheric Magnetic Reconnection in a Weakly Ionized
  Reacting Plasma}},}\ }\href {\doibase 10.1088/0004-637X/760/2/109} {\bibfield
   {journal} {\bibinfo  {journal} {\apj}\ }\textbf {\bibinfo {volume} {760}},\
  \bibinfo {eid} {109} (\bibinfo {year} {2012})},\ \Eprint
  {http://arxiv.org/abs/1210.1807} {arXiv:1210.1807 [physics.plasm-ph]}
  \BibitemShut {NoStop}%
\bibitem [{\citenamefont {{Leake}}, \citenamefont {{Lukin}},\ and\
  \citenamefont {{Linton}}(2013)}]{2013PhPl...20f1202L}%
  \BibitemOpen
  \bibfield  {author} {\bibinfo {author} {\bibfnamefont {J.~E.}\ \bibnamefont
  {{Leake}}}, \bibinfo {author} {\bibfnamefont {V.~S.}\ \bibnamefont
  {{Lukin}}}, \ and\ \bibinfo {author} {\bibfnamefont {M.~G.}\ \bibnamefont
  {{Linton}}},\ }\bibfield  {title} {\enquote {\bibinfo {title} {{Magnetic
  reconnection in a weakly ionized plasma}},}\ }\href {\doibase
  10.1063/1.4811140} {\bibfield  {journal} {\bibinfo  {journal} {Physics of
  Plasmas}\ }\textbf {\bibinfo {volume} {20}},\ \bibinfo {eid} {061202}
  (\bibinfo {year} {2013})},\ \Eprint {http://arxiv.org/abs/1302.3287}
  {arXiv:1302.3287 [physics.plasm-ph]} \BibitemShut {NoStop}%
\bibitem [{\citenamefont {{Tajima}}\ and\ \citenamefont
  {{Sakai}}(1986)}]{1986ITPS...14..929T}%
  \BibitemOpen
  \bibfield  {author} {\bibinfo {author} {\bibfnamefont {T.}~\bibnamefont
  {{Tajima}}}\ and\ \bibinfo {author} {\bibfnamefont {J.~I.}\ \bibnamefont
  {{Sakai}}},\ }\bibfield  {title} {\enquote {\bibinfo {title} {{Explosive
  coalescence of magnetic islands.}}}\ }\href {\doibase
  10.1109/TPS.1986.4316643} {\bibfield  {journal} {\bibinfo  {journal} {IEEE
  Transactions on Plasma Science}\ }\textbf {\bibinfo {volume} {14}},\ \bibinfo
  {pages} {929--933} (\bibinfo {year} {1986})}\BibitemShut {NoStop}%
\bibitem [{\citenamefont {{Vernazza}}, \citenamefont {{Avrett}},\ and\
  \citenamefont {{Loeser}}(1981)}]{1981ApJS...45..635V}%
  \BibitemOpen
  \bibfield  {author} {\bibinfo {author} {\bibfnamefont {J.~E.}\ \bibnamefont
  {{Vernazza}}}, \bibinfo {author} {\bibfnamefont {E.~H.}\ \bibnamefont
  {{Avrett}}}, \ and\ \bibinfo {author} {\bibfnamefont {R.}~\bibnamefont
  {{Loeser}}},\ }\bibfield  {title} {\enquote {\bibinfo {title} {{Structure of
  the solar chromosphere. III. Models of the EUV brightness components of the
  quiet sun.}}}\ }\href {\doibase 10.1086/190731} {\bibfield  {journal}
  {\bibinfo  {journal} {\apjs}\ }\textbf {\bibinfo {volume} {45}},\ \bibinfo
  {pages} {635--725} (\bibinfo {year} {1981})}\BibitemShut {NoStop}%
\bibitem [{\citenamefont {{Pneuman}}, \citenamefont {{Solanki}},\ and\
  \citenamefont {{Stenflo}}(1986)}]{1986A&A...154..231P}%
  \BibitemOpen
  \bibfield  {author} {\bibinfo {author} {\bibfnamefont {G.~W.}\ \bibnamefont
  {{Pneuman}}}, \bibinfo {author} {\bibfnamefont {S.~K.}\ \bibnamefont
  {{Solanki}}}, \ and\ \bibinfo {author} {\bibfnamefont {J.~O.}\ \bibnamefont
  {{Stenflo}}},\ }\bibfield  {title} {\enquote {\bibinfo {title} {{Structure
  and merging of solar magnetic fluxtubes}},}\ }\href@noop {} {\bibfield
  {journal} {\bibinfo  {journal} {\aap}\ }\textbf {\bibinfo {volume} {154}},\
  \bibinfo {pages} {231--242} (\bibinfo {year} {1986})}\BibitemShut {NoStop}%
\bibitem [{\citenamefont {Khomenko}, \citenamefont {Collados},\ and\
  \citenamefont {Felipe}(2008)}]{Khomenko2008}%
  \BibitemOpen
  \bibfield  {author} {\bibinfo {author} {\bibfnamefont {E.}~\bibnamefont
  {Khomenko}}, \bibinfo {author} {\bibfnamefont {M.}~\bibnamefont {Collados}},
  \ and\ \bibinfo {author} {\bibfnamefont {T.}~\bibnamefont {Felipe}},\
  }\bibfield  {title} {\enquote {\bibinfo {title} {Nonlinear numerical
  simulations of magneto-acoustic wave propagation in small-scale flux
  tubes},}\ }\href {\doibase 10.1007/s11207-008-9133-8} {\bibfield  {journal}
  {\bibinfo  {journal} {Solar Physics}\ }\textbf {\bibinfo {volume} {251}},\
  \bibinfo {pages} {589--611} (\bibinfo {year} {2008})}\BibitemShut {NoStop}%
\bibitem [{\citenamefont {{Singh}}\ \emph {et~al.}(2015)\citenamefont
  {{Singh}}, \citenamefont {{Hillier}}, \citenamefont {{Isobe}},\ and\
  \citenamefont {{Shibata}}}]{2015PASJ...67...96S}%
  \BibitemOpen
  \bibfield  {author} {\bibinfo {author} {\bibfnamefont {K.~A.~P.}\
  \bibnamefont {{Singh}}}, \bibinfo {author} {\bibfnamefont {A.}~\bibnamefont
  {{Hillier}}}, \bibinfo {author} {\bibfnamefont {H.}~\bibnamefont {{Isobe}}},
  \ and\ \bibinfo {author} {\bibfnamefont {K.}~\bibnamefont {{Shibata}}},\
  }\bibfield  {title} {\enquote {\bibinfo {title} {{Nonlinear instability and
  intermittent nature of magnetic reconnection in solar chromosphere}},}\
  }\href {\doibase 10.1093/pasj/psv066} {\bibfield  {journal} {\bibinfo
  {journal} {\pasj}\ }\textbf {\bibinfo {volume} {67}},\ \bibinfo {eid} {96}
  (\bibinfo {year} {2015})},\ \Eprint {http://arxiv.org/abs/1602.01999}
  {arXiv:1602.01999 [astro-ph.SR]} \BibitemShut {NoStop}%
\bibitem [{\citenamefont {{Zweibel}}\ \emph {et~al.}(2011)\citenamefont
  {{Zweibel}}, \citenamefont {{Lawrence}}, \citenamefont {{Yoo}}, \citenamefont
  {{Ji}}, \citenamefont {{Yamada}},\ and\ \citenamefont
  {{Malyshkin}}}]{2011PhPl...18k1211Z}%
  \BibitemOpen
  \bibfield  {author} {\bibinfo {author} {\bibfnamefont {E.~G.}\ \bibnamefont
  {{Zweibel}}}, \bibinfo {author} {\bibfnamefont {E.}~\bibnamefont
  {{Lawrence}}}, \bibinfo {author} {\bibfnamefont {J.}~\bibnamefont {{Yoo}}},
  \bibinfo {author} {\bibfnamefont {H.}~\bibnamefont {{Ji}}}, \bibinfo {author}
  {\bibfnamefont {M.}~\bibnamefont {{Yamada}}}, \ and\ \bibinfo {author}
  {\bibfnamefont {L.~M.}\ \bibnamefont {{Malyshkin}}},\ }\bibfield  {title}
  {\enquote {\bibinfo {title} {{Magnetic reconnection in partially ionized
  plasmas}},}\ }\href {\doibase 10.1063/1.3656960} {\bibfield  {journal}
  {\bibinfo  {journal} {Physics of Plasmas}\ }\textbf {\bibinfo {volume}
  {18}},\ \bibinfo {eid} {111211} (\bibinfo {year} {2011})}\BibitemShut
  {NoStop}%
\bibitem [{\citenamefont {{Smith}}\ and\ \citenamefont
  {{Sakai}}(2008)}]{2008A&A...486..569S}%
  \BibitemOpen
  \bibfield  {author} {\bibinfo {author} {\bibfnamefont {P.~D.}\ \bibnamefont
  {{Smith}}}\ and\ \bibinfo {author} {\bibfnamefont {J.~I.}\ \bibnamefont
  {{Sakai}}},\ }\bibfield  {title} {\enquote {\bibinfo {title} {{Chromospheric
  magnetic reconnection: two-fluid simulations of coalescing current loops}},}\
  }\href {\doibase 10.1051/0004-6361:200809624} {\bibfield  {journal} {\bibinfo
   {journal} {\aap}\ }\textbf {\bibinfo {volume} {486}},\ \bibinfo {pages}
  {569--575} (\bibinfo {year} {2008})},\ \Eprint
  {http://arxiv.org/abs/0804.2086} {arXiv:0804.2086 [astro-ph]} \BibitemShut
  {NoStop}%
\bibitem [{\citenamefont {{Sakai}}\ and\ \citenamefont
  {{Smith}}(2009)}]{2009ApJ...691L..45S}%
  \BibitemOpen
  \bibfield  {author} {\bibinfo {author} {\bibfnamefont {J.~I.}\ \bibnamefont
  {{Sakai}}}\ and\ \bibinfo {author} {\bibfnamefont {P.~D.}\ \bibnamefont
  {{Smith}}},\ }\bibfield  {title} {\enquote {\bibinfo {title} {{Two-Fluid
  Simulations of Coalescing Penumbra Filaments Driven by Neutral-Hydrogen
  Flows}},}\ }\href {\doibase 10.1088/0004-637X/691/1/L45} {\bibfield
  {journal} {\bibinfo  {journal} {\apjl}\ }\textbf {\bibinfo {volume} {691}},\
  \bibinfo {pages} {L45--L48} (\bibinfo {year} {2009})}\BibitemShut {NoStop}%
\bibitem [{\citenamefont {{Hillier}}, \citenamefont {{Takasao}},\ and\
  \citenamefont {{Nakamura}}(2016)}]{2016A&A...591A.112H}%
  \BibitemOpen
  \bibfield  {author} {\bibinfo {author} {\bibfnamefont {A.}~\bibnamefont
  {{Hillier}}}, \bibinfo {author} {\bibfnamefont {S.}~\bibnamefont
  {{Takasao}}}, \ and\ \bibinfo {author} {\bibfnamefont {N.}~\bibnamefont
  {{Nakamura}}},\ }\bibfield  {title} {\enquote {\bibinfo {title} {{The
  formation and evolution of reconnection-driven, slow-mode shocks in a
  partially ionised plasma}},}\ }\href {\doibase 10.1051/0004-6361/201628215}
  {\bibfield  {journal} {\bibinfo  {journal} {\aap}\ }\textbf {\bibinfo
  {volume} {591}},\ \bibinfo {eid} {A112} (\bibinfo {year} {2016})},\ \Eprint
  {http://arxiv.org/abs/1602.01112} {arXiv:1602.01112 [astro-ph.SR]}
  \BibitemShut {NoStop}%
\bibitem [{\citenamefont {{Braginskii}}(1965)}]{1965RvPP....1..205B}%
  \BibitemOpen
  \bibfield  {author} {\bibinfo {author} {\bibfnamefont {S.~I.}\ \bibnamefont
  {{Braginskii}}},\ }\bibfield  {title} {\enquote {\bibinfo {title} {{Transport
  Processes in a Plasma}},}\ }\href@noop {} {\bibfield  {journal} {\bibinfo
  {journal} {Reviews of Plasma Physics}\ }\textbf {\bibinfo {volume} {1}},\
  \bibinfo {pages} {205} (\bibinfo {year} {1965})}\BibitemShut {NoStop}%
\bibitem [{\citenamefont {{Meier}}\ and\ \citenamefont
  {{Shumlak}}(2012)}]{2012PhPl...19g2508M}%
  \BibitemOpen
  \bibfield  {author} {\bibinfo {author} {\bibfnamefont {E.~T.}\ \bibnamefont
  {{Meier}}}\ and\ \bibinfo {author} {\bibfnamefont {U.}~\bibnamefont
  {{Shumlak}}},\ }\bibfield  {title} {\enquote {\bibinfo {title} {{A general
  nonlinear fluid model for reacting plasma-neutral mixtures}},}\ }\href
  {\doibase 10.1063/1.4736975} {\bibfield  {journal} {\bibinfo  {journal}
  {Physics of Plasmas}\ }\textbf {\bibinfo {volume} {19}},\ \bibinfo {eid}
  {072508} (\bibinfo {year} {2012})}\BibitemShut {NoStop}%
\bibitem [{\citenamefont {{Draine}}(1986)}]{1986MNRAS.220..133D}%
  \BibitemOpen
  \bibfield  {author} {\bibinfo {author} {\bibfnamefont {B.~T.}\ \bibnamefont
  {{Draine}}},\ }\bibfield  {title} {\enquote {\bibinfo {title}
  {{Multicomponent, reacting MHD flows}},}\ }\href {\doibase
  10.1093/mnras/220.1.133} {\bibfield  {journal} {\bibinfo  {journal} {\mnras}\
  }\textbf {\bibinfo {volume} {220}},\ \bibinfo {pages} {133--148} (\bibinfo
  {year} {1986})}\BibitemShut {NoStop}%
\bibitem [{\citenamefont {{Zank}}\ \emph {et~al.}(2018)\citenamefont {{Zank}},
  \citenamefont {{Adhikari}}, \citenamefont {{Zhao}}, \citenamefont
  {{Mostafavi}}, \citenamefont {{Zirnstein}},\ and\ \citenamefont
  {{McComas}}}]{2018ApJ...869...23Z}%
  \BibitemOpen
  \bibfield  {author} {\bibinfo {author} {\bibfnamefont {G.~P.}\ \bibnamefont
  {{Zank}}}, \bibinfo {author} {\bibfnamefont {L.}~\bibnamefont {{Adhikari}}},
  \bibinfo {author} {\bibfnamefont {L.~L.}\ \bibnamefont {{Zhao}}}, \bibinfo
  {author} {\bibfnamefont {P.}~\bibnamefont {{Mostafavi}}}, \bibinfo {author}
  {\bibfnamefont {E.~J.}\ \bibnamefont {{Zirnstein}}}, \ and\ \bibinfo {author}
  {\bibfnamefont {D.~J.}\ \bibnamefont {{McComas}}},\ }\bibfield  {title}
  {\enquote {\bibinfo {title} {{The Pickup Ion-mediated Solar Wind}},}\ }\href
  {\doibase 10.3847/1538-4357/aaebfe} {\bibfield  {journal} {\bibinfo
  {journal} {\apj}\ }\textbf {\bibinfo {volume} {869}},\ \bibinfo {eid} {23}
  (\bibinfo {year} {2018})}\BibitemShut {NoStop}%
\bibitem [{\citenamefont {{Hillier}}(2019)}]{2019PhPl...26h2902H}%
  \BibitemOpen
  \bibfield  {author} {\bibinfo {author} {\bibfnamefont {A.}~\bibnamefont
  {{Hillier}}},\ }\bibfield  {title} {\enquote {\bibinfo {title} {{Ion-neutral
  decoupling in the nonlinear Kelvin-Helmholtz instability: Case of
  field-aligned flow}},}\ }\href {\doibase 10.1063/1.5103248} {\bibfield
  {journal} {\bibinfo  {journal} {Physics of Plasmas}\ }\textbf {\bibinfo
  {volume} {26}},\ \bibinfo {eid} {082902} (\bibinfo {year} {2019})},\ \Eprint
  {http://arxiv.org/abs/1907.12507} {arXiv:1907.12507 [astro-ph.SR]}
  \BibitemShut {NoStop}%
\bibitem [{\citenamefont {Fadeev}, \citenamefont {Kvabtskhava},\ and\
  \citenamefont {Komarov}(1965)}]{Fadeev_1965}%
  \BibitemOpen
  \bibfield  {author} {\bibinfo {author} {\bibfnamefont {V.~M.}\ \bibnamefont
  {Fadeev}}, \bibinfo {author} {\bibfnamefont {I.~F.}\ \bibnamefont
  {Kvabtskhava}}, \ and\ \bibinfo {author} {\bibfnamefont {N.~N.}\ \bibnamefont
  {Komarov}},\ }\bibfield  {title} {\enquote {\bibinfo {title} {Self-focusing
  of local plasma currents},}\ }\href {\doibase 10.1088/0029-5515/5/3/003}
  {\bibfield  {journal} {\bibinfo  {journal} {Nuclear Fusion}\ }\textbf
  {\bibinfo {volume} {5}},\ \bibinfo {pages} {202--209} (\bibinfo {year}
  {1965})}\BibitemShut {NoStop}%
\bibitem [{\citenamefont {{Arber}}, \citenamefont {{Botha}},\ and\
  \citenamefont {{Brady}}(2009)}]{2009ApJ...705.1183A}%
  \BibitemOpen
  \bibfield  {author} {\bibinfo {author} {\bibfnamefont {T.~D.}\ \bibnamefont
  {{Arber}}}, \bibinfo {author} {\bibfnamefont {G.~J.~J.}\ \bibnamefont
  {{Botha}}}, \ and\ \bibinfo {author} {\bibfnamefont {C.~S.}\ \bibnamefont
  {{Brady}}},\ }\bibfield  {title} {\enquote {\bibinfo {title} {{Effect of
  Solar Chromospheric Neutrals on Equilibrium Field Structures}},}\ }\href
  {\doibase 10.1088/0004-637X/705/2/1183} {\bibfield  {journal} {\bibinfo
  {journal} {\apj}\ }\textbf {\bibinfo {volume} {705}},\ \bibinfo {pages}
  {1183--1188} (\bibinfo {year} {2009})}\BibitemShut {NoStop}%
\bibitem [{\citenamefont {{Harris}}(1962)}]{1962NCim...23..115H}%
  \BibitemOpen
  \bibfield  {author} {\bibinfo {author} {\bibfnamefont {E.~G.}\ \bibnamefont
  {{Harris}}},\ }\bibfield  {title} {\enquote {\bibinfo {title} {{On a plasma
  sheath separating regions of oppositely directed magnetic field}},}\ }\href
  {\doibase 10.1007/BF02733547} {\bibfield  {journal} {\bibinfo  {journal} {Il
  Nuovo Cimento}\ }\textbf {\bibinfo {volume} {23}},\ \bibinfo {pages}
  {115--121} (\bibinfo {year} {1962})}\BibitemShut {NoStop}%
\bibitem [{\citenamefont {{Brandenburg}}\ and\ \citenamefont
  {{Zweibel}}(1994)}]{1994ApJ...427L..91B}%
  \BibitemOpen
  \bibfield  {author} {\bibinfo {author} {\bibfnamefont {A.}~\bibnamefont
  {{Brandenburg}}}\ and\ \bibinfo {author} {\bibfnamefont {E.~G.}\ \bibnamefont
  {{Zweibel}}},\ }\bibfield  {title} {\enquote {\bibinfo {title} {{The
  formation of sharp structures by ambipolar diffusion}},}\ }\href {\doibase
  10.1086/187372} {\bibfield  {journal} {\bibinfo  {journal} {The Astrophysical
  Journal Letters}\ }\textbf {\bibinfo {volume} {427}},\ \bibinfo {pages}
  {L91--L94} (\bibinfo {year} {1994})}\BibitemShut {NoStop}%
\bibitem [{\citenamefont {{Brandenburg}}\ and\ \citenamefont
  {{Zweibel}}(1995)}]{1995ApJ...448..734B}%
  \BibitemOpen
  \bibfield  {author} {\bibinfo {author} {\bibfnamefont {A.}~\bibnamefont
  {{Brandenburg}}}\ and\ \bibinfo {author} {\bibfnamefont {E.~G.}\ \bibnamefont
  {{Zweibel}}},\ }\bibfield  {title} {\enquote {\bibinfo {title} {{Effects of
  Pressure and Resistivity on the Ambipolar Diffusion Singularity: Too Little,
  Too Late}},}\ }\href {\doibase 10.1086/176001} {\bibfield  {journal}
  {\bibinfo  {journal} {The Astrophysical Journal}\ }\textbf {\bibinfo {volume}
  {448}},\ \bibinfo {pages} {734} (\bibinfo {year} {1995})}\BibitemShut
  {NoStop}%
\bibitem [{\citenamefont {{Parker}}(1963)}]{1963ApJS....8..177P}%
  \BibitemOpen
  \bibfield  {author} {\bibinfo {author} {\bibfnamefont {E.~N.}\ \bibnamefont
  {{Parker}}},\ }\bibfield  {title} {\enquote {\bibinfo {title} {{The
  Solar-Flare Phenomenon and the Theory of Reconnection and Annihiliation of
  Magnetic Fields.}}}\ }\href {\doibase 10.1086/190087} {\bibfield  {journal}
  {\bibinfo  {journal} {\apjs}\ }\textbf {\bibinfo {volume} {8}},\ \bibinfo
  {pages} {177} (\bibinfo {year} {1963})}\BibitemShut {NoStop}%
\bibitem [{\citenamefont {{Vishniac}}\ and\ \citenamefont
  {{Lazarian}}(1999)}]{1999ApJ...511..193V}%
  \BibitemOpen
  \bibfield  {author} {\bibinfo {author} {\bibfnamefont {E.~T.}\ \bibnamefont
  {{Vishniac}}}\ and\ \bibinfo {author} {\bibfnamefont {A.}~\bibnamefont
  {{Lazarian}}},\ }\bibfield  {title} {\enquote {\bibinfo {title}
  {{Reconnection in the Interstellar Medium}},}\ }\href {\doibase
  10.1086/306643} {\bibfield  {journal} {\bibinfo  {journal} {\apj}\ }\textbf
  {\bibinfo {volume} {511}},\ \bibinfo {pages} {193--203} (\bibinfo {year}
  {1999})},\ \Eprint {http://arxiv.org/abs/astro-ph/9712067}
  {arXiv:astro-ph/9712067 [astro-ph]} \BibitemShut {NoStop}%
\bibitem [{\citenamefont {{Heitsch}}\ and\ \citenamefont
  {{Zweibel}}(2003)}]{2003ApJ...583..229H}%
  \BibitemOpen
  \bibfield  {author} {\bibinfo {author} {\bibfnamefont {F.}~\bibnamefont
  {{Heitsch}}}\ and\ \bibinfo {author} {\bibfnamefont {E.~G.}\ \bibnamefont
  {{Zweibel}}},\ }\bibfield  {title} {\enquote {\bibinfo {title} {{Fast
  Reconnection in a Two-Stage Process}},}\ }\href {\doibase 10.1086/345082}
  {\bibfield  {journal} {\bibinfo  {journal} {\apj}\ }\textbf {\bibinfo
  {volume} {583}},\ \bibinfo {pages} {229--244} (\bibinfo {year} {2003})},\
  \Eprint {http://arxiv.org/abs/astro-ph/0205103} {arXiv:astro-ph/0205103
  [astro-ph]} \BibitemShut {NoStop}%
\bibitem [{\citenamefont {{Alvarez Laguna}}\ \emph {et~al.}(2017)\citenamefont
  {{Alvarez Laguna}}, \citenamefont {{Lani}}, \citenamefont {{Mansour}},
  \citenamefont {{Deconinck}},\ and\ \citenamefont
  {{Poedts}}}]{2017ApJ...842..117A}%
  \BibitemOpen
  \bibfield  {author} {\bibinfo {author} {\bibfnamefont {A.}~\bibnamefont
  {{Alvarez Laguna}}}, \bibinfo {author} {\bibfnamefont {A.}~\bibnamefont
  {{Lani}}}, \bibinfo {author} {\bibfnamefont {N.~N.}\ \bibnamefont
  {{Mansour}}}, \bibinfo {author} {\bibfnamefont {H.}~\bibnamefont
  {{Deconinck}}}, \ and\ \bibinfo {author} {\bibfnamefont {S.}~\bibnamefont
  {{Poedts}}},\ }\bibfield  {title} {\enquote {\bibinfo {title} {{Effect of
  Radiation on Chromospheric Magnetic Reconnection: Reactive and Collisional
  Multi-fluid Simulations}},}\ }\href {\doibase 10.3847/1538-4357/aa7554}
  {\bibfield  {journal} {\bibinfo  {journal} {\apj}\ }\textbf {\bibinfo
  {volume} {842}},\ \bibinfo {eid} {117} (\bibinfo {year} {2017})}\BibitemShut
  {NoStop}%
\bibitem [{\citenamefont {{Shibayama}}\ \emph {et~al.}(2015)\citenamefont
  {{Shibayama}}, \citenamefont {{Kusano}}, \citenamefont {{Miyoshi}},
  \citenamefont {{Nakabou}},\ and\ \citenamefont
  {{Vekstein}}}]{2015PhPl...22j0706S}%
  \BibitemOpen
  \bibfield  {author} {\bibinfo {author} {\bibfnamefont {T.}~\bibnamefont
  {{Shibayama}}}, \bibinfo {author} {\bibfnamefont {K.}~\bibnamefont
  {{Kusano}}}, \bibinfo {author} {\bibfnamefont {T.}~\bibnamefont {{Miyoshi}}},
  \bibinfo {author} {\bibfnamefont {T.}~\bibnamefont {{Nakabou}}}, \ and\
  \bibinfo {author} {\bibfnamefont {G.}~\bibnamefont {{Vekstein}}},\ }\bibfield
   {title} {\enquote {\bibinfo {title} {{Fast magnetic reconnection supported
  by sporadic small-scale Petschek-type shocks}},}\ }\href {\doibase
  10.1063/1.4934652} {\bibfield  {journal} {\bibinfo  {journal} {Physics of
  Plasmas}\ }\textbf {\bibinfo {volume} {22}},\ \bibinfo {eid} {100706}
  (\bibinfo {year} {2015})}\BibitemShut {NoStop}%
\bibitem [{\citenamefont {{Zenitani}}\ and\ \citenamefont
  {{Miyoshi}}(2011)}]{2011PhPl...18b2105Z}%
  \BibitemOpen
  \bibfield  {author} {\bibinfo {author} {\bibfnamefont {S.}~\bibnamefont
  {{Zenitani}}}\ and\ \bibinfo {author} {\bibfnamefont {T.}~\bibnamefont
  {{Miyoshi}}},\ }\bibfield  {title} {\enquote {\bibinfo {title}
  {{Magnetohydrodynamic structure of a plasmoid in fast reconnection in
  low-beta plasmas}},}\ }\href {\doibase 10.1063/1.3554655} {\bibfield
  {journal} {\bibinfo  {journal} {Physics of Plasmas}\ }\textbf {\bibinfo
  {volume} {18}},\ \bibinfo {pages} {022105--022105} (\bibinfo {year}
  {2011})},\ \Eprint {http://arxiv.org/abs/1101.2255} {arXiv:1101.2255
  [physics.space-ph]} \BibitemShut {NoStop}%
\bibitem [{\citenamefont {{Drazin}}\ and\ \citenamefont
  {{Reid}}(1981)}]{1981STIA...8217950D}%
  \BibitemOpen
  \bibfield  {author} {\bibinfo {author} {\bibfnamefont {P.~G.}\ \bibnamefont
  {{Drazin}}}\ and\ \bibinfo {author} {\bibfnamefont {W.~H.}\ \bibnamefont
  {{Reid}}},\ }\bibfield  {title} {\enquote {\bibinfo {title} {{Hydrodynamic
  stability}},}\ }\href@noop {} {\bibfield  {journal} {\bibinfo  {journal}
  {NASA STI/Recon Technical Report A}\ }\textbf {\bibinfo {volume} {82}},\
  \bibinfo {pages} {17950} (\bibinfo {year} {1981})}\BibitemShut {NoStop}%
\bibitem [{\citenamefont {Drazin}(2015)}]{DRAZIN2015343}%
  \BibitemOpen
  \bibfield  {author} {\bibinfo {author} {\bibfnamefont {P.}~\bibnamefont
  {Drazin}},\ }\bibfield  {title} {\enquote {\bibinfo {title} {Dynamical
  meteorology | kelvin–helmholtz instability},}\ }in\ \href {\doibase
  https://doi.org/10.1016/B978-0-12-382225-3.00190-0} {\emph {\bibinfo
  {booktitle} {Encyclopedia of Atmospheric Sciences (Second Edition)}}},\
  \bibinfo {editor} {edited by\ \bibinfo {editor} {\bibfnamefont {G.~R.}\
  \bibnamefont {North}}, \bibinfo {editor} {\bibfnamefont {J.}~\bibnamefont
  {Pyle}}, \ and\ \bibinfo {editor} {\bibfnamefont {F.}~\bibnamefont {Zhang}}}\
  (\bibinfo  {publisher} {Academic Press},\ \bibinfo {address} {Oxford},\
  \bibinfo {year} {2015})\ \bibinfo {edition} {second edition}\ ed.,\ pp.\
  \bibinfo {pages} {343 -- 346}\BibitemShut {NoStop}%
\bibitem [{\citenamefont {{Hillier}}\ and\ \citenamefont
  {{Polito}}(2018)}]{2018ApJ...864L..10H}%
  \BibitemOpen
  \bibfield  {author} {\bibinfo {author} {\bibfnamefont {A.}~\bibnamefont
  {{Hillier}}}\ and\ \bibinfo {author} {\bibfnamefont {V.}~\bibnamefont
  {{Polito}}},\ }\bibfield  {title} {\enquote {\bibinfo {title} {{Observations
  of the Kelvin-Helmholtz Instability Driven by Dynamic Motions in a Solar
  Prominence}},}\ }\href {\doibase 10.3847/2041-8213/aad9a5} {\bibfield
  {journal} {\bibinfo  {journal} {\apjl}\ }\textbf {\bibinfo {volume} {864}},\
  \bibinfo {eid} {L10} (\bibinfo {year} {2018})},\ \Eprint
  {http://arxiv.org/abs/1808.02286} {arXiv:1808.02286 [astro-ph.SR]}
  \BibitemShut {NoStop}%
\bibitem [{\citenamefont {{Snow, B.}}\ and\ \citenamefont {{Hillier,
  A.}}(2019)}]{refId0}%
  \BibitemOpen
  \bibfield  {author} {\bibinfo {author} {\bibnamefont {{Snow, B.}}}\ and\
  \bibinfo {author} {\bibnamefont {{Hillier, A.}}},\ }\bibfield  {title}
  {\enquote {\bibinfo {title} {Intermediate shock sub-structures within a
  slow-mode shock occurring in partially ionised plasma},}\ }\href {\doibase
  10.1051/0004-6361/201935326} {\bibfield  {journal} {\bibinfo  {journal}
  {A\&A}\ }\textbf {\bibinfo {volume} {626}},\ \bibinfo {pages} {A46} (\bibinfo
  {year} {2019})}\BibitemShut {NoStop}%
\bibitem [{\citenamefont {{Pucci}}\ \emph {et~al.}(2020)\citenamefont
  {{Pucci}}, \citenamefont {{Singh}}, \citenamefont {{Tenerani}},\ and\
  \citenamefont {{Velli}}}]{2020arXiv200603957P}%
  \BibitemOpen
  \bibfield  {author} {\bibinfo {author} {\bibfnamefont {F.}~\bibnamefont
  {{Pucci}}}, \bibinfo {author} {\bibfnamefont {K.~A.~P.}\ \bibnamefont
  {{Singh}}}, \bibinfo {author} {\bibfnamefont {A.}~\bibnamefont {{Tenerani}}},
  \ and\ \bibinfo {author} {\bibfnamefont {M.}~\bibnamefont {{Velli}}},\
  }\bibfield  {title} {\enquote {\bibinfo {title} {{Tearing modes in partially
  ionized plasmas}},}\ }\href@noop {} {\bibfield  {journal} {\bibinfo
  {journal} {arXiv e-prints}\ ,\ \bibinfo {eid} {arXiv:2006.03957}} (\bibinfo
  {year} {2020})},\ \Eprint {http://arxiv.org/abs/2006.03957} {arXiv:2006.03957
  [astro-ph.SR]} \BibitemShut {NoStop}%
\bibitem [{\citenamefont {{Vranjes}}\ and\ \citenamefont
  {{Krstic}}(2013)}]{2013A&A...554A..22V}%
  \BibitemOpen
  \bibfield  {author} {\bibinfo {author} {\bibfnamefont {J.}~\bibnamefont
  {{Vranjes}}}\ and\ \bibinfo {author} {\bibfnamefont {P.~S.}\ \bibnamefont
  {{Krstic}}},\ }\bibfield  {title} {\enquote {\bibinfo {title} {{Collisions,
  magnetization, and transport coefficients in the lower solar atmosphere}},}\
  }\href {\doibase 10.1051/0004-6361/201220738} {\bibfield  {journal} {\bibinfo
   {journal} {\aap}\ }\textbf {\bibinfo {volume} {554}},\ \bibinfo {eid} {A22}
  (\bibinfo {year} {2013})},\ \Eprint {http://arxiv.org/abs/1304.4010}
  {arXiv:1304.4010 [astro-ph.SR]} \BibitemShut {NoStop}%
\bibitem [{\citenamefont {{Leake}}, \citenamefont {{Arber}},\ and\
  \citenamefont {{Khodachenko}}(2005)}]{2005A&A...442.1091L}%
  \BibitemOpen
  \bibfield  {author} {\bibinfo {author} {\bibfnamefont {J.~E.}\ \bibnamefont
  {{Leake}}}, \bibinfo {author} {\bibfnamefont {T.~D.}\ \bibnamefont
  {{Arber}}}, \ and\ \bibinfo {author} {\bibfnamefont {M.~L.}\ \bibnamefont
  {{Khodachenko}}},\ }\bibfield  {title} {\enquote {\bibinfo {title}
  {{Collisional dissipation of Alfv{\'e}n waves in a partially ionised solar
  chromosphere}},}\ }\href {\doibase 10.1051/0004-6361:20053427} {\bibfield
  {journal} {\bibinfo  {journal} {\aap}\ }\textbf {\bibinfo {volume} {442}},\
  \bibinfo {pages} {1091--1098} (\bibinfo {year} {2005})},\ \Eprint
  {http://arxiv.org/abs/astro-ph/0510265} {arXiv:astro-ph/0510265 [astro-ph]}
  \BibitemShut {NoStop}%
\bibitem [{\citenamefont {{Khomenko}}\ and\ \citenamefont {{Collados
  Vera}}(2012)}]{2012ASPC..463..281K}%
  \BibitemOpen
  \bibfield  {author} {\bibinfo {author} {\bibfnamefont {E.}~\bibnamefont
  {{Khomenko}}}\ and\ \bibinfo {author} {\bibfnamefont {M.}~\bibnamefont
  {{Collados Vera}}},\ }\bibfield  {title} {\enquote {\bibinfo {title}
  {{Simulations of Chromospheric Heating by Ambipolar Diffusion}},}\ }in\
  \href@noop {} {\emph {\bibinfo {booktitle} {Second ATST-EAST Meeting:
  Magnetic Fields from the Photosphere to the Corona.}}},\ \bibinfo {series}
  {Astronomical Society of the Pacific Conference Series}, Vol.\ \bibinfo
  {volume} {463},\ \bibinfo {editor} {edited by\ \bibinfo {editor}
  {\bibfnamefont {T.~R.}\ \bibnamefont {{Rimmele}}}, \bibinfo {editor}
  {\bibfnamefont {A.}~\bibnamefont {{Tritschler}}}, \bibinfo {editor}
  {\bibfnamefont {F.}~\bibnamefont {{W{\"o}ger}}}, \bibinfo {editor}
  {\bibfnamefont {M.}~\bibnamefont {{Collados Vera}}}, \bibinfo {editor}
  {\bibfnamefont {H.}~\bibnamefont {{Socas-Navarro}}}, \bibinfo {editor}
  {\bibfnamefont {R.}~\bibnamefont {{Schlichenmaier}}}, \bibinfo {editor}
  {\bibfnamefont {M.}~\bibnamefont {{Carlsson}}}, \bibinfo {editor}
  {\bibfnamefont {T.}~\bibnamefont {{Berger}}}, \bibinfo {editor}
  {\bibfnamefont {A.}~\bibnamefont {{Cadavid}}}, \bibinfo {editor}
  {\bibfnamefont {P.~R.}\ \bibnamefont {{Gilbert}}}, \bibinfo {editor}
  {\bibfnamefont {P.~R.}\ \bibnamefont {{Goode}}}, \ and\ \bibinfo {editor}
  {\bibfnamefont {M.}~\bibnamefont {{Kn{\"o}lker}}}}\ (\bibinfo {year} {2012})\
  p.\ \bibinfo {pages} {281},\ \Eprint {http://arxiv.org/abs/1202.2252}
  {arXiv:1202.2252 [astro-ph.SR]} \BibitemShut {NoStop}%
\bibitem [{\citenamefont {{Khomenko}}\ and\ \citenamefont
  {{Collados}}(2012)}]{2012ApJ...747...87K}%
  \BibitemOpen
  \bibfield  {author} {\bibinfo {author} {\bibfnamefont {E.}~\bibnamefont
  {{Khomenko}}}\ and\ \bibinfo {author} {\bibfnamefont {M.}~\bibnamefont
  {{Collados}}},\ }\bibfield  {title} {\enquote {\bibinfo {title} {{Heating of
  the Magnetized Solar Chromosphere by Partial Ionization Effects}},}\ }\href
  {\doibase 10.1088/0004-637X/747/2/87} {\bibfield  {journal} {\bibinfo
  {journal} {The Astrophysical Journal}\ }\textbf {\bibinfo {volume} {747}},\
  \bibinfo {eid} {87} (\bibinfo {year} {2012})},\ \Eprint
  {http://arxiv.org/abs/1112.3374} {arXiv:1112.3374 [astro-ph.SR]} \BibitemShut
  {NoStop}%
\bibitem [{\citenamefont {{Pandey}}\ and\ \citenamefont
  {{Wardle}}(2008)}]{2008MNRAS.385.2269P}%
  \BibitemOpen
  \bibfield  {author} {\bibinfo {author} {\bibfnamefont {B.~P.}\ \bibnamefont
  {{Pandey}}}\ and\ \bibinfo {author} {\bibfnamefont {M.}~\bibnamefont
  {{Wardle}}},\ }\bibfield  {title} {\enquote {\bibinfo {title} {{Hall
  magnetohydrodynamics of partially ionized plasmas}},}\ }\href {\doibase
  10.1111/j.1365-2966.2008.12998.x} {\bibfield  {journal} {\bibinfo  {journal}
  {\mnras}\ }\textbf {\bibinfo {volume} {385}},\ \bibinfo {pages} {2269--2278}
  (\bibinfo {year} {2008})},\ \Eprint {http://arxiv.org/abs/0707.2688}
  {arXiv:0707.2688 [astro-ph]} \BibitemShut {NoStop}%
\bibitem [{\citenamefont {{Murphy}}\ and\ \citenamefont
  {{Lukin}}(2015)}]{2015ApJ...805..134M}%
  \BibitemOpen
  \bibfield  {author} {\bibinfo {author} {\bibfnamefont {N.~A.}\ \bibnamefont
  {{Murphy}}}\ and\ \bibinfo {author} {\bibfnamefont {V.~S.}\ \bibnamefont
  {{Lukin}}},\ }\bibfield  {title} {\enquote {\bibinfo {title} {{Asymmetric
  Magnetic Reconnection in Weakly Ionized Chromospheric Plasmas}},}\ }\href
  {\doibase 10.1088/0004-637X/805/2/134} {\bibfield  {journal} {\bibinfo
  {journal} {\apj}\ }\textbf {\bibinfo {volume} {805}},\ \bibinfo {eid} {134}
  (\bibinfo {year} {2015})},\ \Eprint {http://arxiv.org/abs/1504.01425}
  {arXiv:1504.01425 [astro-ph.SR]} \BibitemShut {NoStop}%
\bibitem [{\citenamefont {{Ni}}\ \emph
  {et~al.}(2018{\natexlab{a}})\citenamefont {{Ni}}, \citenamefont {{Lukin}},
  \citenamefont {{Murphy}},\ and\ \citenamefont {{Lin}}}]{2018ApJ...852...95N}%
  \BibitemOpen
  \bibfield  {author} {\bibinfo {author} {\bibfnamefont {L.}~\bibnamefont
  {{Ni}}}, \bibinfo {author} {\bibfnamefont {V.~S.}\ \bibnamefont {{Lukin}}},
  \bibinfo {author} {\bibfnamefont {N.~A.}\ \bibnamefont {{Murphy}}}, \ and\
  \bibinfo {author} {\bibfnamefont {J.}~\bibnamefont {{Lin}}},\ }\bibfield
  {title} {\enquote {\bibinfo {title} {{Magnetic Reconnection in Strongly
  Magnetized Regions of the Low Solar Chromosphere}},}\ }\href {\doibase
  10.3847/1538-4357/aa9edb} {\bibfield  {journal} {\bibinfo  {journal} {\apj}\
  }\textbf {\bibinfo {volume} {852}},\ \bibinfo {eid} {95} (\bibinfo {year}
  {2018}{\natexlab{a}})},\ \Eprint {http://arxiv.org/abs/1712.00582}
  {arXiv:1712.00582 [astro-ph.SR]} \BibitemShut {NoStop}%
\bibitem [{\citenamefont {{Ni}}\ \emph
  {et~al.}(2018{\natexlab{b}})\citenamefont {{Ni}}, \citenamefont {{Lukin}},
  \citenamefont {{Murphy}},\ and\ \citenamefont {{Lin}}}]{2018PhPl...25d2903N}%
  \BibitemOpen
  \bibfield  {author} {\bibinfo {author} {\bibfnamefont {L.}~\bibnamefont
  {{Ni}}}, \bibinfo {author} {\bibfnamefont {V.~S.}\ \bibnamefont {{Lukin}}},
  \bibinfo {author} {\bibfnamefont {N.~A.}\ \bibnamefont {{Murphy}}}, \ and\
  \bibinfo {author} {\bibfnamefont {J.}~\bibnamefont {{Lin}}},\ }\bibfield
  {title} {\enquote {\bibinfo {title} {{Magnetic reconnection in the low solar
  chromosphere with a more realistic radiative cooling model}},}\ }\href
  {\doibase 10.1063/1.5018351} {\bibfield  {journal} {\bibinfo  {journal}
  {Physics of Plasmas}\ }\textbf {\bibinfo {volume} {25}},\ \bibinfo {eid}
  {042903} (\bibinfo {year} {2018}{\natexlab{b}})},\ \Eprint
  {http://arxiv.org/abs/1804.05631} {arXiv:1804.05631 [astro-ph.SR]}
  \BibitemShut {NoStop}%
\bibitem [{\citenamefont {{Snow}}\ and\ \citenamefont
  {{Hillier}}(2020)}]{2020arXiv201006303S}%
  \BibitemOpen
  \bibfield  {author} {\bibinfo {author} {\bibfnamefont {B.}~\bibnamefont
  {{Snow}}}\ and\ \bibinfo {author} {\bibfnamefont {A.}~\bibnamefont
  {{Hillier}}},\ }\bibfield  {title} {\enquote {\bibinfo {title} {{Collisional
  ionisation, recombination and ionisation potential in two-fluid slow-mode
  shocks: analytical and numerical results}},}\ }\href@noop {} {\bibfield
  {journal} {\bibinfo  {journal} {arXiv e-prints}\ ,\ \bibinfo {eid}
  {arXiv:2010.06303}} (\bibinfo {year} {2020})},\ \Eprint
  {http://arxiv.org/abs/2010.06303} {arXiv:2010.06303 [astro-ph.SR]}
  \BibitemShut {NoStop}%
\bibitem [{\citenamefont {{Dorman}}\ and\ \citenamefont
  {{Kulsrud}}(1995)}]{1995ApJ...449..777D}%
  \BibitemOpen
  \bibfield  {author} {\bibinfo {author} {\bibfnamefont {V.~L.}\ \bibnamefont
  {{Dorman}}}\ and\ \bibinfo {author} {\bibfnamefont {R.~M.}\ \bibnamefont
  {{Kulsrud}}},\ }\bibfield  {title} {\enquote {\bibinfo {title}
  {{One-dimensional Merging of Magnetic Fields with Cooling}},}\ }\href
  {\doibase 10.1086/176097} {\bibfield  {journal} {\bibinfo  {journal} {\apj}\
  }\textbf {\bibinfo {volume} {449}},\ \bibinfo {pages} {777} (\bibinfo {year}
  {1995})}\BibitemShut {NoStop}%
\bibitem [{\citenamefont {{Uzdensky}}\ and\ \citenamefont
  {{McKinney}}(2011)}]{2011PhPl...18d2105U}%
  \BibitemOpen
  \bibfield  {author} {\bibinfo {author} {\bibfnamefont {D.~A.}\ \bibnamefont
  {{Uzdensky}}}\ and\ \bibinfo {author} {\bibfnamefont {J.~C.}\ \bibnamefont
  {{McKinney}}},\ }\bibfield  {title} {\enquote {\bibinfo {title} {{Magnetic
  reconnection with radiative cooling. I. Optically thin regime}},}\ }\href
  {\doibase 10.1063/1.3571602} {\bibfield  {journal} {\bibinfo  {journal}
  {Physics of Plasmas}\ }\textbf {\bibinfo {volume} {18}},\ \bibinfo {pages}
  {042105--042105} (\bibinfo {year} {2011})},\ \Eprint
  {http://arxiv.org/abs/1007.0774} {arXiv:1007.0774 [astro-ph.HE]} \BibitemShut
  {NoStop}%
\bibitem [{\citenamefont {{Leenaarts}}\ \emph {et~al.}(2007)\citenamefont
  {{Leenaarts}}, \citenamefont {{Carlsson}}, \citenamefont {{Hansteen}},\ and\
  \citenamefont {{Rutten}}}]{2007A&A...473..625L}%
  \BibitemOpen
  \bibfield  {author} {\bibinfo {author} {\bibfnamefont {J.}~\bibnamefont
  {{Leenaarts}}}, \bibinfo {author} {\bibfnamefont {M.}~\bibnamefont
  {{Carlsson}}}, \bibinfo {author} {\bibfnamefont {V.}~\bibnamefont
  {{Hansteen}}}, \ and\ \bibinfo {author} {\bibfnamefont {R.~J.}\ \bibnamefont
  {{Rutten}}},\ }\bibfield  {title} {\enquote {\bibinfo {title}
  {{Non-equilibrium hydrogen ionization in 2D simulations of the solar
  atmosphere}},}\ }\href {\doibase 10.1051/0004-6361:20078161} {\bibfield
  {journal} {\bibinfo  {journal} {\aap}\ }\textbf {\bibinfo {volume} {473}},\
  \bibinfo {pages} {625--632} (\bibinfo {year} {2007})},\ \Eprint
  {http://arxiv.org/abs/0709.3751} {arXiv:0709.3751 [astro-ph]} \BibitemShut
  {NoStop}%
\bibitem [{\citenamefont {{Golding}}, \citenamefont {{Carlsson}},\ and\
  \citenamefont {{Leenaarts}}(2014)}]{2014ApJ...784...30G}%
  \BibitemOpen
  \bibfield  {author} {\bibinfo {author} {\bibfnamefont {T.~P.}\ \bibnamefont
  {{Golding}}}, \bibinfo {author} {\bibfnamefont {M.}~\bibnamefont
  {{Carlsson}}}, \ and\ \bibinfo {author} {\bibfnamefont {J.}~\bibnamefont
  {{Leenaarts}}},\ }\bibfield  {title} {\enquote {\bibinfo {title} {{Detailed
  and Simplified Nonequilibrium Helium Ionization in the Solar Atmosphere}},}\
  }\href {\doibase 10.1088/0004-637X/784/1/30} {\bibfield  {journal} {\bibinfo
  {journal} {\apj}\ }\textbf {\bibinfo {volume} {784}},\ \bibinfo {eid} {30}
  (\bibinfo {year} {2014})},\ \Eprint {http://arxiv.org/abs/1401.7562}
  {arXiv:1401.7562 [astro-ph.SR]} \BibitemShut {NoStop}%
\bibitem [{\citenamefont {{Mart{\'\i}nez-Sykora}}\ \emph
  {et~al.}(2017)\citenamefont {{Mart{\'\i}nez-Sykora}}, \citenamefont {{De
  Pontieu}}, \citenamefont {{Carlsson}}, \citenamefont {{Hansteen}},
  \citenamefont {{N{\'o}brega-Siverio}},\ and\ \citenamefont
  {{Gudiksen}}}]{2017ApJ...847...36M}%
  \BibitemOpen
  \bibfield  {author} {\bibinfo {author} {\bibfnamefont {J.}~\bibnamefont
  {{Mart{\'\i}nez-Sykora}}}, \bibinfo {author} {\bibfnamefont {B.}~\bibnamefont
  {{De Pontieu}}}, \bibinfo {author} {\bibfnamefont {M.}~\bibnamefont
  {{Carlsson}}}, \bibinfo {author} {\bibfnamefont {V.~H.}\ \bibnamefont
  {{Hansteen}}}, \bibinfo {author} {\bibfnamefont {D.}~\bibnamefont
  {{N{\'o}brega-Siverio}}}, \ and\ \bibinfo {author} {\bibfnamefont {B.~V.}\
  \bibnamefont {{Gudiksen}}},\ }\bibfield  {title} {\enquote {\bibinfo {title}
  {{Two-dimensional Radiative Magnetohydrodynamic Simulations of Partial
  Ionization in the Chromosphere. II. Dynamics and Energetics of the Low Solar
  Atmosphere}},}\ }\href {\doibase 10.3847/1538-4357/aa8866} {\bibfield
  {journal} {\bibinfo  {journal} {\apj}\ }\textbf {\bibinfo {volume} {847}},\
  \bibinfo {eid} {36} (\bibinfo {year} {2017})},\ \Eprint
  {http://arxiv.org/abs/1708.06781} {arXiv:1708.06781 [astro-ph.SR]}
  \BibitemShut {NoStop}%
\bibitem [{\citenamefont {{Rutten}}(2019)}]{2019SoPh..294..165R}%
  \BibitemOpen
  \bibfield  {author} {\bibinfo {author} {\bibfnamefont {R.~J.}\ \bibnamefont
  {{Rutten}}},\ }\bibfield  {title} {\enquote {\bibinfo {title}
  {{Non-Equilibrium Spectrum Formation Affecting Solar Irradiance}},}\ }\href
  {\doibase 10.1007/s11207-019-1535-2} {\bibfield  {journal} {\bibinfo
  {journal} {\solphys}\ }\textbf {\bibinfo {volume} {294}},\ \bibinfo {eid}
  {165} (\bibinfo {year} {2019})},\ \Eprint {http://arxiv.org/abs/1908.04624}
  {arXiv:1908.04624 [astro-ph.SR]} \BibitemShut {NoStop}%
\end{thebibliography}%

\end{document}